\documentclass[11pt,a4paper]{article}
\usepackage[utf8]{inputenc}
\usepackage{authblk}
\usepackage{natbib}
\usepackage{threeparttable}
\RequirePackage{graphicx}
\usepackage[
	breaklinks=true,	
	colorlinks=true,	
	citecolor=blue, 
	linkcolor=blue,
	urlcolor=blue,	
	extension=pdf,		
	english]{hyperref}
\usepackage{amsmath}
\usepackage{amsfonts}
\usepackage{amssymb}
\usepackage{float}
\usepackage{booktabs}
\usepackage{multirow}
\usepackage{longtable}
\usepackage{threeparttablex}
\usepackage{doi}
\usepackage{setspace}
\usepackage{amsthm}
\usepackage{algorithm2e}
\usepackage{cleveref}
\usepackage{enumitem}

\newcommand{\convd}{\xrightarrow{D}}

\newcommand{\vect}[1]{\text{vec}(#1)}

\newcommand{\specnorm}[1]{\| #1 \|_{\text{sp}}}
\newcommand{\superref}[2]{\hyperref[#1]{#2}}
\newcommand{\frob}[1]{\|#1\|_F}
\newcommand{\bs}[1]{\boldsymbol{#1}}

\usepackage[a4paper, margin=2.2cm]{geometry}

\numberwithin{equation}{section}
\setcitestyle{authoryear,open={(},close={)}}

\newtheorem{Theorem}{Theorem}[section]
\crefname{Theorem}{Theorem}{Theorems}
\theoremstyle{plain}
\newtheorem{Corollary}[Theorem]{Corollary}
\crefname{Corollary}{Corollary}{Corollaries}
\theoremstyle{definition}
\newtheorem{Assumption}{Assumption}
\crefname{Assumption}{Assumption}{Assumptions}
\theoremstyle{remark} 
\newtheorem{Remark}{Remark}[section]
\theoremstyle{plain}
\newtheorem{Lemma}{Lemma}
\crefname{Lemma}{Lemma}{Lemmas}

\usepackage[dvipsnames,svgnames]{xcolor}

\title{\LARGE Estimation of Latent Group Structures in Time-Varying Panel Data Models}

\author[1]{\large Paul Haimerl\thanks{Correspondence to \href{mailto:paul.haimerl@econ.au.dk}{\texttt{paul.haimerl@econ.au.dk}}}}
\author[2]{\large Stephan Smeekes\thanks{\href{mailto:s.smeekes@maastrichtuniversity.nl}{\texttt{s.smeekes@maastrichtuniversity.nl}}}}
\author[2]{\large Ines Wilms\thanks{\href{mailto:i.wilms@maastrichtuniversity.nl}{\texttt{i.wilms@maastrichtuniversity.nl}} \\ 
We are grateful to Phillip Heiler, Eric Hillebrand, Morten Ø.\ Nielsen, and Martin Schumann for valuable feedback. We also thank the participants of the National Econometrics Study Group (NESG 2024), the Econometric Models of Climate Change conference (EMCC VIII), the Aarhus Workshop in Econometrics (AWE V), the Workshop in Time Series Econometrics (WTSE $XV_t$), the ZeST seminar at Bielefeld University, and the department seminar at Aarhus University for insightful comments. We thank Ali Mehrabani and Xia Wang for providing helpful code. Ines Wilms is supported by a grant from the Dutch Research Council (NWO), research program Vidi under the grant number VI.Vidi.211.032.}}

\affil[1]{\normalsize Department of Economics and Business Economics, CoRE Center for Research in Energy: Economics and Markets, Aarhus University, Universitetsbyen 51, 8000 Aarhus N, Denmark}
\affil[2]{\normalsize Department of Quantitative Economics, Maastricht University, Tongersestraat 53, 6211 ML Maastricht, The Netherlands}

\date{November, 2025}

\begin{document}
\maketitle
\pagenumbering{Roman}


\paragraph{Abstract.} We consider panel data models where coefficients change smoothly over time and follow a latent group structure, being homogeneous within but heterogeneous across groups. To jointly estimate the group membership and group-specific coefficient trajectories, we propose \textit{FUSE-TIME}, a pairwise adaptive group fused-Lasso estimator combined with polynomial spline sieves. We establish consistency, derive the asymptotic distributions of the penalized sieve estimator and its post-selection version, and show oracle efficiency. Monte Carlo experiments demonstrate strong finite-sample performance in terms of estimation accuracy and group identification. An application to the CO\textsubscript{2} intensity of GDP highlights the relevance of addressing both cross-sectional heterogeneity and time-variance in empirical exercises.




\pagenumbering{arabic}
\doublespacing


\section{Introduction}
\label{sec:Intro}

In this paper, we study panel data models in which slope coefficients evolve smoothly over time and vary across an unobserved group structure, being homogeneous within groups but heterogeneous across groups. We propose a penalized sieve approach that leverages the pairwise adaptive group fused-Lasso (\textit{PAGFL}; \citealp{qian2016shrinkage,qian2016shrinkageET}; \citealp{mehrabani2023}) to simultaneously estimate the time-varying coefficients and the latent grouping. We term our resulting estimator \textit{FUSE-TIME}--Fused Unobserved group Spline Estimation of TIME varying coefficients--and derive its asymptotic properties when the number of groups, the cross-sectional dimension, and the time dimension jointly diverge: establishing estimation consistency, clustering consistency, and deriving the asymptotic distribution.

This paper builds on and extends two strands of the literature. The first strand studies panel data models with latent group structures, where coefficients are homogeneous within groups but vary across groups. Prior studies recover the unobserved groupings either by applying clustering algorithms such as $K$-means to preliminary estimates or by directly embedding the grouping into the estimation procedure through penalization schemes \citep{bonhomme2015grouped,ke2015homogeneity,su2016identifying,mehrabani2023,lumsdaine2023estimation,mehrabani2025}. The second strand develops time-varying panel models, which describe temporal heterogeneity either through smoothly changing coefficients \citep{huang2004polynomial,cai2007trending,su2012sieve,robinson2012nonparametric} or through discrete structural breaks \citep{bai2010common,qian2016shrinkage}. 

We contribute to these strands by proposing a single convex optimization problem that jointly identifies the latent group structure and estimates the smooth group-specific coefficient trajectories. Our approach does not require any prior specification of the number of groups or group compositions. Instead, the number of groups and their compositions are determined automatically in a data-driven way through a penalized sieve estimation procedure that adds a \textit{PAGFL} penalty \citep{qian2016shrinkage,mehrabani2023} to the objective function. The penalty collects all pairwise differences of individual coefficient vectors, shrinking them to zero if two units belong to the same group. Following \citet{huang2004polynomial,su2019sieve}, we approximate the time-varying coefficients using polynomial basis splines (B-splines), providing a flexible yet parsimonious representation of smooth temporal change. We propose a post-selection variant of our estimator, and introduce a consistent BIC-type information criterion (IC) to select the tuning parameter that regulates the number of groups and their compositions. 

We provide primitive conditions for both the penalized sieve estimator and the post-selection variant of \textit{FUSE-TIME} to achieve the oracle property, being asymptotically equivalent to an infeasible oracle estimator based on the true grouping. Our theoretical framework allows the number of groups, the cross-sectional dimension, and the time dimension to diverge jointly when deriving the asymptotic distributions of the estimators. \textit{FUSE-TIME} thus provides a general framework for analyzing panels with both cross-sectional and temporal heterogeneity that is also robust to missing values and strikes a good balance between capturing heterogeneity and maintaining parsimony.

We also study the finite-sample performance of our estimator using both simulated and real data. Overall, our proposed estimator performs well across a variety of simulation settings in terms of recovering the latent group structure and estimating the time-varying coefficients accurately. \textit{FUSE-TIME} outperforms the method of \cite{su2019sieve}, which employs the Classifier Lasso (\textit{C-Lasso}; \citealp{su2016identifying}), while also being computationally more efficient. Additionally, we illustrate the merits of the \textit{FUSE-TIME} estimator by analyzing trends in the carbon dioxide (CO\textsubscript{2}) intensity of Gross Domestic Product (GDP), the amount of CO\textsubscript{2} emitted per unit of GDP produced. In a panel dataset of 92 countries spanning 1960 to 2023, the model identifies five groups, each with a unique trend of CO\textsubscript{2} intensity.

The plan for the remainder of this paper is as follows. Section \ref{sec:ModelandEstimation} introduces the model along with our estimation procedure. The asymptotic theory is developed in Section \ref{sec:Asymptotics}. Section \ref{sec:Simulation} investigates the finite-sample performance via a simulation study. Section \ref{sec:Empirical} presents the empirical illustration. The \hyperref[sec:Conclusion]{final section} concludes. We provide a software implementation in our companion R-package \texttt{PAGFL} \citep{haimerl2025pagfl}. Replication files of the simulation study and empirical application are available at \href{https://github.com/Paul-Haimerl/replication-tv-pagfl}{GitHub}.\footnote{R-Notebooks and a Dockerfile with replication material are published at \href{https://github.com/Paul-Haimerl/replication-tv-pagfl}{\texttt{github.com/Paul-Haimerl/replication-fuse-time}}.}

\textbf{Notation}. Throughout, we denote vectors by boldface small letters and matrices by boldface capital letters. For a real matrix $\bs{A}$, the Frobenius norm is written as $\frob{\bs{A}}=\sqrt{\text{tr}(\bs{A}\bs{A}^\prime)}$. $\mu_{\max}(\bs{A})$ and $\mu_{\min}(\bs{A})$ denote the largest and smallest eigenvalues, respectively. $\specnorm{\bs{A}}= \sqrt{\mu_{\max} (\bs{A}\bs{A}^\prime)}$ gives the spectral norm. $\vect{\bs{A}}$ describes a column-wise vectorization of $\bs{A}$. $\|\bs{a}\|_2$ denotes the Euclidean norm for a real vector $\bs{a}$. The operators $\overset{P}{\to}$, $\overset{D}{\to}$, and \text{plim} signal convergence in probability, convergence in distribution, and the probability limit. $\otimes$ represents the Kronecker product. The superscript zero, or index when appropriate, marks a true quantity. $\bs{I}_a$ and $\bs{0}_a$ denote an $a \times a$ identity matrix and an $a \times 1$ vector of zeros. `With probability approaching one' is abbreviated as w.p.a.1. For two sequences of positive (random) numbers $a_n$ and $b_n$, $a_n \lesssim b_n$ indicates that $a_n / b_n$ is (stochastically) bounded and $a_n \asymp b_n$ signals that both $a_n \lesssim b_n$ and $b_n \lesssim a_n$ hold.


\section{Model and Estimation}
\label{sec:ModelandEstimation}
\label{sec:pagfl}

In this section, we introduce the time-varying panel data model (Section \ref{sec:Model}) and our penalized sieve estimation procedure (Section \ref{sec:sieve}).


\subsection{The Model}
\label{sec:Model}
Consider the time-varying panel data model
\begin{equation}
    \label{eq:DGP}
    y_{it} = \gamma_i^0 + \bs{\beta}_{it}^{0 \prime} \bs{x}_{it} + \epsilon_{it}, \quad i = 1, \dots, N, \quad t = 1, \dots, T,
\end{equation}
where
$y_{it}$ is the scalar response, $\bs{x}_{it}$ is a $p \times 1$ vector of explanatory variables, the $\gamma_i^0$'s denote the unobserved individual fixed effects, and $\epsilon_{it}$ represents a zero-mean idiosyncratic error. The superscript 0 is used to denote a true parameter value. The fixed effects $\gamma_i^0$ may correlate with some elements of $\bs{x}_{it}$ and are independently distributed across individuals. We assume that the $p$-dimensional vector of slope coefficients $\bs{\beta}_{it}^0$ varies smoothly over time:
\begin{equation} \label{eq:beta-smooth}
    \bs{\beta}_{it}^0 = \bs{\beta}_{i}^0 \left( \frac{t}{T} \right).
\end{equation}
Furthermore, we assume that the time-varying slope parameters adhere to the following unknown group structure
\begin{equation}
    \label{eq:beta_group}
    \bs{\beta}_{i}^0 \left(\frac{t}{T}\right) = \sum_{k = 1}^{K_0} \bs{\alpha}_k^0 \left(\frac{t}{T}\right) \bs{1} \left\{i \in G_k^0 \right\},
\end{equation}
where $\bs{\alpha}_k^0 (t/T)$ is a $p \times 1$ vector of group-specific time-varying functional coefficients and $ \bs{1}\{\cdot\}$ denotes the indicator function. The latent group structure $\mathcal{G}_{K}^0=\{G_1^0, \dots, G_{K_0}^0\}$ partitions the cross-sectional units into disjoint sets; $\cup_{k=1}^{K_0} G_k^0 = \{1, \dots, N\}$
and $G_j^0 \cap G_k^0 = \emptyset$ for any $j \neq k$. We denote the cardinality of group $G_k^0$ as $N_k$.

\begin{Remark}
    The group-specific coefficients $\bs{\alpha}_k^0 (\cdot)$ are smooth functions of $t/T$ (see Assumption \hyperref[line:A1]{1(vi)}, discussed in Section \ref{sec:Asymptotics}). As a consequence, the data generating process (DGP) in \eqref{eq:DGP} captures smooth temporal variation rather than discrete structural breaks in the slope parameters. Furthermore, we assume discrete rather than continuous cross-sectional heterogeneity: unit-specific variation can be represented by a finite number of latent processes, namely the grouping in \eqref{eq:beta_group}. We refer to \citet{bonhomme2022discretizing} for a treatment of discrete groupings when cross-sectional heterogeneity is continuous. Our framework thus nests several previously studied models as special cases, including time-constant panel data models with latent group structures \citep{su2016identifying,sarafidis2015partially,wang2018homogeneity} and time-varying panel data models \citep{cai2007trending,robinson2012nonparametric,vogt2020multiscale}.
\end{Remark}


\subsection{Penalized Sieve Estimation of Time-Varying Coefficients}
\label{sec:sieve}

Similarly to \citet{huang2004polynomial,su2019sieve}, we approximate the $p \times 1$ functional coefficient vector $\bs{\beta}_{i}^0 (t/T)$ using an $M$-dimensional vector $\bs{b}(t/T)$ of polynomial spline basis functions:
\begin{equation}
    \label{eq:Splines}
    \bs{\beta}_i^0 \left( \frac{t}{T} \right) =  \bs{\Pi}^{0 \prime}_i \bs{b} \left( \frac{t}{T} \right) + \bs{\eta}_{it},
\end{equation}
where $\bs{\Pi}^0_i = (\bs{\pi}^0_{i1}, \dots, \bs{\pi}^0_{ip})$ denotes a $M \times p$ matrix of B-spline control points, and $\bs{\eta}_{it} = \bs{\beta}_i^0 (t/T) - \bs{\Pi}^{0 \prime}_i \bs{b} (t/T)$ is a $p \times 1$ sieve approximation error. B-splines offer two decisive advantages over typically employed kernel estimators. First, spline functions are computationally efficient and numerically stable while maintaining good approximation properties. Second, B-splines can be expressed as linear combinations of the basis functions $\bs{b}(t/T)$. The possibility of separating time-varying basis functions $\bs{b}(t/T)$, which do not have to be estimated, from individual time-constant control points $\bs{\Pi}^0_i$, which are estimated, makes B-splines particularly convenient to use in conjunction with the \textit{PAGFL} penalty.

Note that $\bs{\Pi}_i^{0 \prime} \bs{b}(t/T)$ is a function in the sieve-space $\mathbb{B}_M$ spanned by the $M$ basis functions in $\bs{b}(t/T)$. By increasing the polynomial degree $d$ (polynomial order $d+1$) and the number of interior knots $M^*$ as $T$ diverges, we extend $M = M^* + d + 1$ along with $\mathbb{B}_M$ and obtain arbitrarily close approximations of $\bs{\beta}_{i}^{0} (t/T)$.
Appendix \ref{sec:Appendix_sieve} contains full details on the spline basis functions.

Plugging \eqref{eq:Splines} into model \eqref{eq:DGP} yields 
\begin{equation}
    \label{eq:DGP_spline}
    \begin{split}
    y_{it} &= \gamma_i^0 + \left[\bs{\Pi}^{0 \prime}_i \bs{b} \left( \frac{t}{T} \right) + \bs{\eta}_{it} \right]^\prime \bs{x}_{it} + \epsilon_{it}\\
    y_{it} &= \gamma_i^0 + \bs{\pi}_i^{0 \prime} \bs{z}_{it} + u_{it},
    \end{split}
\end{equation}
where $\bs{\pi}_i^0 = \text{vec} (\bs{\Pi}^{0}_i)$ is a $Mp \times 1$ coefficient vector, $\bs{z}_{it} = \bs{x}_{it} \otimes \bs{b}(t/T)$ is a $Mp \times 1$ vector of regressors, and $u_{it}$ collects the idiosyncratic error $\epsilon_{it}$ and the sieve approximation error $\bs{\eta}_{it}^\prime \bs{x}_{it}$. To obtain an estimate of the time-varying slope parameters $\bs{\beta}_{i}^0 (t/T)$ in model \eqref{eq:DGP}, we estimate the time-constant vector $\bs{\pi}_i^{0}$ in model \eqref{eq:DGP_spline} and take $\hat{\bs{\beta}}_{i} (t/T) = \hat{\bs{\Pi}}_i^\prime \bs{b}(t/T)$, where $\hat{\bs{\pi}}_i = \vect{\hat{\bs{\Pi}}_i}$.

We jointly identify the latent group structure in \eqref{eq:beta_group} and the time-varying coefficients $\bs{\beta}_{i}(t/T)$ through {a \textit{penalized} sieve technique}, leveraging the \textit{PAGFL} penalty \citep{qian2016shrinkage,mehrabani2023}. In what follows, we introduce this estimation procedure in the context of the time-varying panel data model \eqref{eq:DGP}. Beyond this baseline, our proposal can readily accommodate extensions, as discussed in Appendix \ref{sec:Extensions}.

First, we concentrate out the individual fixed effects $\gamma_i^0$ in \eqref{eq:DGP_spline} using $\tilde{y}_{it} = y_{it} - T^{-1} \sum_{t = 1}^{T} y_{it}$, with $\tilde{\bs{z}}_{it}$ 
defined analogously. Let $\bs{\pi} = (\bs{\pi}_1^\prime, \dots, \bs{\pi}_N^\prime)^\prime$ collect all individual parameters. We then take as objective function
\begin{equation}
    \label{eq:Obj_penalty}
    \mathcal{F}_{NT}(\bs{\pi}, \lambda) = \frac{1}{T} \sum_{i = 1}^{N} \sum_{t=1}^{T} \left(\tilde{y}_{it} - \bs{\pi}_i^\prime \tilde{\bs{z}}_{it} \right)^2 + \frac{\lambda}{N} \sum_{i = 1}^{N - 1} \sum_{j=i+1}^{N} \dot{\omega}_{ij} \left\| \bs{\pi}_i - \bs{\pi}_j \right\|_2,
\end{equation}
where the first part is the usual sum of squared residuals and the second part is the \textit{PAGFL} penalty term with tuning parameter $\lambda>0$ and adaptive penalty weights $\dot{\omega}_{ij}$. The penalty encourages sparsity in the differences of all $N(N-1)/2$ pairs of individual coefficient vectors. For large $\lambda$, some of these differences will be shrunken to exactly zero, implying that the corresponding cross-sectional units feature identical parameter estimates. The weights are set to $\dot{\omega}_{ij} = \| \dot{\bs{\pi}}_i - \dot{\bs{\pi}}_j \|_2^{-\kappa}$, where $\dot{\bs{\pi}}_i$ represents an initial consistent estimate $\dot{\bs{\pi}} = \arg \min_{\pi} T^{-1} \sum_{i = 1}^{N} \sum_{t=1}^{T} \left(\tilde{y}_{it} - \bs{\pi}_i^\prime \tilde{\bs{z}}_{it} \right)^2$ and $\kappa$ is specified by the user; for simplicity and in line with the standard choice in the adaptive Lasso literature, we maintain $\kappa=2$ throughout the paper.

Objective \eqref{eq:Obj_penalty} generalizes the objective function in \citet[eq. 2.6]{mehrabani2023} to our \textit{FUSE-TIME} approach. The penalized sieve estimator (\textit{PSE}) $\hat{\bs{\pi}}$ of \textit{FUSE-TIME} is then obtained by minimizing \eqref{eq:Obj_penalty}, namely
$$
    \hat{\bs{\pi}}_\lambda = \arg \min_{\pi} \mathcal{F}_{NT}(\bs{\pi}, \lambda).
$$
Cross-sectional units with identical slope estimates are assigned to the same 
cluster by collecting all $\hat{K}$ unique subvectors $\hat{\bs{\pi}}_i$ of $\hat{\bs{\pi}}_\lambda$ in the vector $\hat{\bs{\xi}} = (\hat{\bs{\xi}}_1^\prime, \dots, \hat{\bs{\xi}}_{\hat{K}}^\prime)^\prime$ and defining the set $\hat{G}_k = \{ i: \hat{\bs{\pi}}_i = \hat{\bs{\xi}}_k, \; 1 \leq i \leq N \}$ for each $k = 1, \dots, \hat{K}$. Subsequently, the group-specific \textit{PSE} coefficient function equals $\hat{\bs{\alpha}}_{k} (t/T) = \hat{\bs{\Xi}}^\prime_k \bs{b}(t/T)$, where $\hat{\bs{\xi}}_k = \vect{\hat{\bs{\Xi}}_k}$. As such, parameter estimation and identification of the latent group structure are performed simultaneously. The total number of clusters $\hat{K}$ is determined by the tuning parameter $\lambda$. We propose to select $\lambda$ using the consistent information criterion (IC) introduced in Section \ref{sec:Asymptotics}. In the following, unless required, we suppress the dependence of $\hat{\bs{\pi}}$ on $\lambda$ to lighten the notation.

\begin{Remark}
    The penalty term in criterion \eqref{eq:Obj_penalty} is time-invariant and, therefore, also the group structure. Nevertheless, cross-sectional units switching groups do not lead to inconsistent estimates of the time-varying coefficients but only to
    a larger number of groups $\hat{K}$, since each distinct combination of the time of the switch, origin group, and destination group implies one excess group. Alternatively, one can define the penalty in \eqref{eq:Obj_penalty} over each of the $M$ rows in
    $\bs{\Pi}_i$ individually, implying groupings specific to each basis function and thus allowing for time-varying group membership. Such an extension is further discussed in Appendix \ref{sec:Extension_coefgroups}.
\end{Remark}

Objective function \eqref{eq:Obj_penalty} is convex in $\bs{\pi}$. To solve for $\bs{\pi}$, we employ a computationally efficient Alternating Direction Method of Multipliers (\textit{ADMM}) algorithm, adapted from \citet[sec. 5]{mehrabani2023} and detailed in Appendix \ref{sec:Appendix_Algo}. An open-source implementation of this algorithm is provided in our companion R-package \texttt{PAGFL} \citep{haimerl2025pagfl}.

\begin{Remark}
    \textit{FUSE-TIME} shows similarities with the proposal in \citet{su2019sieve}. However, they use a different approach to identify the latent group structure, namely the \textit{C-Lasso} of \citet{su2016identifying}. The \textit{C-Lasso} shrinks $N$ individual coefficients towards $K$ group-level coefficients and thus requires an additional tuning parameter to explicitly determine $K$. Additionally, the \textit{C-Lasso} objective function is not convex, though decomposable into convex subproblems. Objective function \eqref{eq:Obj_penalty}, in contrast, is convex. 
\end{Remark}

Finally, to mitigate the bias coming from the fusion penalty term in \eqref{eq:Obj_penalty}, we propose a post-selection fused-Lasso estimator of \textit{FUSE-TIME}, labeled \textit{post-Lasso} estimator for brevity, given the estimated group pattern $\hat{\mathcal{G}} = \{\hat{G_1}, \dots, \hat{G}_{\hat{K}} \}$:
\begin{equation*}
    \label{eq:post-Lasso}
    \hat{\bs{\xi}}^p_{\hat{G}_k} = \left( \sum_{i \in \hat{G}_k} \sum_{t = 1}^{T} \tilde{\bs{z}}_{it} \tilde{\bs{z}}_{it}^\prime \right)^{-1} \sum_{i \in \hat{G}_k} \sum_{t = 1}^{T} \tilde{\bs{z}}_{it} \tilde{y}_{it}, \quad k = 1, \dots, \hat{K}.
\end{equation*}
Subsequently, the final \textit{post-Lasso} time-varying coefficient estimates follow as $\hat{\bs{\alpha}}^p_{\hat{G}_k } (t/T) = \hat{\bs{\Xi}}^{p\prime}_{\hat{G}_k} \bs{b}(t/T)$, using $\hat{\bs{\xi}}^{p}_{\hat{G}_k} = \vect{\hat{\bs{\Xi}}^{p}_{\hat{G}_k}}$.


\section{Asymptotic Properties}
\label{sec:Asymptotics}

In this section, we study the consistency of the coefficient estimates and the clustering procedure (Sections \ref{subsec:theor:prelim} and \ref{sec:Classification}). Furthermore, we establish the limiting distribution of the \textit{PSE} and the \textit{post-Lasso} estimator (Section \ref{sec:Limit_dist}),  and we propose a consistent BIC criterion to select the tuning parameter (Section \ref{sec:lambda}). Formal proofs appear in Appendix \ref{sec:Proofs}.

\subsection{Preliminary Convergence Rates} \label{subsec:theor:prelim}

Let $\mathbb{C}^{\theta}[0,1]$ denote the space of functions that are $\theta$-times continuously differentiable on the unit interval. Moreover, let $\bs{x}_{it}^{(2)} = \bs{x}_{it}$ if no intercept is included in $\bs{x}_{it}$ and $\bs{x}_{it} = (1, \bs{x}_{it}^{(2)\prime})^\prime$  otherwise.

\phantomsection
\label{line:A1}
\begin{Assumption}
\label{ass:A1}
    \begin{enumerate}[label=(\roman*)]
        \item[]
        \item\label{ass:A1_1} There exists a positive constant $ c_{\epsilon \epsilon} < \infty$ such that $N^{-1} \sum_{i=1}^{N} \sum_{j=1}^{N} \left| \max_{t} E ( \epsilon_{it} \epsilon_{jt} ) \right| \leq c_{\epsilon \epsilon}$.\footnote{We abbreviate $\max_{1 \leq i \leq N}$, $\max_{1 \leq t \leq T}$, and $\max_{1 \leq i \leq N, 1 \leq t \leq T}$ with $\max_{i}$, $\max_{t}$, and $\max_{i,t}$, respectively. $\min_{i}$, $\min_{t}$, and $\min_{i,t}$ are defined likewise.}
        \item\label{ass:A1_2} $\left\{ (\bs{x}_{it}^{(2)}, \epsilon_{it}), \; t = 1, \dots, T \right\}$ is a strong mixing process with geometric decay such that the mixing coefficient $\phi(j)$ satisfies $\phi(j) \leq c_\phi \varphi^j$ for some $c_\phi < \infty$, $0 \leq \varphi < 1$, and each $i = 1, \dots, N$.
        \item\label{ass:A1_3} There exist two positive constants $c_x < \infty$ and $c_\epsilon < \infty$ such that $\max_{i,t} E \| \bs{x}_{it} \|^q_2 \leq c_x$ and $\max_{i,t} E | \epsilon_{it} |^q \leq c_\epsilon$ for some $q \geq 6$, when $\bs{x}_{it} \neq 1$.
        \item\label{ass:A1_4} There exist a lower and upper bound $0 < \underbar{c}_{x x} \leq \bar{c}_{x x} < \infty$ such that $\underbar{c}_{x x} \leq \min_{i,t} \mu_{\min} \left( Var(\bs{x}_{it}^{(2)}) \right) \leq \max_{i,t} \mu_{\max} \left( E(\bs{x}_{it}^{(2)} \bs{x}_{it}^{(2) \prime}) \right) \leq \bar{c}_{x x}$.
        \item\label{ass:A1_5} $\lim_{N \to \infty} N_k/N \in [0, 1)$ for each $k = 1, \dots, K_0$ and $N_k > 1$ for all $k \in \{k : \lim_{N \to \infty} N_k/N = 0, 1 \leq k \leq K_0\}$.
        \item\label{ass:A1_6} $\bs{\alpha}_k^0(t/T) \in \mathbb{C}^{\theta} [0,1]$ for each $k = 1, \dots, K_0$ and some $1 < \theta \leq d+2$.
    \end{enumerate}
\end{Assumption}

\cref{ass:A1}\ref{ass:A1_1} is standard in the factor model literature and limits cross-sectional dependence \citep[cf.][]{bai2002determining}. This condition is trivially satisfied if the idiosyncratic errors are independently distributed, an assumption frequently made for similar panel data models with latent group structures \citep[cf.][]{su2016identifying,qian2016shrinkage,su2019sieve}. 

\cref{ass:A1}\ref{ass:A1_2}-\ref{ass:A1_4} loosely follows \citet[Assumption 1(ii)-(iv)]{su2019sieve}. \cref{ass:A1}\ref{ass:A1_2} imposes a strong mixing process, which nests popular and extensively studied time series processes with geometrically decaying innovations, such as common ARMA, GARCH, Markov-switching, or threshold autoregressive models. Serial correlation as well as conditional heteroscedasticity in the error process are allowed. As \citet{su2019sieve} point out, in dynamic panels Assumption \hyperref[line:A1]{1(ii)} implies the fixed effects $\gamma_i$ to be nonrandom. If the fixed effects are treated as random variables, the strong mixing condition in Assumption \hyperref[line:A1]{1(ii)} must hold conditional on each realized $\gamma_i$, with the distribution of $\gamma_i$ being bounded uniformly in $i$ \citep[cf.][Condition 3]{hahn2011bias}.

Assumptions \hyperref[line:A1]{1(i)-(ii)} can also be replaced by similar primitive \citep[][Assumption 1(d)-(f)]{bonhomme2015grouped} or higher-level assumptions \citep[][Assumption A.1(i)-(iii)]{mehrabani2023} that limit the degree of dependence across time and the cross-section such that a central limit theorem can be applied.

Assumptions \hyperref[line:A1]{1(iii)-(iv)} place common moment conditions on the regressors and innovations. The conditions on the regressors in Assumptions \hyperref[line:A1]{1(iii)-(iv)} are redundant if $\bs{x}_{it}$ is deterministic. Assumption \hyperref[line:A1]{1(v)} allows group sizes to remain constant or to diverge at a speed slower than $N$. Assumption \hyperref[line:A1]{1(vi)} imposes smoothness on the true functional coefficients and ensures that they can be well approximated by splines of polynomial degree $d$.

Let $J_{\min}$ denote the minimum group separability in the sieve-space $\mathbb{B}_M$, $J_{\min} = \min_{i \in G_k^0, j \notin G_k^0} \| \bs{\pi}_i^0 - \bs{\pi}_j^0 \|_2$.

\phantomsection
\label{line:A2}
\begin{Assumption}
\label{ass:A2}
    \begin{enumerate}[label=(\roman*)]
        \item[]
        \item $\lim_{T \to \infty} M T^{-1/2} = 0$.
        \item $\lim_{T \to \infty} (MT^{-1/2} + M^{-\theta+1/2})^{-1} J_{\min} = \infty$.
        \item $\text{plim}_{T \to \infty} (T^{-1/2} + M^{-\theta - 1/2})^{-1} \lambda J_{\min}^{-\kappa} = c_\lambda$ for some constant $c_\lambda > 0$.
        \item[(iv)] $\text{plim}_{(N,T) \to \infty} (MT^{-1/2} + M^{-\theta + 1/2})^{-\kappa-1} M^{1/2} \lambda N_k/N = \infty$ for each $k = 1, \dots, K_0$.
    \end{enumerate}
\end{Assumption}
Assumption \hyperref[line:A1]{2(i)} controls the size of the spline basis system. Assumption \hyperref[line:A2]{2(ii)} determines the rate at which the minimum group separability in $\mathbb{B}_{M}$ may shrink to zero. Assumption \hyperref[line:A2]{2(iii)} governs the speed at which the tuning parameter $\lambda$ must shrink to zero. Assumption \hyperref[line:A2]{2(iv)} places conditions on the relative rates of the number of coefficients to be estimated, coefficient convergence, and $\lambda$ so that group-specific trajectories can still be consistently estimated. In sum, $N$, $T$, $M$, and $K_0$ diverge to infinity, whereas $\lambda$ and $J_{\min}$ tend to zero in the limit.

\begin{Theorem}
    \label{sec:Theo_1}
    Given that Assumptions \hyperref[line:A1]{1} and \hyperref[line:A2]{2(i)-(iii)} are satisfied, for $i = 1, \dots, N$, we have
    \begin{enumerate}[label=(\roman*)]
        \item\label{sec:Theo_1_1} $\| \hat{\bs{\pi}}_i - \bs{\pi}_i^0 \|_2 = O_p(MT^{- 1/2} + M^{-\theta + 1/2})$,
        \item\label{sec:Theo_1_2} $N^{-1} \sum_{i = 1}^{N} \| \hat{\bs{\pi}}_i - \bs{\pi}_i^0 \|_2^2 = O_p(M^2 T^{-1} + M^{-2\theta +1})$.
    \end{enumerate}
\end{Theorem}

Theorem \hyperref[sec:Theo_1]{3.1} establishes pointwise and mean-square convergence of $\hat{\bs{\pi}}_i$. The first term in the rates of Theorem \hyperref[sec:Theo_1]{3.1} reflects the stochastic error. The second term corresponds to the asymptotic bias of the sieve technique. Increasing the complexity of the spline system $M$ involves a bias-variance trade-off. On the one hand, the larger $M$, the slower the convergence due to the increased size of the coefficient vector.\footnote{Note that $O_p(MT^{-1/2} + M^{- \theta + 1/2}) = o_p(1)$ by Assumptions \hyperref[line:A2]{1(vi)} and \hyperref[line:A2]{2(i)}.} On the other, increasing $M$ reduces the asymptotic sieve bias. Moreover, the greater the order of continuous differentiability of the true coefficient functions $\theta$, the faster the sieve bias shrinks in $M$.

\begin{Corollary}
    \label{sec:Coro_1}
    Given that Assumptions \hyperref[line:A1]{1} and \hyperref[line:A2]{2(i)-(iii)} are satisfied, for $i = 1, \dots, N$, we have
    \begin{enumerate}[label=(\roman*)]
        \item\label{sec:Coro_1_1} $\sup_{v \in [0,1]} \| \hat{\bs{\beta}}_i (v) - \bs{\beta}_i^0 (v) \|_2 = O_p(MT^{- 1/2} + M^{-\theta + 1/2})$, 
        \item\label{sec:Coro_1_2} $\int_{0}^{1} \| \hat{\bs{\beta}}_i (v) - \bs{\beta}_i^0 (v) \|_2^2 \,dv = O_p(M T^{-1} + M^{-2\theta}) $.
    \end{enumerate}
\end{Corollary}

Corollary \hyperref[sec:Coro_1]{3.2} relates the results of Theorem \hyperref[sec:Theo_1]{3.1} to the actual functional coefficients. Note that the true coefficient functions are time-continuous. As a consequence, we report the supremum and integral over the unit interval. Interestingly, the pointwise rates of $\hat{\bs{\pi}}_i$ and $\hat{\bs{\beta}}_i(v)$ match, whereas the $L_2$ rate of $\hat{\bs{\beta}}_i(v)$ is faster in $M$. This result is due to the boundedness of the B-splines (see \hyperref[sec:Lemma_statements]{Lemma A.1(ii)}).

\subsection{Clustering Consistency}
\label{sec:Classification}

This subsection establishes clustering consistency.

\begin{Theorem}
    \label{sec:Theo_2}
    Given that Assumptions \hyperref[line:A1]{1} and \hyperref[line:A2]{2} are satisfied,
    $\Pr (\| \hat{\bs{\pi}}_i - \hat{\bs{\pi}}_j \|_2 = 0 \; \forall \, i,j \in G_k^0, \; 1 \leq k \leq K_0 ) \to 1$, as $(N,T) \to \infty$.
\end{Theorem}

Theorem \hyperref[sec:Theo_2]{3.3} establishes that, in the limit, all cross-sectional units belonging to the same group $G_k^0$ are jointly assigned to the same group; no other units are assigned to this group.

\begin{Corollary}
    \label{sec:Coro_2}
    Given that Assumptions \hyperref[line:A1]{1} and \hyperref[line:A2]{2} are satisfied,
    \begin{enumerate}[label=(\roman*)]
        \item\label{sec:Coro_2_1} $\lim_{(N,T) \to \infty} \Pr \left( \hat{K} = K_0 \right) = 1$,
        \item\label{sec:Coro_2_2} $\lim_{(N,T) \to \infty} \Pr \left( \hat{\mathcal{G}}_{K_0} = \mathcal{G}^0_{K_0} \right) = 1$.
    \end{enumerate}
\end{Corollary}

Based on Theorem \hyperref[sec:Theo_2]{3.3}, Corollary \hyperref[sec:Coro_2]{3.4} shows that the correct number of groups and group structure will be derived asymptotically. This result is intuitive since, given Theorem \hyperref[sec:Theo_2]{3.3}, $\hat{\bs{\pi}}$ can only hold $K_0$ distinct individual subvectors and all homogeneous individuals are assigned to the same group as $(N,T) \to \infty$. Since the latent grouping will be identified w.p.a.1, Corollary \hyperref[sec:Coro_2]{3.4} motivates the oracle property of the procedure. This implies that the estimation procedure is asymptotically equivalent to an infeasible oracle estimator based on the true grouping.

\subsection{Limiting Distribution of the \textit{PSE} and \textit{Post-Lasso} Estimators}
\label{sec:Limit_dist}

In the following, we derive the asymptotic distribution of the \textit{PSE} and the \textit{post-Lasso}.

\phantomsection
\label{line:A3}
\begin{Assumption}
\label{ass:A3}
    $ \lim_{(N,T) \to \infty} (N_k T M)^{1/2} \lambda J_{\min}^{-\kappa} = 0$ for each $k = 1, \dots, K_0$.
\end{Assumption}

The penalty term in \eqref{eq:Obj_penalty} grows quadratically in $N$. Hence, Assumption \hyperref[line:A3]{3} strengthens Assumption \hyperref[line:A2]{2(iii)} and imposes conditions for the penalty term to vanish asymptotically, resulting in coinciding limiting distributions for the \textit{PSE} and the \textit{post-Lasso}. To this end, Assumption \hyperref[line:A3]{3} specifies a larger group separation or a faster rate at which $\lambda$ converges to zero. Assumption \hyperref[line:A3]{3} is only relevant for the \textit{PSE} and need not hold for the asymptotic properties of the \textit{post-Lasso}.

Let $\bs{\epsilon}_i = (\epsilon_{i1}, \dots, \epsilon_{iT})^\prime$.
\phantomsection
\label{line:A4}
\begin{Assumption}
\label{ass:A4}
    \begin{description}
        \item[]
        \item[(i)] There exists a positive constant $\bar{c}_{\epsilon \epsilon} < \infty$ such that $\lim_{(N,T) \to \infty} \max_{i \in G_k^0} \mu_{max} \left( E(\bs{\epsilon}_i \bs{\epsilon}_i^\prime) \right) \leq \bar{c}_{\epsilon \epsilon}$ for each $k = 1, \dots, K_0$.
        \item[(ii)] $\lim_{(N,T)\to \infty} NT M^{-2\theta} = 0$.
    \end{description}
\end{Assumption}

Assumption \hyperref[line:A4]{4(i)} imposes a mild restriction on the error process, enabling the use of the Lindeberg condition to derive the limiting distribution in Theorem \hyperref[sec:Theo_3]{3.5}. Assumption \hyperref[line:A4]{4(ii)} provides an additional regularity condition that precludes the sieve approximation error from dominating the limiting distributions of the \textit{PSE} and the \textit{post-Lasso}. Let $\hat{\bs{\mathcal{Q}}}_{G_k^0, \tilde{z} \tilde{z}} = \sum_{i \in G_k^0} \hat{\bs{Q}}_{i, \tilde{z} \tilde{z}}$ with $\hat{\bs{Q}}_{i, \tilde{z} \tilde{z}} = T^{-1} \sum_{t = 1}^{T} \tilde{\bs{z}}_{it} \tilde{\bs{z}}_{it}^\prime$ and $\tilde{\bs{Z}}_i = (\tilde{\bs{z}}_{i1}, \dots, \tilde{\bs{z}}_{iT})$. Furthermore, define the variance-covariance matrix $\hat{\bs{\Omega}}_{G_k^0} = \hat{\bs{\nu}}_{G_k^0}^{\prime} \hat{\bs{\mathcal{E}}}_{G_k^0} \hat{\bs{\nu}}_{G_k^0}$ with
$$
    \hat{\bs{\nu}}_{G_k^0} = \left( M N_k^{-1} \hat{\bs{\mathcal{Q}}}_{G_k^0, \tilde{z} \tilde{z}} \right)^{-1} \left( \bs{I}_p \otimes \bs{b} (v) \right),
$$
$$
    \hat{\bs{\mathcal{E}}}_{G_k^0} = \frac{M}{N_k T} \sum_{i \in G_k^0} \tilde{\bs{Z}}_{i}^\prime E(\bs{\epsilon}_i \bs{\epsilon}_i^\prime) \tilde{\bs{Z}}_{i},
$$
and $\bs{q}_{G_k^0} = \sqrt{M / (N_k T)} \sum_{i \in G_k^0} \sum_{t = 1}^{T} E(\tilde{\bs{z}}_{it} \tilde{\epsilon}_{it})$.

\begin{Theorem}
    \label{sec:Theo_3}
    Given that Assumptions \hyperref[line:A1]{1}-\hyperref[line:A4]{4} are satisfied,
    \begin{enumerate}[label=(\roman*)]
        \item\label{sec:Theo_3_1} $ \sqrt{N_k T / M} \hat{\bs{\Omega}}_{G_k^0}^{-1/2} \left(\hat{\bs{\alpha}}_{k} (t/T) - \bs{\alpha}^0_k (t/T) \right) - \hat{\bs{\mathcal{E}}}_{G_k^0}^{-1/2} \bs{q}_{G_k^0} \convd N \left( 0, \bs{I}_p \right)$,
        \item\label{sec:Theo_3_2} $ \sqrt{N_k T / M} \hat{\bs{\Omega}}_{G_k^0}^{-1/2} (\hat{\bs{\alpha}}^p_{\hat{G}_k} (t/T) - \bs{\alpha}^0_k (t/T) ) - \hat{\bs{\mathcal{E}}}_{G_k^0}^{-1/2} \bs{q}_{G_k^0} \convd N \left( 0, \bs{I}_p \right)$.
    \end{enumerate}
\end{Theorem}

$\bs{q}_{G_k^0}$ equals zero in the case of strictly exogenous regressors. However, commonly referred to as the Nickell bias, $\bs{q}_{G_k^0}$ is nonzero and of order $O \left( \sqrt{N_k / T} \right)$ for dynamic panel data models \citep{nickell1981biases,phillips2007bias}. The bias emerges when $T$ remains fixed or grows slower than $N_k$. The within-transformation nets out the fixed effect $\gamma_i$ but simultaneously induces a contemporaneous correlation between the error term and the autoregressive regressor, thus biasing the coefficient estimate if $T$ does not grow fast enough relative to the number of individuals that feed into the group-specific autoregressive coefficient function \citep{nickell1981biases,kiviet1995bias,hahn2011bias,su2016identifying}.

The \textit{PSE} and the \textit{post-Lasso} estimators are asymptotically equivalent to the infeasible oracle estimator with true grouping structure, hence both achieve oracle efficiency. Additionally, the B-splines yield arbitrarily close approximation of the true coefficient functions $\bs{\alpha}_k^0 (t/T)$. Given Assumption \hyperref[line:A3]{3}, both estimators feature the same asymptotic distribution. Nevertheless, despite their equivalence in the limit, we recommend using the \textit{post-Lasso} in finite-sample applications. The penalty term of the \textit{PSE} can lead to non-negligible bias, particularly in small samples. The simulation study in Section \ref{sec:Simulation} corroborates this finding.

Given a consistent estimator $\hat{\bs{\mathcal{E}}}_{\hat{G}_k}$ and exploiting the oracle property in Corollary \hyperref[sec:Coro_2]{3.4}, the variance of the limiting distribution can be estimated consistently. Potential techniques to derive $\hat{\bs{\mathcal{E}}}_{\hat{G}_k}$ are numerous. We follow the literature on heteroscedasticity and autocorrelation robust estimation of covariance matrices and take
\begin{equation*}
    \label{eq:V_estimation}
    \hat{\bs{\mathcal{E}}}_{\hat{G}_k} = \hat{\bs{\mathcal{E}}}_{\hat{G}_k}^{(0)} + \sum_{h = 1}^{H_T} w(h, H_T) \left( \hat{\bs{\mathcal{E}}}_{\hat{G}_k}^{(h)} + \hat{\bs{\mathcal{E}}}_{\hat{G}_k}^{(h) \prime} \right),
\end{equation*}
where $H_T$ is the window size, $\hat{\bs{\mathcal{E}}}_{\hat{G}_k}^{(h)} = N_k^{-1} \sum_{i \in \hat{G}_k} T^{-1} \sum_{t = h + 1}^{T} \bs{\tilde{z}}_{it} \tilde{z}_{it-1}^\prime \hat{\epsilon}_{it} \hat{\epsilon}_{i t-1}$, and $\hat{\epsilon}_{it} = \tilde{y}_{it} - \hat{\bs{\alpha}}_{\hat{G}_k}^{p \prime} (t/T) \tilde{\bs{x}}_{it}$ \citep{newey1987simple,hahn2002asymptotically,pesaran2006estimation,muller2014hac}. $w(h, H_T)$ denotes a weighting function subject to $\sup_{h} | w(h, H_T) | < \infty$ and $\lim_{T \to \infty} w(h, H_T) = 1$. It is common to specify $w(h,H_T) = 1 - h / (H_T + 1) \bs{1} \{ h \leq H_T \}$ \citep{newey1987simple}. A rigorous derivation of the consistency of $\hat{\bs{\mathcal{E}}}_{\hat{G}_k}$ and required primitive conditions are beyond the scope of this paper and are left for future work \citep[see, e.g.,][Theorem 4.3 for further reference]{su2012sieve}.

\begin{Remark}
It is important to highlight that the limiting distributions in \cref{sec:Theo_3} hold conditional on the true group structure, rather than uniformly for all possible groupings. This limitation is ubiquitous in the literature on latent-class panel data models that involve a model selection step \citep{leeb2008sparse}.
\end{Remark}

\subsection{Selecting the Tuning Parameter $\lambda$}
\label{sec:lambda}

In the remainder of this section, we make the dependence of $\hat{K}_\lambda$ and $\hat{G}_{k,\lambda}$ on $\lambda$ explicit. We choose the tuning parameter $\lambda$ that minimizes the following BIC-type criterion 
\begin{equation}
    \label{eq:IC}
    IC(\lambda) = \ln \left( \hat{\sigma}^2_{\hat{\mathcal{G}}_{\hat{K}, \lambda}} \right) + \rho_{NT} p M \hat{K}_{\lambda},
\end{equation}
where $\hat{\sigma}^2_{\hat{\mathcal{G}}_{\hat{K}, \lambda}} = (NT)^{-1} \sum_{k = 1}^{\hat{K}_{\lambda}} \sum_{i \in \hat{G}_{k, \lambda}} \sum_{t = 1}^{T} \left(\tilde{y}_{it} - \hat{\bs{\alpha}}_{\hat{G}_{k, \lambda}}^{p \prime} (t/T) \tilde{\bs{x}}_{it} \right)^2$ and $\rho_{NT}$ represents a tuning parameter \citep[see][for similar approaches]{qian2016shrinkage,su2019sieve,mehrabani2023}. Define the set $\Lambda = [0, \lambda_{\max}]$ for some sufficiently large $\lambda_{\max} < \infty$ and partition $\Lambda$ into $\Lambda_{0,NT}$, $\Lambda_{-, NT}$, and $\Lambda_{+, NT}$, such that $\Lambda_{0,NT} = \{ \lambda \in \Lambda: \hat{K}_{\lambda} = K_0 \}$, $\Lambda_{-,NT} = \{ \lambda \in \Lambda: \hat{K}_{\lambda} < K_0 \}$, and $\Lambda_{+,NT} = \{ \lambda \in \Lambda: \hat{K}_{\lambda} > K_0 \}$.\footnote{We index the three subsets of $\Lambda$ with $NT$ to make it apparent that $\hat{K}$ is obtained from a random sample of dimension $(N,T)$.} In addition, assume that all $\lambda \in \Lambda_0$ comply with the regularity conditions stated in Assumptions \hyperref[line:A2]{2}-\hyperref[line:A3]{3}. Denote the set of all $K$-partitions of $\{1, \dots, N\}$ as $\mathbb{G}$.

\phantomsection
\label{line:A5}
\begin{Assumption}
\label{ass:A5}
    $\text{plim}_{(N,T) \to \infty} \min_{1 \leq K < K_0} \inf_{\mathcal{G}_{K} \in \mathbb{G}} \hat{\sigma}^2_{\mathcal{G}_{K}} = \underline{\sigma}^2 > \sigma_0^2$.
\end{Assumption}
Assumption \hyperref[line:A5]{5} is standard in the literature and implies that the mean squared error (\textit{MSE}) of an underfitted model is asymptotically larger than $\sigma_0^2$, the \textit{MSE} of the true model.

\phantomsection
\label{line:A6}
\begin{Assumption}
\label{ass:A6}
    \begin{description}
        \item[]
        \item[(i)] $\lim_{(N,T) \to \infty} \rho_{NT} M K_0 = 0$.
        \item[(ii)] $\lim_{(N,T) \to \infty} T \rho_{NT} = \infty$.
    \end{description}
\end{Assumption}
Assumption \hyperref[line:A6]{6} collects conditions such that the penalty term in \eqref{eq:IC} either dominates the \textit{MSE} or vanishes in the limit, depending on whether $\hat{K} > K_0$ or $\hat{K} \leq K_0$, respectively.

\begin{Theorem}
    \label{sec:Theo_4}
    Given that Assumptions \hyperref[line:A1]{1}-\hyperref[line:A6]{6} are satisfied, \\ $\Pr \left(\inf_{\lambda \in \Lambda_{-,NT} \cup \Lambda_{+,NT}} IC(\lambda) > \sup_{\lambda \in \Lambda_{0,NT}} IC(\lambda) \right) \to 1$ as $(N,T) \to \infty$.
\end{Theorem}
From Theorem \hyperref[sec:Theo_4]{3.6} follows that neither an underfitted nor an overfitted model maximizes the IC as $N$ and $T$ diverge to infinity. Consequently, the IC in \eqref{eq:IC} uncovers the true model asymptotically. 

When applying the IC, a practitioner must select $\rho_{NT}$. After some preliminary experiments and in line with previous literature, we recommend specifying $\rho_{NT} = c_\lambda \log(NT) (NT)^{-1/2}$ with $c_\lambda = 0.04$ \citep[cf.][sec. 6]{mehrabani2023}.


\section{Monte Carlo Simulation Study}
\label{sec:Simulation}

This section examines the finite-sample properties of \textit{FUSE-TIME} using several Monte Carlo simulation studies. Section \ref{sec:Simulation_DGP} describes the DGPs, followed by implementation details and evaluation metrics in Section \ref{sec:Simulation_implementation}. In Section \ref{sec:Simulation_results}, we present the results.

\subsection{The Data Generating Processes}
\label{sec:Simulation_DGP}

\begin{figure}[t]
    \centering
    \includegraphics[width=15cm]{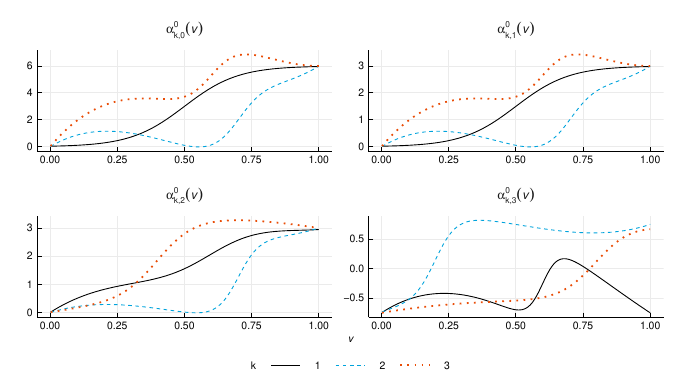}
    \caption[caption]{\small Simulated group-specific coefficient functions for DGP 1 (top left), DGP 2 (top right and bottom left) and DGP 3 (bottom right)}.
    \label{fig:sim_alpha}
\end{figure}

We consider three DGPs for all $(N,T)$ combinations of $N=\{50, 100\}$ and $T=\{50, 100\}$. The cross-sectional individuals are sampled from $K_0=3$ groups with the proportions $0.3$, $0.3$, and $0.4$. Throughout the simulation study, we draw the individual fixed effects $\gamma_i^0$ and the idiosyncratic innovations $\epsilon_{it}$ from mutually independent standard normal distributions. Following \citet[sec. 6]{su2019sieve}, the DGPs are constructed as follows:
\\
\textbf{DGP 1}: Trending panel data model \\
$$y_{it} = \gamma_i^0 + \beta_{i,0}^0 \left( \frac{t}{T} \right) + \epsilon_{it},$$
with the coefficient functions
$$
    \beta_{i,0}^0(v)= \begin{cases}
        \alpha_{1,0}^0(v)=6 F(v; 0.5, 0.1)                                  & \text { if } i \in G_1^0 \\
        \alpha_{2,0}^0(v)=6\left[2 v-6 v^2+4 v^3 + F(v ; 0.7, 0.05) \right] & \text { if } i \in G_2^0 \\ \alpha_{3,0}^0(v)=6\left[4 v-8 v^2+4 v^3 + F(v ; 0.6, 0.05) \right] & \text { if } i \in G_3^0.
    \end{cases}
$$
$F(v; a, b) = \left[1 + \exp(- (v - a) / b) \right]^{-1} $ denotes a cumulative logistic function.
\\
\textbf{DGP 2}: Trending panel with an exogenous regressor\\
$$y_{it} = \gamma_i^0 + \beta_{i,1}^0 \left( \frac{t}{T} \right) + \beta_{i,2}^0 \left( \frac{t}{T} \right) x_{it} + \epsilon_{it},$$
that augments DGP 1 with a scalar exogenous explanatory variable $x_{it}$ which is generated from a  standard normal distribution. Let $\beta_{i,1}^0 (t/T) = 0.5 \beta_{i,0}^0(t/T)$ (and hence $\alpha^0_{i,1}(t/T) = 0.5 \alpha_{i,0}^0(t/T)$), where $\beta_{i,0}^0$ and $\alpha_{i,0}^0$ are as defined in DGP 1, and
$$
    \beta_{i,2}^0(v)= \begin{cases}
        \alpha_{1,2}^0(v)=3 \left[2v - 4v^2 + 2v^3 + F(v; 0.6, 0.1) \right]  & \text { if } i \in G_1^0 \\
        \alpha_{2,2}^0(v)=3 \left[v - 3v^2 + 2v^3 + F(v ; 0.7, 0.04) \right] & \text { if } i \in G_2^0 \\ \alpha_{3,2}^0(v)=3 \left[0.5v - 0.5v^2 + F(v ; 0.4, 0.07) \right] & \text { if } i \in G_3^0.
    \end{cases}
$$
\\
\textbf{DGP 3}: Dynamic panel data model
$$y_{it} = \gamma_i^0 + \beta_{i,3}^0 \left( \frac{t}{T} \right) y_{it-1} + \epsilon_{it},$$
featuring a time-varying autoregressive functional relationship. In order to comply with the strong mixing condition in Assumption \hyperref[line:A1]{A.1(ii)}, $\sup_{v \in [0,1]} | \alpha_{k,3}^0(v) | < 1$ for all $k = 1, \dots, K_0$, the autoregressive coefficient is taken as
$$
    \beta_{i,3}^0(v)= \begin{cases}
        \alpha_{1,3}^0(v)=1.5 \left[-0.5 + 2v - 5v^2 + 2v^3 + F(v; 0.6, 0.03) \right] & \text { if } i \in G_1^0 \\
        \alpha_{2,3}^0(v)=1.5 \left[-0.5 + v - 3v^2 + 2v^3 + F(v ; 0.2, 0.04) \right] & \text { if } i \in G_2^0 \\ \alpha_{3,3}^0(v)=1.5 \left[-0.5 + 0.5 v - 0.5v^2 + F(v ; 0.8, 0.07) \right] & \text { if } i \in G_{3}^0.
    \end{cases}
$$
Figure \ref{fig:sim_alpha} presents the sample paths of the simulated coefficient functions across the three DGPs.

\subsection{Implementation and Evaluation}
\label{sec:Simulation_implementation} 

The \textit{FUSE-TIME} estimator is obtained as described in Section \ref{sec:sieve}. We select the number of interior knots $M^*$ of the spline system $\bs{b}(t/T)$ by taking $M^* = \max \left\{ \left\lfloor (NT)^{1/7} - \log(p)\right\rfloor, 1 \right\} $, where $\left\lfloor \cdot \right\rfloor$ rounds to the lower integer. When $T$ is small, large $M^*$ tend to overfit the individual time-varying coefficients and pollute the grouping. Conversely, a small $M^*$ caps the flexibility of the time-varying coefficients and may complicate the group differentiation in large panels with many latent groups. Furthermore, we set $d=3$ since cubic splines offer a good trade-off between flexibility and parsimony. This gives a $M^* + d + 1 = M$-dimensional vector of spline basis functions.

We compute the \textit{FUSE-TIME} estimator for a grid of $\lambda$ values; the grids are specified in Table \ref{tab:lambda_values} of Appendix \ref{sec:Appendix_Sim} for the three DGPs. The IC in equation \eqref{eq:IC} selects the $\lambda$ tuning parameter. We hereby set $\rho_{NT} = c_\lambda \log(NT) (NT)^{-1/2}$ with $c_\lambda = 0.04$ as described in Section \ref{sec:lambda}. In large samples, the results do not vary substantially in $c_\lambda$.

To evaluate performance, we inspect grouping and estimation accuracy. The clustering performance is evaluated across $n_{\text{sim}}$ Monte Carlo experiments according to the following criteria:
\begin{enumerate}[label=(\roman*)]
    \item Frequency of $\hat{K} = K_0$, $n_{\text{sim}}^{-1} \sum_{j=1}^{n_{\text{sim}}}\bs{1} \{ \hat{K}_j = K_{0} \}$.
    \item Frequency of $\hat{\mathcal{G}}_{\hat{K}} = \mathcal{G}^0$, $n_{\text{sim}}^{-1} \sum_{j=1}^{n_{\text{sim}}}\bs{1} \{ \hat{\mathcal{G}}_{\hat{K},j} = \mathcal{G}^0_j \}$.
    \item Adjusted Rand Index ($ARI$). The $ARI$ ranges from minus one to one and measures the similarity between two groupings; one signals total agreement, minus one total disagreement, and zero reflects random assignment. We report the average $ARI$ over all Monte Carlo iterations $ARI = n_{\text{sim}}^{-1} \sum_{j = 1}^{n_{\text{sim}}} ARI_j(\hat{\mathcal{G}}_{\hat{K},j}, \mathcal{G}^0_j)$.
    \item Average $\hat{K}$, $\bar{K} = n_{\text{sim}}^{-1} \sum_{j=1}^{n_{\text{sim}}} \hat{K}_j$.
\end{enumerate}
The root mean square error (\textit{RMSE}) quantifies the estimation accuracy of the time-varying coefficient functions and is computed as
$$RMSE \left(\hat{\alpha}_l \right) = \frac{1}{N} \sum_{i = 1}^{N} \sqrt{ \frac{1}{T} \sum_{t = 1}^{T} \left[ \hat{\alpha}_{i,l} \left( \frac{t}{T} \right) - \alpha_{i,l}^0 \left( \frac{t}{T} \right) \right]^2},$$ where $\hat{\alpha}_{i,l} (t/T) = \hat{\alpha}_{k,l} (t/T)$ if $i \in \hat{G}_k$ and $\alpha_{i,l}^0 (t/T) = \alpha_{j,l}^0 (t/T)$ if $i \in G_j^0$, for $l = \{0, 1, 2, 3 \}$.

When reporting results on estimation accuracy, note that we explicitly distinguish between the proposed \textit{PSE} and the \textit{post-Lasso} estimator as defined in Section \ref{sec:sieve}; whereas throughout the rest of the paper we refer to the latter as the \textit{FUSE-TIME} estimator.Their performance is compared with two benchmarks. The first is the infeasible oracle estimator that is based on the true latent grouping $\mathcal{G}_0$, averaged across all $n_{\text{sim}}$ experiments. The second is the \textit{time-varying C-Lasso} by \citet{su2019sieve}. Appendix \ref{sec:Appendix_C_lasso} lists details on the implementation of the \textit{time-varying C-Lasso} and Appendix \ref{sec:Appendix_implementation_speed} benchmarks the computational cost of the \textit{FUSE-TIME} and \textit{C-Lasso} software implementations; the computational gains of the former are sizable.

\subsection{Simulation Study Results}
\label{sec:Simulation_results}

We simulate all DGPs $n_{\text{sim}} = 300$ times and apply \textit{FUSE-TIME} as well as the \textit{time-varying C-Lasso} procedures. Table \ref{tab:sim_grouping} reports metrics on the grouping performance; note that there is no distinction between \textit{PSE} and \textit{post-Lasso} in terms of grouping performance, hence we report the results of our proposal under \textit{FUSE-TIME} in Table \ref{tab:sim_grouping}.
\begin{table}[t]
    \centering
    \caption{Clustering accuracy}
    \begin{tabular}{ccccccccccc}
        \hline \hline
                                  & \multirow{2}[1]{*}{N} & \multirow{2}[1]{*}{T} & \multicolumn{2}{c}{Freq. $\hat{K} = K_0$} & \multicolumn{2}{c}{Freq. $\hat{\mathcal{G}}_{\hat{K}} = \mathcal{G}^0_{K_0}$} & \multicolumn{2}{c}{ARI} & \multicolumn{2}{c}{$\bar{K}$}                                                                         \\
        \cmidrule(lr){4-5} \cmidrule(lr){6-7} \cmidrule(lr){8-9} \cmidrule(lr){10-11}
                                  &                       &                       & \textit{FT}                            & \textit{C-Lasso}                                                              & \textit{FT}          & \textit{C-Lasso}              & \textit{FT} & \textit{C-Lasso} & \textit{FT} & \textit{C-Lasso} \\
        \hline
        \multirow{4}[2]{*}{DGP 1} & 50                    & 50                    & 1.000                                     & 0.947                                                                         & 0.960                   & 0.867                         & 0.997          & 0.974            & 3.000          & 2.980            \\
                                  & 50                    & 100                   & 1.000                                     & 1.000                                                                         & 1.000                   & 1.000                         & 1.000          & 1.000            & 3.000          & 3.000            \\
                                  & 100                   & 50                    & 1.000                                     & 0.900                                                                         & 0.943                   & 0.667                         & 0.998          & 0.940            & 3.000          & 2.900            \\
                                  & 100                   & 100                   & 1.000                                     & 1.000                                                                         & 1.000                   & 1.000                         & 1.000          & 1.000            & 3.000          & 3.000            \\
        \hline
        \multirow{4}[2]{*}{DGP 2} & 50                    & 50                    & 0.937                                     & 0.430                                                                         & 0.623                   & 0.140                         & 0.949          & 0.691            & 2.957          & 2.430            \\
                                  & 50                    & 100                   & 1.000                                     & 0.357                                                                         & 0.983                   & 0.320                         & 0.999          & 0.725            & 3.000          & 2.357            \\
                                  & 100                   & 50                    & 0.943                                     & 0.523                                                                         & 0.487                   & 0.023                         & 0.951          & 0.618            & 2.957          & 2.523            \\
                                  & 100                   & 100                   & 1.000                                     & 0.973                                                                         & 0.977                   & 0.540                         & 0.999          & 0.960            & 3.000          & 2.973            \\
        \hline
        \multirow{4}[2]{*}{DGP 3} & 50                    & 50                    & 0.713                                     & 0.000                                                                         & 0.120                   & 0.000                         & 0.838          & 0.503            & 3.280          & 2.000            \\
                                  & 50                    & 100                   & 0.937                                     & 0.000                                                                         & 0.677                   & 0.000                         & 0.974          & 0.541            & 3.063          & 2.000            \\
                                  & 100                   & 50                    & 0.750                                     & 0.000                                                                         & 0.047                   & 0.000                         & 0.829          & 0.485            & 2.810          & 2.000            \\
                                  & 100                   & 100                   & 0.993                                     & 0.000                                                                         & 0.633                   & 0.000                         & 0.983          & 0.542            & 2.993          & 2.000            \\
        \hline \hline
    \end{tabular}%
    \label{tab:sim_grouping}%
    \begin{tablenotes}
        \small
        \item
        \textit{Notes}: Frequency of obtaining the correct number of groups $\hat{K} = K_0$ and the correct grouping $\hat{\mathcal{G}}_{\hat{K}} = \mathcal{G}^0_{K_0}$, the ARI, and the average estimated number of total groups $\bar{K}$ based on a Monte Carlo study with 300 replications. \textit{FT}--\textit{FUSE-TIME}--denotes our proposed methodology. \textit{C-Lasso} refers to the benchmark model by \citet{su2019sieve}.
    \end{tablenotes}
\end{table}%

\textit{FUSE-TIME} convincingly outperforms the \textit{C-Lasso} benchmark, particularly in smaller samples and DGPs 2 and 3. \textit{FUSE-TIME}'s clustering accuracy improves quickly with increasing $T$, but not with $N$. This is intuitive since the clustering routine is driven by the estimated individual control points $\hat{\bs{\pi}}_i$, which are not consistent with respect to the cross-sectional dimension (see Theorem \hyperref[sec:Theo_1]{3.1}). Accordingly, when $T$ is small, the estimation of individual coefficient functions is highly noisy, complicating correct group assignment. This property is particularly evident when comparing DGP 1 and DGP 2. The latter involves estimating a second regression curve, which introduces additional uncertainty and subsequently does not lead to the near-perfect clustering observed in DGP 1. Nevertheless, \textit{FUSE-TIME} still shows strong clustering performance for DGP 2. In contrast, the results for the dynamic panel data model of DGP 3 are markedly worse than for the other DGPs, especially when $T = 50$. Similar estimation devices have previously reported decreased performance for dynamic panel data models \citep[see][]{mehrabani2023}. Nevertheless, as indicated by the $ARI$ measure, even when individual cross-sectional units are misclassified, the estimated grouping remains a close approximation of the true unobserved group structure, with \textit{FUSE-TIME} substantially outperforming the C-Lasso: ARIs in the range of 0.83-0.98 for \textit{FUSE-TIME} versus 0.49-0.54 for C-Lasso. 

\begin{figure}[t]
    \centering
    \includegraphics[width=15cm]{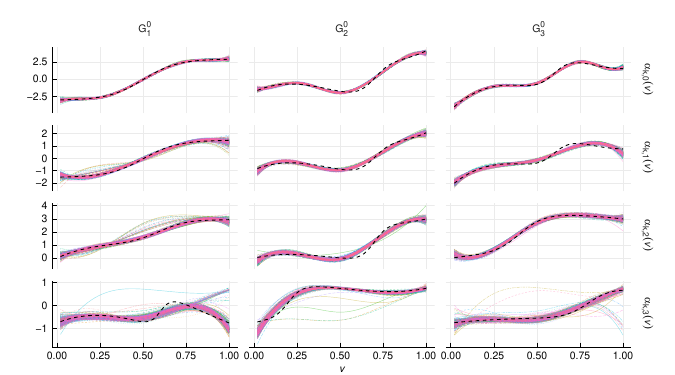}
    \caption[Simulation study]{\small Estimated \textit{FUSE-TIME} \textit{post-Lasso} functional coefficients when $N,T = 50$ and $n_{\text{sim}} = 300$. The true coefficient functions are shown in black (dashed).}
    \label{fig:sim}
\end{figure}

Table \ref{tab:RMSE} reports the \textit{RMSE} for each time-varying coefficient, and split among \textit{PSE} and \textit{post-Lasso} estimators. Figure \ref{fig:sim} provides the estimated sample paths of the functional coefficients for each DGP and $N,T = 50$.
\begin{table}[t]
    \centering
    \caption{\textit{RMSE} of coefficient estimates}
    \begin{tabular}{ccccccccc}
        \hline
        \hline
                                  &                                                 & \multirow{2}[1]{*}{N} & \multirow{2}[1]{*}{T} & \multicolumn{2}{c}{\textit{PSE}} & \multicolumn{2}{c}{\textit{post-Lasso}} & \multirow{2}[1]{*}{oracle}                            \\
        \cmidrule(lr){5-6} \cmidrule(lr){7-8}
                                  &                                                 &                       &                       & \textit{FUSE-TIME}                   & \textit{C-Lasso}                        & \textit{FUSE-TIME}             & \textit{C-Lasso} &       \\
        \hline
        \multirow{4}[2]{*}{DGP 1} & \multirow{4}[2]{*}{$\hat{\alpha}_0 (t/T)$}  & 50                    & 50                    & 0.274                            & 0.915                                   & 0.160                      & 0.311            & 0.159 \\
                                  &                                                 & 50                    & 100                   & 0.197                            & 1.037                                   & 0.146                      & 0.277            & 0.146 \\
                                  &                                                 & 100                   & 50                    & 0.274                            & 1.034                                   & 0.146                      & 0.299            & 0.146 \\
                                  &                                                 & 100                   & 100                   & 0.190                            & 1.095                                   & 0.139                      & 0.271            & 0.139 \\
        \hline
        \multirow{8}[2]{*}{DGP 2} & \multirow{4}[1]{*}{$\hat{\alpha}_1 (t/T)$}  & 50                    & 50                    & 0.263                            & 0.432                                   & 0.154                      & 0.231            & 0.130 \\
                                  &                                                 & 50                    & 100                   & 0.189                            & 0.453                                   & 0.140                      & 0.209            & 0.117 \\
                                  &                                                 & 100                   & 50                    & 0.275                            & 0.433                                   & 0.146                      & 0.226            & 0.117 \\
                                  &                                                 & 100                   & 100                   & 0.189                            & 0.529                                   & 0.134                      & 0.153            & 0.079 \\
        \cmidrule{2-9}            & \multirow{4}[1]{*}{$\hat{\alpha}_2 (t/T)$}  & 50                    & 50                    & 0.306                            & 0.655                                   & 0.153                      & 0.231            & 0.135 \\
                                  &                                                 & 50                    & 100                   & 0.207                            & 0.698                                   & 0.127                      & 0.186            & 0.116 \\
                                  &                                                 & 100                   & 50                    & 0.321                            & 0.662                                   & 0.139                      & 0.220            & 0.119 \\
                                  &                                                 & 100                   & 100                   & 0.206                            & 0.836                                   & 0.121                      & 0.075            & 0.085 \\
        \hline
        \multirow{4}[2]{*}{DGP 3} & \multirow{4}[1]{*}{$\hat{\alpha}_3 (t/T) $} & 50                    & 50                    & 0.217                            & 0.422                                   & 0.145                      & 0.418            & 0.074 \\
                                  &                                                 & 50                    & 100                   & 0.168                            & 0.457                                   & 0.119                      & 0.423            & 0.059 \\
                                  &                                                 & 100                   & 50                    & 0.243                            & 0.413                                   & 0.147                      & 0.416            & 0.060 \\
                                  &                                                 & 100                   & 100                   & 0.153                            & 0.422                                   & 0.052                      & 0.412            & 0.050 \\
        \hline
        \hline
    \end{tabular}%
    \label{tab:RMSE}%
    \begin{tablenotes}
        \small
        \item
        \textit{Notes}: \textit{RMSE} of the \textit{PSE}, the \textit{post-Lasso}, and an infeasible oracle estimator based on a Monte Carlo study with 300 replications. \textit{FUSE-TIME} denotes our proposed methodology. \textit{C-Lasso} refers to the benchmark model by \citet{su2019sieve}. The index in $\hat{\alpha}_l (t/T)$ does not refer to the group structure but to the coefficients $l = \{0, 1, 2, 3\}$ used in DGPs 1-3 (see. Section \ref{sec:Simulation_DGP}).
    \end{tablenotes}
\end{table}%
The \textit{RMSE} results largely align with the clustering performance. This also holds for the comparison with the \textit{time-varying C-Lasso}, which returns considerably poorer results regarding the \textit{RMSE} as well. Unlike clustering accuracy, the \textit{post-Lasso} \textit{RMSE} also improves with $N$ due to the larger number of cross-sectional units available for pooling when estimating group-specific trajectories. Notably, the \textit{post-Lasso} performs well relative to the oracle estimator, even in cases, such as DGP 3, where the clustering results may suggest a poor fit. This finding is likely because misclassified units tend to feature sample paths that are particularly similar to other groups, leading to a minor impact on the \textit{RMSE} compared to the oracle estimation. Precise identification of group-specific coefficients despite several misclassifications has been previously reported by \citet{bonhomme2015grouped}. Table \ref{tab:RMSE} further highlights that the shrinkage penalty leads to a sizeable finite-sample bias in the \textit{PSE}, even though the \textit{PSE} and the \textit{post-Lasso} share the same asymptotic distribution (see Theorem \hyperref[sec:Theo_3]{3.5}). This motivates our usage of the \textit{post-Lasso} for empirical applications. Finally, Figure \ref{fig:sim} corroborates the previous findings: while minor deviations occur, partly due to misclassified units, the estimated trajectories closely follow the true underlying function.

Additionally, we also simulate the three DGPs with either idiosyncratic errors following an $AR(1)$ process with lag polynomial $a(L) = 1 - 0.3L$  or unbalanced panels with missing data, where we generate sample paths that randomly discard 30\% of the observations. Detailed results are available in Appendix \ref{sec:Appendix_Sim}. For unbalanced panels with missing data (see Appendix \ref{sec:Appendix_NA}),  the simulation results largely mirror the ones presented here. Discarding part of the sample has a similar effect to decreasing $T$ in a balanced panel. However, when the innovations follow an $AR(1)$ process (see Appendix \ref{sec:Appendix_corr_errors}), the performance decreases markedly, particularly when $T$ is small. Introducing autocorrelation to the errors increases the estimation uncertainty and thus contaminates the clustering mechanism. However, even when the grouping is correctly estimated, the coefficient trajectories offer a significantly poorer fit than the baseline case. 


\section{Empirical Illustration}
\label{sec:Empirical}

Many major economies pledge to reduce the emission of harmful greenhouse gases, a key driver of global warming. Achieving this objective without impeding economic growth requires emissions per unit of GDP to decrease. CO\textsubscript{2} accounts for approximately 66\% of the anthropogenic contribution to global warming and is a ubiquitous by-product of economic activity and energy production \citep{bennedsen2023neural}. Consequently, understanding the relationship between CO\textsubscript{2} pollution and economic growth is crucial for effective policy and climate action. We contribute by applying our \textit{FUSE-TIME} procedure to identify trends in the CO\textsubscript{2} emission intensity of GDP, the CO\textsubscript{2} emitted per unit of GDP produced.

The CO\textsubscript{2} intensity of GDP is determined by the structural composition of an economy and the \textit{``greenness"} of its energy production \citep{bella2014relationship}. For instance, transitioning from a production-based to a service-oriented economy reduces energy consumption and direct emissions from high-pollution industries. Likewise, shifting from emission-intensive solid fuels, such as coal or biomass, to cleaner carriers like natural gas or renewables, lowers the elasticity between energy demand and associated pollution. The combined effects of economic restructuring and energy source optimization are often credited with driving the decoupling of GDP growth from production-based emissions in high-income countries \citep{jakob2012will,wang2017multi}. This decoupling pattern has inspired an extensive literature on the inverted U-shaped environmental Kuznets curve, which posits a positive elasticity between income and emissions during the early stages of economic development, turning negative as technological progress and greater prosperity take hold \citep[see][among other]{grossman1995economic,wagner2015environmental,bennedsen2023neural}. However, as (log) GDP, an integrated process, clearly does not satisfy the regularity conditions in Assumption \hyperref[line:A1]{1}, we instead study the relationship between emissions and income by estimating trends in CO\textsubscript{2} intensity. Trend functions capture the long-term behavior of CO\textsubscript{2} intensity by smoothing over short-run fluctuations such as business cycles, fuel price volatility, or temporary energy supply disruptions. Moreover, it is well-established that economies exhibit considerable heterogeneity in their developmental trajectories \citep{kose2003international,azomahou2006economic}. Consequently, assuming a common CO\textsubscript{2} intensity trend across many countries is implausible. Nevertheless, much of the existing empirical literature imposes cross-sectional homogeneity or homogeneity conditional on an exogenous grouping \citep{churchill2018environmental,bennedsen2023neural}. We relax this condition. Using our \textit{FUSE-TIME} procedure, different economies only share a common trend if the consistent grouping mechanism identifies homogeneous coefficients, enabling us to exploit the cross-sectional dimension without imposing restrictive homogeneity assumptions. In addition, given the small number of group-specific trend functions relative to the cross-sectional dimension, it becomes feasible to interpret the long-term behavior of CO\textsubscript{2} intensity for a large number of economies.

To estimate trends in the CO\textsubscript{2} intensity of GDP, we employ production-based CO\textsubscript{2} emissions data from the Global Carbon Project \citep{Friedlingstein2024global}.\footnote{Dataset \textit{National fossil carbon emissions v2024}, available at \href{https://globalcarbonbudgetdata.org}{\texttt{globalcarbonbudgetdata.org}}. Accessed November 14}. GDP series are obtained from the World Bank Development Indicators database.\footnote{Series \textit{NY.GDP.MKTP.CD}, available at \href{https://data.worldbank.org/indicator/NY.GDP.MKTP.CD}{\texttt{data.worldbank.org/indicator/NY.GDP.MKTP.CD}}. Accessed November 12, 2024.} Both datasets are in annual frequency and span the time period from 1960 to 2023 for 92 countries. Details on the countries included in the sample and descriptive statistics are provided in Appendix \ref{sec:empirical_appendix}. Using this data, we compute the CO\textsubscript{2} intensity and construct an unbalanced panel where the individual series range from 29 to 64 years in length.

Let $y_{it} = \text{CO}\textsubscript{2}_{it} / \text{GDP}_{it}$, with CO\textsubscript{2} measured in million tonnes and GDP in billion 2024 U.S. dollar. We formulate the time-varying panel data model subject to a latent grouping
\begin{equation*}
    \label{eq:empirical_DGP}
    y_{it} = \gamma_i + \beta_i \left(\frac{t}{T}\right) + \epsilon_{it},
\end{equation*}
where $\beta_i (t/T)$ denotes the trend function of interest (time-varying intercept).

Searching over a dense grid of $d$, $M^*$, and $\lambda$ values, $d = 2$, $M^* = 4$, and $\lambda = 0.72$ yields the lowest IC, implying five latent groups; the entire grid is provided in Appendix \ref{sec:empirical_appendix}. Figure \ref{fig:intensity_trend} presents the estimated group-specific trend functions $\hat{\alpha}^p_{\hat{G}_k} (t/T)$. Figure \ref{fig:world_map} sketches the spatial pattern in the group structure. Table \ref{tab:Group_estim} in Appendix \ref{sec:empirical_appendix} provides a detailed record of which countries are assigned to which group.

\begin{figure}
    \centering
    \includegraphics[width=15cm]{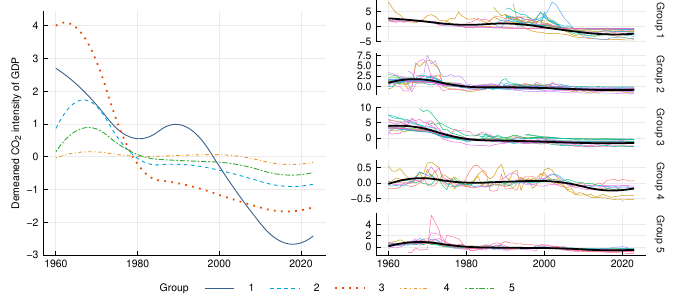}
    \caption[caption]{\small The left panel displays the group-specific \textit{FUSE-TIME} \textit{post-Lasso} trend functions. The right panel contrasts the estimated trend functions (black, thick) with the demeaned data of CO\textsubscript{2} intensity (colored, thin).}
    \label{fig:intensity_trend}
\end{figure}

\begin{figure}
    \centering
    \includegraphics[width=15cm]{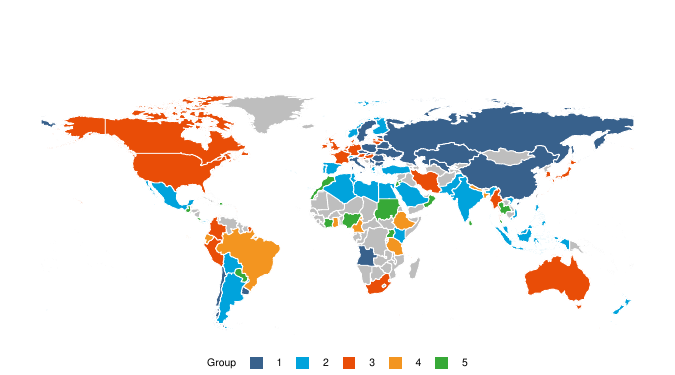}
    \caption[caption]{\small Estimated group structure in the trends of CO\textsubscript{2} intensity.}
    \label{fig:world_map}
\end{figure}

Group 1 predominantly comprises middle- and high-income economies in Eastern Europe and Asia, which have exhibited a substantial and ongoing decline in CO\textsubscript{2} emission intensity since the 1990s. Structural changes and rapid technological advances following the fall of the USSR and the economic liberalization of China may drive this trend. Group 2 largely includes low-to-middle-income economies, with outliers such as Switzerland, Hong Kong, and Singapore. The trend of Group 2 features a much more gradual decrease compared to the first group due to more moderate growth in both income and emissions. Group 3 primarily consists of high-income countries that experienced a sharp decline in emission intensity pre-1980, followed by a continued but more dampened decline thereafter. This pattern reflects a stark reduction in emissions until the 1980s, accompanied by consistent GDP growth throughout the entire sample period, aligning with a transition to service-based economies and the offshoring of emission- and energy-intensive sectors to low-income countries. Group 4 contains low-income and developing economies, characterized by the absence of a trend. In these economies, technological progress has yet to reduce emission intensity significantly or, as in the case of Brazil, income growth is largely driven by the exploitation of natural resources. Group 5, consisting of low-income economies, exhibits a similar but slightly more attenuated trend compared to Group 2.


\section{Conclusion}
\label{sec:Conclusion}

This article introduces a novel penalized sieve estimation technique--\textit{FUSE-TIME}--to estimate panel data models with smoothly time-varying coefficients that are subject to an unobserved group structure. Our approach simultaneously identifies the functional coefficients, the number of latent groups, and group compositions. In addition, we propose a consistent BIC-type criterion to determine the penalty tuning parameter, introduce a \textit{post-Lasso} estimator, and prove oracle-efficiency. Monte Carlo simulation studies confirm strong finite-sample performance in recovering the latent group structure and accurately estimating the time-varying coefficients across a wide range of scenarios. We apply our method to analyze trends in the CO\textsubscript{2} emission intensity of GDP.

An important yet largely unexplored extension regards inference methods for (time-varying) panel data models with a latent group pattern. The estimated grouping is an inherently noisy representation of the true unobserved structure. Subsequently, the clustering introduces uncertainty to the slope coefficients, analogous to the well-established problem of non-uniform convergence of post-selection estimation \citep[see][]{greenshtein2004persistence,leeb2008sparse,adamek2023lasso}. To the best of our knowledge, \citet{dzemski2024confidence} provide the only approach to creating confidence sets of the grouping by inverting poolability tests to date. Consequently, the development of inference methods for penalized panel data models that admit a latent grouping, including confidence sets for the group structure, as well as tests for coefficient poolability and constancy akin to \citet{friedrich2024sieve}, remains an important direction for future research.


\cleardoublepage
\setstretch{0.1}
\bibliography{References}

@article{adamek2023lasso,
  title   = {Lasso inference for high-dimensional time series},
  journal = {Journal of Econometrics},
  volume  = {235},
  number  = {2},
  pages   = {1114-1143},
  year    = {2023},
  issn    = {0304-4076},
  doi     = {https://doi.org/10.1016/j.jeconom.2022.08.008},
  url     = {https://www.sciencedirect.com/science/article/pii/S0304407622001804},
  author  = {Robert Adamek and Stephan Smeekes and Ines Wilms}
}

@article{ando2016panel,
  author  = {Ando, Tomohiro and Bai, Jushan},
  title   = {Panel Data Models with Grouped Factor Structure Under Unknown Group Membership},
  journal = {Journal of Applied Econometrics},
  volume  = {31},
  number  = {1},
  pages   = {163-191},
  doi     = {https://doi.org/10.1002/jae.2467},
  url     = {https://onlinelibrary.wiley.com/doi/abs/10.1002/jae.2467},
  eprint  = {https://onlinelibrary.wiley.com/doi/pdf/10.1002/jae.2467},
  year    = {2016}
}

@article{churchill2018environmental,
  title   = {The Environmental Kuznets Curve in the OECD: 1870–2014},
  journal = {Energy Economics},
  volume  = {75},
  pages   = {389-399},
  year    = {2018},
  issn    = {0140-9883},
  doi     = {https://doi.org/10.1016/j.eneco.2018.09.004},
  url     = {https://www.sciencedirect.com/science/article/pii/S0140988318303736},
  author  = {Sefa {Awaworyi Churchill} and John Inekwe and Kris Ivanovski and Russell Smyth}
}

@article{azomahou2006economic,
  title   = {Economic development and CO2 emissions: A nonparametric panel approach},
  journal = {Journal of Public Economics},
  volume  = {90},
  number  = {6},
  pages   = {1347-1363},
  year    = {2006},
  issn    = {0047-2727},
  doi     = {https://doi.org/10.1016/j.jpubeco.2005.09.005},
  url     = {https://www.sciencedirect.com/science/article/pii/S0047272705001301},
  author  = {Théophile Azomahou and François Laisney and Phu {Nguyen Van}}
}

@article{bai2010common,
  title   = {Common breaks in means and variances for panel data},
  journal = {Journal of Econometrics},
  volume  = {157},
  number  = {1},
  pages   = {78-92},
  year    = {2010},
  note    = {Nonlinear and Nonparametric Methods in Econometrics},
  issn    = {0304-4076},
  doi     = {https://doi.org/10.1016/j.jeconom.2009.10.020},
  url     = {https://www.sciencedirect.com/science/article/pii/S0304407609002735},
  author  = {Jushan Bai}
}

@article{bai2002determining,
  author  = {Bai, Jushan and Ng, Serena},
  title   = {Determining the Number of Factors in Approximate Factor Models},
  journal = {Econometrica},
  volume  = {70},
  number  = {1},
  pages   = {191-221},
  doi     = {https://doi.org/10.1111/1468-0262.00273},
  url     = {https://onlinelibrary.wiley.com/doi/abs/10.1111/1468-0262.00273},
  eprint  = {https://onlinelibrary.wiley.com/doi/pdf/10.1111/1468-0262.00273},
  year    = {2002}
}

@article{bella2014relationship,
  title   = {The relationship among CO2 emissions, electricity power consumption and GDP in OECD countries},
  journal = {Journal of Policy Modeling},
  volume  = {36},
  number  = {6},
  pages   = {970-985},
  year    = {2014},
  issn    = {0161-8938},
  doi     = {https://doi.org/10.1016/j.jpolmod.2014.08.006},
  url     = {https://www.sciencedirect.com/science/article/pii/S016189381400088X},
  author  = {Giovanni Bella and Carla Massidda and Paolo Mattana}
}

@article{bennedsen2023neural,
  title   = {A neural network approach to the environmental Kuznets curve},
  journal = {Energy Economics},
  volume  = {126},
  pages   = {106985},
  year    = {2023},
  issn    = {0140-9883},
  doi     = {https://doi.org/10.1016/j.eneco.2023.106985},
  url     = {https://www.sciencedirect.com/science/article/pii/S0140988323004838},
  author  = {Mikkel Bennedsen and Eric Hillebrand and Sebastian Jensen}
}

@book{bernstein2009matrix,
  title     = {Matrix Mathematics, Theory, Facts, and Formulas with Application to Linear Systems Theory},
  author    = {Bernstein, Dennis S},
  publisher = {Princeton University Press},
  location  = {Princeton, NJ},
  year      = {2009}
}

@article{bonhomme2015grouped,
  author   = {Bonhomme, Stéphane and Manresa, Elena},
  title    = {Grouped Patterns of Heterogeneity in Panel Data},
  journal  = {Econometrica},
  volume   = {83},
  number   = {3},
  pages    = {1147-1184},
  doi      = {https://doi.org/10.3982/ECTA11319},
  url      = {https://onlinelibrary.wiley.com/doi/abs/10.3982/ECTA11319},
  eprint   = {https://onlinelibrary.wiley.com/doi/pdf/10.3982/ECTA11319},
  year     = {2015}
}

@article{bonhomme2022discretizing,
  author   = {Bonhomme, Stéphane and Lamadon, Thibaut and Manresa, Elena},
  title    = {Discretizing Unobserved Heterogeneity},
  journal  = {Econometrica},
  volume   = {90},
  number   = {2},
  pages    = {625-643},
  doi      = {https://doi.org/10.3982/ECTA15238},
  url      = {https://onlinelibrary.wiley.com/doi/abs/10.3982/ECTA15238},
  eprint   = {https://onlinelibrary.wiley.com/doi/pdf/10.3982/ECTA15238},
  year     = {2022}
}

@article{cai2007trending,
  title   = {Trending time-varying coefficient time series models with serially correlated errors},
  journal = {Journal of Econometrics},
  volume  = {136},
  number  = {1},
  pages   = {163-188},
  year    = {2007},
  issn    = {0304-4076},
  doi     = {https://doi.org/10.1016/j.jeconom.2005.08.004},
  url     = {https://www.sciencedirect.com/science/article/pii/S0304407605002058},
  author  = {Zongwu Cai}
}

@article{cashin2014differential,
  title   = {The differential effects of oil demand and supply shocks on the global economy},
  journal = {Energy Economics},
  volume  = {44},
  pages   = {113-134},
  year    = {2014},
  issn    = {0140-9883},
  doi     = {https://doi.org/10.1016/j.eneco.2014.03.014},
  url     = {https://www.sciencedirect.com/science/article/pii/S0140988314000619},
  author  = {Paul Cashin and Kamiar Mohaddes and Maziar Raissi and Mehdi Raissi}
}

@book{de2001practical,
  title     = {A practical guide to splines},
  author    = {De Boor, Carl},
  year      = {2001},
  publisher = {Springer-Verlag New York}
}

@article{diebold2008global,
  title   = {Global yield curve dynamics and interactions: A dynamic Nelson–Siegel approach},
  journal = {Journal of Econometrics},
  volume  = {146},
  number  = {2},
  pages   = {351-363},
  year    = {2008},
  note    = {Honoring the research contributions of Charles R. Nelson},
  issn    = {0304-4076},
  doi     = {https://doi.org/10.1016/j.jeconom.2008.08.017},
  url     = {https://www.sciencedirect.com/science/article/pii/S0304407608001127},
  author  = {Francis X. Diebold and Canlin Li and Vivian Z. Yue}
}

@article{dung2017direct,
  title     = {A direct method to solve optimal knots of {B}-spline curves: An application for non-uniform {B}-spline curves fitting},
  author    = {Dung, Van Than and Tjahjowidodo, Tegoeh},
  journal   = {PloS one},
  volume    = {12},
  number    = {3},
  pages     = {e0173857},
  year      = {2017},
  publisher = {Public Library of Science San Francisco, CA USA}
}

@article{dzemski2024confidence,
  author   = {Dzemski, Andreas and Okui, Ryo},
  title    = {Confidence set for group membership},
  journal  = {Quantitative Economics},
  volume   = {15},
  number   = {2},
  pages    = {245-277},
  doi      = {https://doi.org/10.3982/QE1975},
  url      = {https://onlinelibrary.wiley.com/doi/abs/10.3982/QE1975},
  eprint   = {https://onlinelibrary.wiley.com/doi/pdf/10.3982/QE1975},
  year     = {2024}
}

@article{freyaldenhoven2022factor,
  title   = {Factor models with local factors — Determining the number of relevant factors},
  journal = {Journal of Econometrics},
  volume  = {229},
  number  = {1},
  pages   = {80-102},
  year    = {2022},
  issn    = {0304-4076},
  doi     = {https://doi.org/10.1016/j.jeconom.2021.04.006},
  url     = {https://www.sciencedirect.com/science/article/pii/S0304407621001275},
  author  = {Simon Freyaldenhoven}
}

@article{Friedlingstein2024global,
  author  = {Friedlingstein, P. and O'Sullivan, M. and Jones, M. W. and Andrew, R. M. and Hauck, J. and Landsch\"utzer, P. and Le Qu\'er\'e, C. and Li, H. and Luijkx, I. T. and Olsen, A. and Peters, G. P. and Peters, W. and Pongratz, J. and Schwingshackl, C. and Sitch, S. and Canadell, J. G. and Ciais, P. and Jackson, R. B. and Alin, S. R. and Arneth, A. and Arora, V. and Bates, N. R. and Becker, M. and Bellouin, N. and Berghoff, C. F. and Bittig, H. C. and Bopp, L. and Cadule, P. and Campbell, K. and Chamberlain, M. A. and Chandra, N. and Chevallier, F. and Chini, L. P. and Colligan, T. and Decayeux, J. and Djeutchouang, L. and Dou, X. and Duran Rojas, C. and Enyo, K. and Evans, W. and Fay, A. and Feely, R. A. and Ford, D. J. and Foster, A. and Gasser, T. and Gehlen, M. and Gkritzalis, T. and Grassi, G. and Gregor, L. and Gruber, N. and G\"urses, \"O. and Harris, I. and Hefner, M. and Heinke, J. and Hurtt, G. C. and Iida, Y. and Ilyina, T. and Jacobson, A. R. and Jain, A. and Jarn\'{\i}kov\'a, T. and Jersild, A. and Jiang, F. and Jin, Z. and Kato, E. and Keeling, R. F. and Klein Goldewijk, K. and Knauer, J. and Korsbakken, J. I. and Lauvset, S. K. and Lef\`evre, N. and Liu, Z. and Liu, J. and Ma, L. and Maksyutov, S. and Marland, G. and Mayot, N. and McGuire, P. and Metzl, N. and Monacci, N. M. and Morgan, E. J. and Nakaoka, S.-I. and Neill, C. and Niwa, Y. and N\"utzel, T. and Olivier, L. and Ono, T. and Palmer, P. I. and Pierrot, D. and Qin, Z. and Resplandy, L. and Roobaert, A. and Rosan, T. M. and R\"odenbeck, C. and Schwinger, J. and Smallman, T. L. and Smith, S. and Sospedra-Alfonso, R. and Steinhoff, T. and Sun, Q. and Sutton, A. J. and S\'ef\'erian, R. and Takao, S. and Tatebe, H. and Tian, H. and Tilbrook, B. and Torres, O. and Tourigny, E. and Tsujino, H. and Tubiello, F. and van der Werf, G. and Wanninkhof, R. and Wang, X. and Yang, D. and Yang, X. and Yu, Z. and Yuan, W. and Yue, X. and Zaehle, S. and Zeng, N. and Zeng, J.},
  title   = {Supplemental data of Global Carbon Budget 2024},
  journal = {Earth System Science Data Discussions},
  volume  = {2024},
  year    = {2024},
  pages   = {1--133},
  url     = {https://essd.copernicus.org/preprints/essd-2024-519/},
  doi     = {10.5194/essd-2024-519}
}

@article{friedrich2024sieve,
  title   = {Sieve bootstrap inference for linear time-varying coefficient models},
  journal = {Journal of Econometrics},
  volume  = {239},
  number  = {1},
  pages   = {105345},
  year    = {2024},
  note    = {Climate Econometrics},
  issn    = {0304-4076},
  doi     = {https://doi.org/10.1016/j.jeconom.2022.09.004},
  url     = {https://www.sciencedirect.com/science/article/pii/S0304407622001701},
  author  = {Marina Friedrich and Yicong Lin}
}

@article{greenshtein2004persistence,
  title     = {Persistence in high-dimensional linear predictor selection and the virtue of overparametrization},
  author    = {Greenshtein, Eitan and Ritov, Ya'Acov},
  journal   = {Bernoulli},
  volume    = {10},
  number    = {6},
  pages     = {971--988},
  year      = {2004},
  publisher = {Bernoulli Society for Mathematical Statistics and Probability},
  doi       = {https://doi.org/10.3150/bj/1106314846}
}

@article{grossman1995economic,
  issn      = {00335533, 15314650},
  url       = {http://www.jstor.org/stable/2118443},
  author    = {Gene M. Grossman and Alan B. Krueger},
  journal   = {The Quarterly Journal of Economics},
  number    = {2},
  pages     = {353--377},
  publisher = {Oxford University Press},
  title     = {Economic Growth and the Environment},
  urldate   = {2025-11-14},
  volume    = {110},
  year      = {1995}
}

@article{hahn2002asymptotically,
  issn      = {00129682, 14680262},
  url       = {http://www.jstor.org/stable/3082010},
  author    = {Jinyong Hahn and Guido Kuersteiner},
  journal   = {Econometrica},
  number    = {4},
  pages     = {1639--1657},
  publisher = {[Wiley, Econometric Society]},
  title     = {Asymptotically Unbiased Inference for a Dynamic Panel Model with Fixed Effects When Both n and T Are Large},
  urldate   = {2025-11-14},
  volume    = {70},
  year      = {2002}
}

@article{hahn2011bias,
  title     = {Bias reduction for dynamic nonlinear panel models with fixed effects},
  author    = {Hahn, Jinyong and Kuersteiner, Guido},
  journal   = {Econometric Theory},
  volume    = {27},
  number    = {6},
  pages     = {1152--1191},
  year      = {2011},
  publisher = {Cambridge University Press}
}

@manual{haimerl2025pagfl,
  title  = {PAGFL: Joint Estimation of Latent Groups and Group-Specific Coefficients in Panel Data Models},
  author = {Haimerl, Paul and Smeekes, Stephan and Wilms, Ines and Mehrabani, Ali},
  year   = {2025},
  note   = {R-package version 1.1.3}
}

@article{hansen2002spline,
  author    = {Mark H. Hansen and Charles Kooperberg},
  title     = {{Spline Adaptation in Extended Linear Models (with comments and a rejoinder by the authors}},
  volume    = {17},
  journal   = {Statistical Science},
  number    = {1},
  publisher = {Institute of Mathematical Statistics},
  pages     = {2 -- 51},
  year      = {2002},
  doi       = {10.1214/ss/1023798997},
  url       = {https://doi.org/10.1214/ss/1023798997}
}

@article{huang2004polynomial,
  issn         = {10170405, 19968507},
  howpublished = {\href{http://www.jstor.org/stable/24307415}{jstor.org/stable/24307415}},
  author       = {Jianhua Z. Huang and Colin O. Wu and Lan Zhou},
  journal      = {Statistica Sinica},
  number       = {3},
  pages        = {763--788},
  publisher    = {Institute of Statistical Science, Academia Sinica},
  title        = {POLYNOMIAL SPLINE ESTIMATION AND INFERENCE FOR VARYING COEFFICIENT MODELS WITH LONGITUDINAL DATA},
  urldate      = {2024-05-02},
  volume       = {14},
  year         = {2004}
}

@article{jakob2012will,
  title   = {Will history repeat itself? Economic convergence and convergence in energy use patterns},
  journal = {Energy Economics},
  volume  = {34},
  number  = {1},
  pages   = {95-104},
  year    = {2012},
  issn    = {0140-9883},
  doi     = {https://doi.org/10.1016/j.eneco.2011.07.008},
  url     = {https://www.sciencedirect.com/science/article/pii/S0140988311001381},
  author  = {Michael Jakob and Markus Haller and Robert Marschinski}
}

@article{ke2015homogeneity,
  author    = {Zheng Tracy Ke and Jianqing Fan and Yichao Wu},
  title     = {Homogeneity Pursuit},
  journal   = {Journal of the American Statistical Association},
  volume    = {110},
  number    = {509},
  pages     = {175--194},
  year      = {2015},
  publisher = {ASA Website},
  doi       = {10.1080/01621459.2014.892882},
  note      = {PMID: 26085701},
  url       = {https://doi.org/10.1080/01621459.2014.892882},
  eprint    = {https://doi.org/10.1080/01621459.2014.892882}
}

@article{kiviet1995bias,
  title   = {On bias, inconsistency, and efficiency of various estimators in dynamic panel data models},
  journal = {Journal of Econometrics},
  volume  = {68},
  number  = {1},
  pages   = {53-78},
  year    = {1995},
  issn    = {0304-4076},
  doi     = {https://doi.org/10.1016/0304-4076(94)01643-E},
  url     = {https://www.sciencedirect.com/science/article/pii/030440769401643E},
  author  = {Jan F. Kiviet}
}

@article{kose2003international,
  author  = {Kose, M. Ayhan and Otrok, Christopher and Whiteman, Charles H.},
  title   = {International Business Cycles: World, Region, and Country-Specific Factors},
  journal = {American Economic Review},
  volume  = {93},
  number  = {4},
  year    = {2003},
  month   = {September},
  pages   = {1216–1239},
  doi     = {10.1257/000282803769206278},
  url     = {https://www.aeaweb.org/articles?id=10.1257/000282803769206278}
}

@article{leeb2008sparse,
  title   = {Sparse estimators and the oracle property, or the return of Hodges’ estimator},
  journal = {Journal of Econometrics},
  volume  = {142},
  number  = {1},
  pages   = {201-211},
  year    = {2008},
  issn    = {0304-4076},
  doi     = {https://doi.org/10.1016/j.jeconom.2007.05.017},
  url     = {https://www.sciencedirect.com/science/article/pii/S0304407607001273},
  author  = {Hannes Leeb and Benedikt M. Pötscher}
}

@article{lumsdaine2023estimation,
  title   = {Estimation of panel group structure models with structural breaks in group memberships and coefficients},
  journal = {Journal of Econometrics},
  volume  = {233},
  number  = {1},
  pages   = {45-65},
  year    = {2023},
  issn    = {0304-4076},
  doi     = {https://doi.org/10.1016/j.jeconom.2022.01.001},
  url     = {https://www.sciencedirect.com/science/article/pii/S0304407622000033},
  author  = {Robin L. Lumsdaine and Ryo Okui and Wendun Wang}
}

@article{ma2017,
  author    = {Shujie Ma and Jian Huang},
  title     = {A Concave Pairwise Fusion Approach to Subgroup Analysis},
  journal   = {Journal of the American Statistical Association},
  volume    = {112},
  number    = {517},
  pages     = {410--423},
  year      = {2017},
  publisher = {ASA Website},
  doi       = {10.1080/01621459.2016.1148039},
  url       = {https://doi.org/10.1080/01621459.2016.1148039},
  eprint    = {https://doi.org/10.1080/01621459.2016.1148039}
}

@article{mehrabani2023,
  title   = {Estimation and identification of latent group structures in panel data},
  journal = {Journal of Econometrics},
  volume  = {235},
  number  = {2},
  pages   = {1464-1482},
  year    = {2023},
  issn    = {0304-4076},
  doi     = {https://doi.org/10.1016/j.jeconom.2022.12.002},
  url     = {https://www.sciencedirect.com/science/article/pii/S030440762200207X},
  author  = {Ali Mehrabani}
}

@article{mehrabani2025,
  author    = {Ali Mehrabani and Shahnaz Parsaeian},
  title     = {Shrinkage Estimation and Identification of Latent Group Structures in Panel Data with Interactive Fixed Effects},
  journal   = {Journal of Business \& Economic Statistics},
  year      = {2025},
  note    = {In press},
  publisher = {ASA Website},
  doi       = {10.1080/07350015.2025.2571185},
  url       = {https://doi.org/10.1080/07350015.2025.2571185},
  eprint    = {https://doi.org/10.1080/07350015.2025.2571185}
}

@article{muller2014hac,
  author    = {Ulrich K. Müller},
  title     = {{HAC} Corrections for Strongly Autocorrelated Time Series},
  journal   = {Journal of Business \& Economic Statistics},
  volume    = {32},
  number    = {3},
  pages     = {311--322},
  year      = {2014},
  publisher = {ASA Website},
  doi       = {10.1080/07350015.2014.931238},
  url       = {https://doi.org/10.1080/07350015.2014.931238},
  eprint    = {https://doi.org/10.1080/07350015.2014.931238}
}

@article{newey1987simple,
  issn      = {00129682, 14680262},
  url       = {http://www.jstor.org/stable/1913610},
  author    = {Whitney K. Newey and Kenneth D. West},
  journal   = {Econometrica},
  number    = {3},
  pages     = {703--708},
  publisher = {[Wiley, Econometric Society]},
  title     = {A Simple, Positive Semi-Definite, Heteroskedasticity and Autocorrelation Consistent Covariance Matrix},
  urldate   = {2024-06-25},
  volume    = {55},
  year      = {1987}
}

@article{nickell1981biases,
  issn      = {00129682, 14680262},
  url       = {http://www.jstor.org/stable/1911408},
  author    = {Stephen Nickell},
  journal   = {Econometrica},
  number    = {6},
  pages     = {1417--1426},
  publisher = {[Wiley, Econometric Society]},
  title     = {Biases in Dynamic Models with Fixed Effects},
  urldate   = {2025-11-14},
  volume    = {49},
  year      = {1981}
}

@article{pesaran2006estimation,
  author   = {Pesaran, M. Hashem},
  title    = {Estimation and Inference in Large Heterogeneous Panels with a Multifactor Error Structure},
  journal  = {Econometrica},
  volume   = {74},
  number   = {4},
  pages    = {967-1012},
  doi      = {https://doi.org/10.1111/j.1468-0262.2006.00692.x},
  url      = {https://onlinelibrary.wiley.com/doi/abs/10.1111/j.1468-0262.2006.00692.x},
  eprint   = {https://onlinelibrary.wiley.com/doi/pdf/10.1111/j.1468-0262.2006.00692.x},
  year     = {2006}
}

@article{phillips2007bias,
  title   = {Bias in dynamic panel estimation with fixed effects, incidental trends and cross section dependence},
  journal = {Journal of Econometrics},
  volume  = {137},
  number  = {1},
  pages   = {162-188},
  year    = {2007},
  issn    = {0304-4076},
  doi     = {https://doi.org/10.1016/j.jeconom.2006.03.009},
  url     = {https://www.sciencedirect.com/science/article/pii/S030440760600042X},
  author  = {Peter C.B. Phillips and Donggyu Sul}
}

@article{qian2016shrinkage,
  title   = {Shrinkage estimation of common breaks in panel data models via adaptive group fused Lasso},
  journal = {Journal of Econometrics},
  volume  = {191},
  number  = {1},
  pages   = {86-109},
  year    = {2016},
  issn    = {0304-4076},
  doi     = {https://doi.org/10.1016/j.jeconom.2015.09.004},
  url     = {https://www.sciencedirect.com/science/article/pii/S0304407615002377},
  author  = {Junhui Qian and Liangjun Su}
}

@article{qian2016shrinkageET,
  title     = {Shrinkage estimation of regression models with multiple structural changes},
  author    = {Qian, Junhui and Su, Liangjun},
  journal   = {Econometric Theory},
  volume    = {32},
  number    = {6},
  pages     = {1376--1433},
  year      = {2016},
  publisher = {Cambridge University Press},
  doi       = {10.1017/S0266466615000237}
}

@article{robinson2012nonparametric,
  title   = {Nonparametric trending regression with cross-sectional dependence},
  journal = {Journal of Econometrics},
  volume  = {169},
  number  = {1},
  pages   = {4-14},
  year    = {2012},
  note    = {Recent Advances in Panel Data, Nonlinear and Nonparametric Models: A Festschrift in Honor of Peter C.B. Phillips},
  issn    = {0304-4076},
  doi     = {https://doi.org/10.1016/j.jeconom.2012.01.005},
  url     = {https://www.sciencedirect.com/science/article/pii/S0304407612000061},
  author  = {Peter M. Robinson}
}

@article{sarafidis2015partially,
  author  = {Sarafidis, Vasilis and Weber, Neville},
  title   = {A Partially Heterogeneous Framework for Analyzing Panel Data},
  journal = {Oxford Bulletin of Economics and Statistics},
  volume  = {77},
  number  = {2},
  pages   = {274-296},
  doi     = {https://doi.org/10.1111/obes.12062},
  url     = {https://onlinelibrary.wiley.com/doi/abs/10.1111/obes.12062},
  eprint  = {https://onlinelibrary.wiley.com/doi/pdf/10.1111/obes.12062},
  year    = {2015}
}

@article{scarpiniti2013nonlinear,
  title   = {Nonlinear spline adaptive filtering},
  journal = {Signal Processing},
  volume  = {93},
  number  = {4},
  pages   = {772-783},
  year    = {2013},
  issn    = {0165-1684},
  doi     = {https://doi.org/10.1016/j.sigpro.2012.09.021},
  url     = {https://www.sciencedirect.com/science/article/pii/S0165168412003568},
  author  = {Michele Scarpiniti and Danilo Comminiello and Raffaele Parisi and Aurelio Uncini}
}

@article{su2012sieve,
  title   = {Sieve estimation of panel data models with cross section dependence},
  journal = {Journal of Econometrics},
  volume  = {169},
  number  = {1},
  pages   = {34-47},
  year    = {2012},
  note    = {Recent Advances in Panel Data, Nonlinear and Nonparametric Models: A Festschrift in Honor of Peter C.B. Phillips},
  issn    = {0304-4076},
  doi     = {https://doi.org/10.1016/j.jeconom.2012.01.006},
  url     = {https://www.sciencedirect.com/science/article/pii/S0304407612000073},
  author  = {Liangjun Su and Sainan Jin}
}

@article{su2016identifying,
  author   = {Su, Liangjun and Shi, Zhentao and Phillips, Peter C. B.},
  title    = {Identifying Latent Structures in Panel Data},
  journal  = {Econometrica},
  volume   = {84},
  number   = {6},
  pages    = {2215-2264},
  doi      = {https://doi.org/10.3982/ECTA12560},
  url      = {https://onlinelibrary.wiley.com/doi/abs/10.3982/ECTA12560},
  eprint   = {https://onlinelibrary.wiley.com/doi/pdf/10.3982/ECTA12560},
  year     = {2016}
}

@article{su2019sieve,
  author    = {Liangjun Su and Xia Wang and Sainan Jin},
  title     = {Sieve Estimation of Time-Varying Panel Data Models With Latent Structures},
  journal   = {Journal of Business \& Economic Statistics},
  volume    = {37},
  number    = {2},
  pages     = {334--349},
  year      = {2019},
  publisher = {ASA Website},
  doi       = {10.1080/07350015.2017.1340299},
  url       = {https://doi.org/10.1080/07350015.2017.1340299},
  eprint    = {https://doi.org/10.1080/07350015.2017.1340299}
}

@article{vogt2020multiscale,
  title   = {Multiscale clustering of nonparametric regression curves},
  journal = {Journal of Econometrics},
  volume  = {216},
  number  = {1},
  pages   = {305-325},
  year    = {2020},
  note    = {Annals Issue in honor of George Tiao: Statistical Learning for Dependent Data},
  issn    = {0304-4076},
  doi     = {https://doi.org/10.1016/j.jeconom.2020.01.020},
  url     = {https://www.sciencedirect.com/science/article/pii/S0304407620300269},
  author  = {Michael Vogt and Oliver Linton}
}

@article{wagner2015environmental,
  author  = {Wagner, Martin},
  title   = {The Environmental Kuznets Curve, Cointegration and Nonlinearity},
  journal = {Journal of Applied Econometrics},
  volume  = {30},
  number  = {6},
  pages   = {948-967},
  doi     = {https://doi.org/10.1002/jae.2421},
  url     = {https://onlinelibrary.wiley.com/doi/abs/10.1002/jae.2421},
  eprint  = {https://onlinelibrary.wiley.com/doi/pdf/10.1002/jae.2421},
  year    = {2015}
}

@article{wang2017multi,
  title   = {A Multi-region Structural Decomposition Analysis of Global CO2 Emission Intensity},
  journal = {Ecological Economics},
  volume  = {142},
  pages   = {163-176},
  year    = {2017},
  issn    = {0921-8009},
  doi     = {https://doi.org/10.1016/j.ecolecon.2017.06.023},
  url     = {https://www.sciencedirect.com/science/article/pii/S0921800917303749},
  author  = {H. Wang and B.W. Ang and Bin Su}
}

@article{wang2018homogeneity,
  author  = {Wang, Wuyi and Phillips, Peter C. B. and Su, Liangjun},
  title   = {Homogeneity pursuit in panel data models: Theory and application},
  journal = {Journal of Applied Econometrics},
  volume  = {33},
  number  = {6},
  pages   = {797-815},
  doi     = {https://doi.org/10.1002/jae.2632},
  url     = {https://onlinelibrary.wiley.com/doi/abs/10.1002/jae.2632},
  eprint  = {https://onlinelibrary.wiley.com/doi/pdf/10.1002/jae.2632},
  year    = {2018}
}


\newpage
\doublespacing


\appendix
\counterwithin{figure}{section}
\counterwithin{table}{section}
\numberwithin{Lemma}{section}

\section{Details on the Sieve Estimation of Time-varying Coefficient Functions}
\label{sec:Appendix_sieve}

Consider a B-spline with $M^* > 0$ interior knots that is piece-wise polynomial of degree $d \geq 1$ (polynomial order $d + 1$) on the unit interval. The $M \times 1$ vector $\bs{b} (v)$ holds the common time-varying basis functions $\bs{b}(v) = (b_{-d}(v), \dots, b_{M^*}(v))^\prime$, with $M = M^* + d + 1$ and $v \in [0,1]$. Let $\mathbb{V}_{\text{int}}$ represent an increasing sequence $0 < v_1 < \dots < v_{M^*} < 1$, such that $\mathbb{V}_{\text{int}}$ gives $M^*$ equidistant interior knots of the B-spline, which divide the unit interval into $M^* + 1$ partitions.\footnote{We conjecture that the basic results of this paper also hold for free-knot spline functions, where the distance between interior knots may deviate from $1/(M^* + 1)$. However, altering the theory in this respect and introducing a data-driven knot placement routine is very involved \citep[see][]{hansen2002spline,scarpiniti2013nonlinear,dung2017direct}. A rigorous extension of the theory to free-knot splines is beyond the scope of this paper.} The total set of knots $\mathbb{V}$ extends $\mathbb{V}_{\text{int}}$ to $0 = v_{-d} = \dots = v_0 < v_1 < \dots < v_{M^*} < v_{M^* + 1} = \dots = v_{M} = 1$. The boundary knots $\{v_m\}_{m = -d}^0 \cup \{v_m\}_{m = M^* + 1}^{M}$ coincide at either zero or one and force the final B-spline to pass through the start and end points exactly using the \textit{de Boor recurrence relation}:
\begin{equation*}
    \label{eq:spline_rel}
    b_{m, j}(v) = a_{m, j}(v) b_{m, j-1}(v) + \left[1 - a_{m+1, j}(v) \right] b_{m+1, j-1}(v),
\end{equation*}
with $a_{m, j}(v) = \left[ (v - v_m) / (v_{m+j} - v_m) \right] \bs{1} \{v_{m+j} \neq v_m \}$, $b_{m,0}(v) = \bs{1} \{v_m \leq v < v_{m+1} \}$, and $b_{m,d}(v) = b_m(v)$ \citep[ch. 9, eq. 14]{de2001practical}. Hence, each basis function $b_m(v), \; m = -d, \dots, M^*$, defined on the knots $\{v_m\}_{m = -d}^{M}$, is a convex combination of two lower-order basis functions and vanishes outside the interval $\{ V_m \}_{m = -d}^{M^*}$ with
\begin{equation}
    \label{eq:b_spline_interval}
    V_m  = \left\{\begin{array}{lc}
        \left[v_m, v_{m + d + 1} \right),  & \text { for } m = -d, \dots, M^* - d - 1 \\
        \left[ v_m, v_{m + d + 1} \right], & \text { for } m = M^* - d, \dots, M^*.
    \end{array}\right.
\end{equation}
The space generated by these $M$ polynomial basis functions is denoted as $\mathbb{B}_M$ and each function in $\mathbb{B}_M$ is a B-spline that is piece-wise polynomial of degree $d$ on each sub-interval $\{V_m\}_{m = -d}^{M^*}$ and globally $d - 1$ times continuously differentiable for $d \geq 1$. We refer to \citet{de2001practical} for a textbook treatment of spline functions.

The $M \times 1$ vector of control points $\bs{\pi}_{il}^0$ weights each basis function in $\bs{b}(v)$ and, in turn, constructs a linear combination, the B-spline, that approximates a scalar square-integrable coefficient function $\beta_{il}^0 (v)$
\begin{equation}
    \label{eq:beta_approx}
    \beta_{il}^0 \left( v \right) = \bs{\pi}_{il}^{0 \prime} \bs{b} \left( v \right) + \eta_{il}, \quad l = 1, \dots, p.\footnote{A B-spline can be equivalently expressed as convex combinations of control points, with $B_{m,0}$ acting as indicators. In this case, recursively constructed convex combinations of control points yield the $M^*$ polynomial functions, which continuously tie together at the interior knots \citep[see][pp. 99]{de2001practical}.}
\end{equation}
The sieve approximation error $\eta_{il} (v) = \beta_{il}^0 \left( v \right) - \bs{\pi}_{il}^{0 \prime} \bs{b}(v)$ is zero if $\beta_{il}^0 (v) \in \mathbb{B}_M$ and $M$ is known. However, since $\beta_{il}^0 (v)$ may generally not belong to the linear space $\mathbb{B}_M$, we allow $M$ to increase with the sample size and obtain ever closer approximations of the true underlying coefficient function. $\mathbb{B}_M$ acts as a tractable sieve space, where increasing $M$ is akin to moving to an ever denser sieve. Throughout the paper, we assume the generalizing case $\beta_{il}^0(v) \notin \mathbb{B}_M$.

\begin{figure}[t]
    \centering
    \includegraphics[width=15cm]{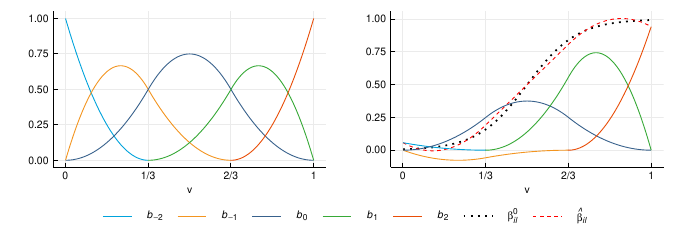}
    \caption[B-Spline illustration]{\small The left panel sketches five polynomial basis functions $b_m(v), \; m = -d, \dots, M^*$ for $d=2$ and $M^* = 2$. The right panel displays the weighted basis functions (solid) and the resulting B-spline (red, dashed), approximating a logistic CDF $\bs{\beta}^0_{il}(v)$ (black, dotted).}
    \label{fig:Splines}
\end{figure}

Figure \ref{fig:Splines} illustrates the approximation of a logistic cumulative distribution function (CDF) using a B-spline with $M^* = 2$ interior knots and polynomial degree $d = 2$, resulting in a system of $M = M^* + d + 1 = 5$ basis functions. The vector of control points $\bs{\pi}^0_{il}$ is estimated by regressing realizations of the function of interest, in this example, a logistic CDF, on $\bs{b}(v)$, using least squares \citep[][sec. 2]{huang2004polynomial}.

This technique readily extends to a $p$-dimensional functional vector $\bs{\beta}_i^0(v)$. Instead of a $M \times 1$ vector of control points $\bs{\pi}_{il}^0$, we now require a $M \times p$ matrix $\bs{\Pi}_i^0$, where each column of $\bs{\Pi}_i^0$ corresponds to one of the $p$ scalar functions in $\bs{\beta}_i^0(v)$. Subsequently,
$$
\bs{\beta}_{i}^0 \left( v \right) = \bs{\Pi}_{i}^{0 \prime} \bs{b} \left( v \right) + \bs{\eta}_{i},
$$
as in \eqref{eq:Splines}.

\section{Proof of the Results in Section \ref{sec:Asymptotics}}
\label{sec:Proofs}

\subsection{Technical Lemmas}
\label{sec:Lemma_statements}
\begin{Lemma}
\label{lma:spline_properties}
Let $b_m(v) > 0$, $m = -d, \dots, M^*$, denote a basis function of degree $d>0$ (order $d+1$), defined on the sequence of knots $\mathbb{V}$ as introduced in Appendix \ref{sec:Appendix_sieve}. 
\begin{enumerate}[label=(\roman*)]
\item\label{lma:spline_properties_1}
Then  $\| \bs{b}(v)\|_2 = \left[ \sum_{j = -d}^{M^*} b_m(v)^2 \right]^{1/2} \leq \left[ \sum_{j = -d}^{M^*} b_m(v) \right]^{1/2} = 1$.
\item\label{lma:spline_properties_2}
$\int_{0}^{1} b_m(v) \, dv = O(M^{-1})$ and $\int_{0}^{1} \| \bs{b}(v) \|_2 \, dv = \int_{0}^{1} \left[ \sum_{-d}^{M^*} b_m(v)^2 \right]^2 \, dv \leq \int_{0}^{1} \left[ \sum_{-d}^{M^*} b_m(v) \right]^2 \, dv = 1$. 
\item\label{lma:spline_properties_3}
There exist two constants $0 < \underbar{c}_{b} \leq \bar{c}_{b} < \infty$ such that $$ \underbar{c}_{b} \leq \mu_{\min} \left(M \int_{0}^{1} \bs{b} (v) \bs{b} (v)^\prime \,dv \right) \leq \mu_{\max} \left(M \int_{0}^{1} \bs{b} (v) \bs{b} (v)^\prime \,dv \right) \leq \bar{c}_{b}. $$ 
\item\label{lma:spline_properties_4}
Given Assumption \hyperref[line:A1]{1(vi)} and $\bs{\beta}_i^0(v) \notin \mathbb{B}_{\mathbb{V}, M}$, there exists a coefficient matrix $\bs{\Pi}_i^0 \in \mathbb{R}^{M \times p}$ such that $\sup_{v \in [0, 1]} \| \bs{\beta}_i^0(v) - \bs{\Pi}_i^{0 \prime} \bs{b} (v) \|_2 = O(M^{-\theta})$, where $\theta \geq 1$.
\end{enumerate}
\end{Lemma}


\begin{Lemma}
\label{lma:Q_zz}
Define $\hat{\bs{Q}}_{i, \tilde{z} \tilde{z}} = T^{-1} \sum_{t = 1}^T \tilde{\bs{z}}_{it} \tilde{\bs{z}}_{it}^\prime$ and $\hat{\bs{\mathcal{Q}}}_{i, \tilde{z} \tilde{z}} = \sum_{i \in G_k^0} T^{-1} \sum_{t = 1}^T \tilde{\bs{z}}_{it} \tilde{\bs{z}}_{it}^\prime$. Given Assumptions \hyperref[line:A1]{1(i)-(ii)}, \hyperref[line:A1]{1(iv)}, and \hyperref[line:A1]{1(vi)}, there exist two constants $0 < \underbar{c}_{\tilde{z} \tilde{z}} \leq \bar{c}_{\tilde{z} \tilde{z}} < \infty$ such that
\begin{enumerate}[label=(\roman*)]
    \item\label{lma:Q_zz_1} $\Pr \left( \underbar{c}_{\tilde{z} \tilde{z}} \leq \min_{i} \mu_{\min}(M \hat{\bs{Q}}_{i, \tilde{z} \tilde{z}}) \leq \max_{i} \mu_{\max}(M \hat{\bs{Q}}_{i, \tilde{z} \tilde{z}}) \leq \bar{c}_{\tilde{z} \tilde{z}} \right) = 1 - o(N^{-1}), $
    \item\label{lma:Q_zz_2} $\Pr \left( \underbar{c}_{\tilde{z} \tilde{z}} \leq \min_{i}  \mu_{\min}(N_k^{-1} M \hat{\bs{\mathcal{Q}}}_{i, \tilde{z} \tilde{z}}) \leq \max_{i}  \mu_{\max}(N_k^{-1} M \hat{\bs{\mathcal{Q}}}_{i, \tilde{z} \tilde{z}}) \leq \bar{c}_{\tilde{z} \tilde{z}} \right) = 1 - o(N^{-1}).$
\end{enumerate}
\end{Lemma}


\begin{Lemma}
\label{lma:comp_error}
Define $\hat{\bs{q}}_{i, \tilde{z} \tilde{u}} = T^{-1} \sum_{t = 1}^{T} \tilde{\bs{z}}_{it} \tilde{u}_{it}$, with the composite error $\tilde{u}_{it} = \tilde{\epsilon}_{it} + \tilde{\varrho}_{it}$, $\varrho_{it} = \bs{\eta}_{it}^\prime \bs{x}_{it}$.
Given Assumption \hyperref[line:A1]{1},
\begin{enumerate}[label=(\roman*)]
    \item\label{lma:comp_error_1} $\| \hat{\bs{q}}_{i, \tilde{z} \tilde{u}} \|_2 = O_p(T^{-1/2} M^{-\theta - 1/2})$,
    \item\label{lma:comp_error_2} $N^{-1} \sum_{i=1}^{N} \| \hat{\bs{q}}_{i, \tilde{z} \tilde{u}} \|_2^2 = O_p(T^{-1} + M^{-2 \theta - 1})$.
\end{enumerate}
\end{Lemma}


\begin{Lemma}
\label{lma:prelim_estimate}
Recall the preliminary estimate $\dot{\bs{\pi}} = \arg \min_{\bs{\pi}} T^{-1} \sum_{i = 1}^{N} \sum_{t=1}^{T} \left(\tilde{y}_{it} - \bs{\pi}_i^\prime \tilde{\bs{z}}_{it} \right)^2$. Given Assumptions \hyperref[line:A1]{1} and \hyperref[line:A2]{2(i)}, $\| \dot{\bs{\pi}}_i - \bs{\pi}_i^0 \|_2 = O_p(MT^{-1/2} + M^{-\theta - 1/2})$.
\end{Lemma}


\begin{Lemma}
\label{lma:adaptive_weight}
    Suppose that Assumptions \hyperref[line:A1]{1} and \hyperref[line:A2]{2(i)-(ii)} hold. Then 
    \begin{enumerate}[label=(\roman*)]
        \item\label{lma:adaptive_weight_1} $\min_{i,j \in G_k^0 } \dot{\omega}_{ij} = \|\dot{\bs{\pi}}_i -\dot{\bs{\pi}}_j \|_2^{-\kappa} = O_p((MT^{-1/2} + M^{-\theta + 1/2})^{-\kappa})$,
        \item\label{lma:adaptive_weight_2} $\max_{i \in G_k^0, j \notin G_k^0} \dot{\omega}_{ij} = O_p(J_{\min}^{-\kappa})$.
    \end{enumerate}
\end{Lemma}


\begin{Lemma}
\label{lma:dist}
    Suppose that Assumption \hyperref[line:A1]{1} is satisfied. Let $\bs{b}_c = \bs{c} \otimes \bs{b} (v) $ for some nonrandom $p \times 1$ vector $\bs{c}$ with $\| \bs{c} \|_2 = 1$. 
    \begin{enumerate}[label=(\roman*)]
        \item\label{lma:dist_1} $ \| \bs{b}_c \|_2 = 1$.
        \item\label{lma:dist_2} $s_{c, \hat{G}_k} = O_p(1)$, with \\$ s_{c, \hat{G}_k}^2 = M (N_k T)^{-1} \bs{b}_c^\prime \left(M N_k^{-1} \hat{\bs{\mathcal{Q}}}_{G_k^0, \tilde{z} \tilde{z}} \right)^{-1} \sum_{i \in G_k^0} \left( \tilde{\bs{Z}}_i^\prime E(\bs{\epsilon}_i \bs{\epsilon}_i^{\prime}) \tilde{\bs{Z}}_i \right) \left(M N_k^{-1} \hat{\bs{\mathcal{Q}}}_{G_k^0, \tilde{z} \tilde{z}} \right)^{-1} \bs{b}_c,$ and the individual terms as defined in the \hyperref[sec:3.5]{proof} of \cref{sec:Theo_3}.
    \end{enumerate}
\end{Lemma}


\begin{Lemma}
\label{lma:mse}
    Suppose that Assumptions \hyperref[line:A1]{1}-\hyperref[line:A4]{4} hold. Let $\bar{\mathbb{G}}_K = \{\bar{\mathcal{G}}_K = \{\bar{G}_k\}_{k = 1}^K: \nexists \, i,j \in \bar{G}_K \; \text{where} \; i \in G_l^0, \, j \notin G_l^0, \, 1 \leq l \leq K_0\}$ with $K_0<K \leq K_{\max} \leq N$.\footnote{$\bar{\mathbb{G}}_K$ denotes the set of all $K>K_0$ partitions over $N$ such that there exists excess groups, where no two heterogeneous individuals are pooled together. Subsequently, $\mathcal{G}^0$ can be retrieved just by merging certain groups of each $\bar{\mathcal{G}}_K \in \bar{\mathbb{G}}_K$.} Then, $\max_{K_0 < K\leq K_{\max}} \; \sup_{\bar{\mathcal{G}}_K \in \bar{\mathbb{G}}_K} |\hat{\sigma}^2_{\bar{\mathcal{G}}_K} - \hat{\sigma}^2_{\mathcal{G}^0} | = O_p ( T^{-1} M )$. The \textit{MSE} $\sigma^2_{\mathcal{G}_K}$ is defined as in Theorem \hyperref[sec:3.6]{3.6}.
\end{Lemma}

\subsection{Proofs of Theorems, Corollaries, and Lemmas}
\label{sec:Theorems}


\begin{proof}[{\textbf{Proof of \Cref{sec:Theo_1}}}]\label{sec:3.1}
\label{sec:3.1.1}
\label{sec:3.1.2}

Define $\bs{a}_i = \bs{\pi}_i - \bs{\pi}^0_i$. Recall the unit-specific criterion function $\mathcal{F}^*_{NT,i}(\bs{\pi}_i) = T^{-1}\sum_{t = 1}^T \left[ \tilde{y}_{it} - \bs{\pi}_i^\prime \tilde{\bs{z}}_{it} \right]^2$ and $\mathcal{F}_{NT,i}(\bs{\pi}_i, \lambda) = \mathcal{F}^*_{NT,i}(\bs{\pi}_i) + \lambda N^{-1} \sum_{j = 1, j\neq i}^N \dot{\omega}_{ij} \| \bs{\pi}_i - \bs{\pi}_j \|_2$ following \eqref{eq:Obj_penalty}.

\paragraph{\textnormal{\textit{Part (i):}}} Recognize that
\begin{equation}
    \label{eq:Q_diff}
    \begin{split}
         & \mathcal{F}^*_{NT,i}(\bs{\pi}_i) - \mathcal{F}^*_{NT,i}(\bs{\pi}_i^0) = \frac{1}{T} \sum_{t = 1}^T \left[ \tilde{y}_{it} - \bs{\pi}_i^\prime \tilde{\bs{z}}_{it} \right]^2 - \frac{1}{T}\sum_{t = 1}^T \left[\tilde{y}_{it} - \bs{\pi}_i^{0 \prime} \tilde{\bs{z}}_{it} \right]^2                                  \\
         & = \frac{1}{T} \sum_{t = 1}^T \left[\tilde{y}_{it} - (\bs{a}_i + \bs{\pi}_i^{0})^\prime \tilde{\bs{z}}_{it} \right]^2 - \frac{1}{T} \sum_{t = 1}^T \tilde{u}_{it}^2 = \frac{1}{T} \sum_{t = 1}^T \left[ \tilde{u}_{it} - \bs{a}_i^\prime \tilde{\bs{z}}_{it} \right]^2 - \frac{1}{T} \sum_{t = 1}^T \tilde{u}_{it}^2 \\
         & = \frac{1}{T} \sum_{t = 1}^T \tilde{u}_{it}^2 - \frac{1}{T} \sum_{t = 1}^T \left[ 2 \tilde{u}_{it} \bs{a}_i^\prime \tilde{\bs{z}}_{it} \right] + \frac{1}{T} \sum_{t = 1}^T \left[\bs{a}_i^\prime \tilde{\bs{z}}_{it}\right]^2 - \frac{1}{T} \sum_{t = 1}^T \tilde{u}_{it}^2                                       \\
         & = -2 \bs{a}_i^\prime \frac{1}{T} \sum_{t = 1}^T \left( \tilde{\bs{z}}_{it} \tilde{u}_{it} \right) + \bs{a}_i^\prime \frac{1}{T} \sum_{t = 1}^T \left( \tilde{\bs{z}}_{it} \tilde{\bs{z}}_{it}^\prime \right) \bs{a}_i                                                                                              \\
         & = -2 \bs{a}_i^\prime \hat{\bs{q}}_{i, \tilde{z}\tilde{u}} + \bs{a}_i^\prime \hat{\bs{Q}}_{i, \tilde{z}\tilde{z}} \bs{a}_i.
    \end{split}
\end{equation}
Notice that $\mathcal{F}_{NT,i}(\hat{\bs{\pi}}_i, \lambda) \leq \mathcal{F}_{NT,i}(\bs{\pi}_i^0, \lambda)$ holds trivially since $\hat{\bs{\pi}}_i$ estimates $\bs{\pi}_i^0$ by minimizing $\mathcal{F}_{NT,i}(\bs{\pi}_i, \lambda)$. Using the above decomposition \eqref{eq:Q_diff},
\begin{equation}
    \label{eq:T31_ineq}
    \begin{split}
        0 & \geq \mathcal{F}_{NT,i}(\hat{\bs{\pi}}_i, \lambda) - \mathcal{F}_{NT,i}(\bs{\pi}_i^0, \lambda)                                                                                                                                                                                                                 \\
          & = -2 \hat{\bs{a}}_i^\prime \hat{\bs{q}}_{i, \tilde{z}\tilde{u}} + \hat{\bs{a}}_i^\prime \hat{\bs{Q}}_{i, \tilde{z}\tilde{z}} \hat{\bs{a}}_i + \frac{\lambda}{N} \sum_{j = 1, j \neq i}^{N} \dot{\omega}_{ij} \left( \| \hat{\bs{\pi}}_i - \hat{\bs{\pi}}_j \|_2 - \| \bs{\pi}_i^0 - \bs{\pi}_j^0 \|_2 \right).
    \end{split}
\end{equation}

Let $i \in G_k^0$. Focusing on the penalty term in \eqref{eq:T31_ineq},
\begin{equation*}
    \label{eq:T31_penalty}
    \begin{split}
         & \frac{\lambda}{N} \sum_{j = 1, j \neq i}^{N} \dot{\omega}_{ij} \left( \| \hat{\bs{\pi}}_i - \hat{\bs{\pi}}_j \|_2 - \| \bs{\pi}_i^0 - \bs{\pi}_j^0 \|_2 \right) \\
         & \geq \frac{\lambda}{N} \sum_{j = 1, j \neq i}^{N} \dot{\omega}_{ij} \| \hat{\bs{\pi}}_i - \hat{\bs{\pi}}_j - \bs{\pi}_i^0 + \bs{\pi}_j^0 \|_2                   \\
         & \geq \frac{\lambda}{N} \sum_{j \notin G_k^0} \dot{\omega}_{ij} \| \hat{\bs{a}}_i - \hat{\bs{a}}_j \|_2.
    \end{split}
\end{equation*}
The first inequality holds due to the triangle inequality and the second inequality since all other individuals of group $k$ are discarded. Multiplying with -1 and employing the triangle inequality again gives
\begin{equation}
    \label{eq:T31_penalty2}
    \begin{split}
         & \frac{\lambda}{N} \sum_{j \notin G_k^0} \dot{\omega}_{ij} \| \hat{\bs{a}}_i - \hat{\bs{a}}_j \|_2   \geq - \frac{\lambda}{N} \sum_{j \notin G_k^0} \dot{\omega}_{ij} \| \hat{\bs{a}}_i + \hat{\bs{a}}_j \|_2 \geq - \frac{\lambda}{N} \sum_{j \notin G_k^0} \dot{\omega}_{ij} \left( \| \hat{\bs{a}}_i \|_2+ \| \hat{\bs{a}}_j \|_2 \right).
    \end{split}
\end{equation}

Plugging \eqref{eq:T31_penalty2} back into \eqref{eq:T31_ineq} and averaging across all individuals results in
\begin{equation}
    \label{eq:T31_penalty3}
    \begin{split}
        0 & \geq \frac{1}{N}  \sum_{i = 1}^{N} \left[ -2 \hat{\bs{a}}_i^\prime \hat{\bs{q}}_{i, \tilde{z}\tilde{u}} \right] + \frac{1}{N} \sum_{i = 1}^{N} \left[ \hat{\bs{a}}_i^\prime \hat{\bs{Q}}_{i, \tilde{z}\tilde{z}} \hat{\bs{a}}_i \right]                                                                                                                        \\
          & \qquad \qquad \qquad \qquad \qquad \qquad - \frac{\lambda}{N^2} \sum_{k = 1}^{K_0} \sum_{i \in G_k^0} \sum_{j \notin G_k^0} \left[ \dot{\omega}_{ij} \left( \| \hat{\bs{a}}_i \|_2+ \| \hat{\bs{a}}_j \|_2 \right) \right]                                                                                                                                       \\
          & \geq \frac{1}{N} \sum_{i = 1}^{N} \left[ -2 \hat{\bs{a}}_i^\prime \hat{\bs{q}}_{i, \tilde{z}\tilde{u}} \right] + \frac{1}{N} \sum_{i = 1}^{N} \left[ \hat{\bs{a}}_i^\prime \hat{\bs{Q}}_{i, \tilde{z}\tilde{z}} \hat{\bs{a}}_i \right]                                                                                                                         \\
          & \qquad \qquad \qquad \qquad \qquad \qquad - \frac{\lambda}{N^2} ( \max_{i \in G_k^0, j \notin G_k^0} \dot{\omega}_{ij} ) \sum_{k = 1}^{K_0} \sum_{i \in G_k^0} \sum_{j \notin G_k^0} \left[ \| \hat{\bs{a}}_i \|_2+ \| \hat{\bs{a}}_j \|_2 \right]                                                                                                                 \\
          & \geq \frac{1}{N}  \sum_{i = 1}^{N} \left[ -2 \hat{\bs{a}}_i^\prime \hat{\bs{q}}_{i, \tilde{z}\tilde{u}} \right] + \frac{1}{N}  \sum_{i = 1}^{N} \left[ \hat{\bs{a}}_i^\prime \hat{\bs{Q}}_{i, \tilde{z}\tilde{z}} \hat{\bs{a}}_i \right] - \frac{2 \lambda}{N} ( \max_{i \in G_k^0, j \notin G_k^0} \dot{\omega}_{ij} ) \sum_{i = 1}^{N} \\| \hat{\bs{a}}_i \|_2.
    \end{split}
\end{equation}

The second inequality in \eqref{eq:T31_penalty3} holds by taking the maximum adaptive penalty. In the following, we show the derivation of the third inequality in \eqref{eq:T31_penalty3} explicitly. Focusing on the third term of the second inequality in \eqref{eq:T31_penalty3},
\begin{equation*}
    \label{eq:T31Excursion}
    \begin{split}
         & \frac{\lambda}{N^2} ( \max_{i \in G_k^0, j \notin G_k^0} \dot{\omega}_{ij} ) \sum_{k = 1}^{K_0} \sum_{i \in G_k^0} \sum_{j \notin G_k^0} \left( \| \hat{\bs{a}}_i \|_2 + \| \hat{\bs{a}}_j \|_2 \right)                                                                                          \\
         & = \frac{\lambda}{N^2} ( \max_{i \in G_k^0, j \notin G_k^0} \dot{\omega}_{ij} ) \left[ \sum_{k = 1}^{K_0} \sum_{i \in G_k^0} \sum_{j \notin G_k^0} \left( \| \hat{\bs{a}}_i \|_2 \right) + \sum_{k = 1}^{K_0} \sum_{i \in G_k^0} \sum_{j \notin G_k^0} \left(\| \hat{\bs{a}}_j \|_2 \right) \right] \\
         & \leq \frac{\lambda}{N^2} ( \max_{i \in G_k^0, j \notin G_k^0} \dot{\omega}_{ij} ) \left[ \sum_{i = 1}^{N} \left( \| \hat{\bs{a}}_i \|_2 (N-1) \right) + \sum_{j = 1}^{N} \left(\| \hat{\bs{a}}_j \|_2 (N-1) \right) \right]                                                                    \\
         & = \frac{\lambda}{N^2} ( \max_{i \in G_k^0, j \notin G_k^0} \dot{\omega}_{ij} ) 2(N-1) \sum_{i = 1}^{N}  \| \hat{\bs{a}}_i \|_2                                                                                                                                                                 \\
         & \leq \frac{2 \lambda}{N} ( \max_{i \in G_k^0, j \notin G_k^0} \dot{\omega}_{ij} ) \sum_{i = 1}^{N} \| \hat{\bs{a}}_i \|_2,
    \end{split}
\end{equation*}
where one is the smallest possible group cardinality, which gives a maximum of $N -1$ summands in the third sum of $ \sum_{k = 1}^{K_0} \sum_{i \in G_k^0} \sum_{j \notin G_k^0}$.

Returning to \eqref{eq:T31_penalty3},
\begin{equation}
    \label{eq:T31semifin}
    \begin{split}
        0 & \geq \frac{1}{N} \sum_{i = 1}^{N} \left[ -2 \hat{\bs{a}}_i^\prime \hat{\bs{q}}_{i, \tilde{z}\tilde{u}} \right] + \frac{1}{N}  \sum_{i = 1}^{N} \left[ \hat{\bs{a}}_i^\prime \hat{\bs{Q}}_{i, \tilde{z}\tilde{z}} \hat{\bs{a}}_i \right] - \frac{2 \lambda}{N} ( \max_{i \in G_k^0, j \notin G_k^0} \dot{\omega}_{ij} ) \sum_{i = 1}^{N} \| \hat{\bs{a}}_i\|_2 \\
          & \geq \frac{1}{N} \sum_{i = 1}^{N} \left[ -2  \| \hat{\bs{a}}_i \|_2 \| \hat{\bs{q}}_{i, \tilde{z} \tilde{u}} \|_2 + \| \hat{\bs{a}}_i \|_2^2 \frac{\underbar{c}_{\tilde{z} \tilde{z}}}{M} - 2 \lambda ( \max_{i \in G_k^0, j \notin G_k^0} \dot{\omega}_{ij} ) \| \hat{\bs{a}}_i \|_2 \right] \\ 
          & = \frac{1}{N} \sum_{i = 1}^{N} \left[ \| \hat{\bs{a}}_i \|_2^2 - \| \hat{\bs{a}}_i \|_2 \frac{2M}{\underbar{c}_{\tilde{z} \tilde{z}}} \left( \| \hat{\bs{q}}_{i, \tilde{z} \tilde{u}} \|_2 + \lambda ( \max_{i \in G_k^0, j \notin G_k^0} \dot{\omega}_{ij} )  \right)  \right].
    \end{split}
\end{equation}
The inequality in \eqref{eq:T31semifin} holds due to the sub-multiplicative property regarding the first summand, and using \cref{lma:Q_zz}\ref{lma:Q_zz_1} to substitute $M \hat{\bs{Q}}_{i, \tilde{z} \tilde{z}}$ with $\underbar{c}_{\tilde{z} \tilde{z}}$ in the second summand. 

Recognize that $\max_{i \in G_k^0, j \notin G_k^0} \dot{\omega}_{ij} = O_p(J_{\min}^{-\kappa})$ as provided by \cref{lma:adaptive_weight}\ref{lma:adaptive_weight_2}, with $\lambda O_p( J_{\min}^{-\kappa}) = O_p(T^{-1/2} + M^{-\theta - 1/2})$ by Assumption \superref{line:A2}{2(iii)}. Furthermore, \cref{lma:comp_error}\ref{lma:comp_error_1} yields $\|\hat{\bs{q}}_{i, \tilde{z}\tilde{u}} \|_2 = O_p(T^{-1/2} + M^{-\theta - 1/2})$. Plugging these rates into \eqref{eq:T31semifin} we obtain
\begin{equation}
    \label{eq:T31semi2fin}
    \begin{split}
        0 & \geq \frac{1}{N} \sum_{i = 1}^{N} \left[ \| \hat{\bs{a}}_i \|_2^2 - \| \hat{\bs{a}}_i \|_2 \frac{4M}{\underbar{c}_{\tilde{z} \tilde{z}}} O_p (T^{-1/2} + M^{-\theta - 1/2}) \right]                                                              \\
          & = \frac{1}{N} \sum_{i = 1}^{N} \left[ O_p(MT^{-1/2} + M^{-\theta + 1/2})^{-2} \| \hat{\bs{a}}_i \|_2^2 - \frac{4}{\underbar{c}_{\tilde{z} \tilde{z}}} \| \hat{\bs{a}}_i \|_2 O_p(MT^{-1/2} + M^{-\theta + 1/2})^{-1} \right],
    \end{split}
\end{equation}
where the second equality is expanded with $(MT^{-1/2} + M^{-\theta + 1/2})^{-2}$. For a sufficiently large $\| \hat{\bs{a}}_i \|_2$, $\| \hat{\bs{a}}_i \|_2 \ll \| \hat{\bs{a}}_i \|_2^2$, in which case $\mathcal{F}_{NT,i}(\hat{\bs{\pi}}_i, \lambda)$ dominates $\mathcal{F}_{NT,i}(\bs{\pi}_i^0, \lambda)$ and the inequality \eqref{eq:T31semi2fin} cannot hold. Since $\mathcal{F}_{NT,i}(\bs{\pi}_i^0, \lambda)$ cannot be minimized and $\mathcal{F}_{NT,i}(\hat{\bs{\pi}}_i, \lambda) \leq \mathcal{F}_{NT,i}(\bs{\pi}_i^0, \lambda)$ must hold, $(MT^{-1/2} + M^{-\theta + 1/2})^{-1} \| \hat{\bs{a}}_i \|_2$ is stochastically bounded and $\|\hat{\bs{a}}_i \|_2 = \|\hat{\bs{\pi}}_i - \bs{\pi}_i^0 \|_2 = O_p(M T^{- 1/2} + M^{-\theta + 1/2})$. Assumption \hyperref[line:A2]{2(i)} ensures that $MT^{-1/2}$ is not explosive and $-\theta + 1/2 < 0$ holds by Assumption \hyperref[line:A1]{1(vi)}. This concludes the proof of Theorem \hyperref[sec:Theo_1]{3.1(i)}.

\paragraph{\textnormal{\textit{Part (ii):}}} Expanding upon the intermediary result in \eqref{eq:T31_penalty3} gives
\begin{equation*}
    \label{eq:T31_2}
    \begin{split}
        0 & \geq \mathcal{F}_{NT,i}(\hat{\bs{\pi}}_i, \lambda) - \mathcal{F}_{NT,i}(\bs{\pi}_i^0, \lambda)                                                                                                                                                                                                                                                                                                                                    \\
          & \geq \frac{1}{N}  \sum_{i = 1}^{N} \left[ -2 \hat{\bs{a}}_i^\prime \hat{\bs{q}}_{i, \tilde{z}\tilde{u}} \right] + \frac{1}{N} \sum_{i = 1}^{N} \left[ \hat{\bs{a}}_i^\prime \hat{\bs{Q}}_{i, \tilde{z}\tilde{z}} \hat{\bs{a}}_i \right] - \frac{2 \lambda}{N} ( \max_{i \in G_k^0, j \notin G_k^0} \dot{\omega}_{ij} ) \sum_{i = 1}^{N} \| \hat{\bs{a}}_i \|_2                                                                      \\
          & \geq -2 \left[ \frac{1}{N} \sum_{i = 1}^{N} \| \hat{\bs{a}}_i \|_2^2 \right]^{1/2} \left[\frac{1}{N} \sum_{i=1}^{N} \| \hat{\bs{q}}_{i, \tilde{z} \tilde{u}} \|_2^2 \right]^{1/2} + \frac{\underbar{c}_{\tilde{z} \tilde{z}}}{M} \frac{1}{N} \sum_{i=1}^{N} \| \hat{\bs{a}}_i \|_2^2 - \frac{2}{\sqrt{N}} \lambda ( \max_{i \in G_k^0, j \notin G_k^0} \dot{\omega}_{ij} ) \left[ \frac{1}{N} \sum_{i = 1}^{N} \| \hat{\bs{a}}_i \|_2^2 \right]^{1/2},
    \end{split}
\end{equation*}
where the last inequality holds because of (i) the Cauchy-Schwarz inequality in the first summand, (ii) by employing the lower bound $\underbar{c}_{\tilde{z} \tilde{z}}$ of the minimum eigenvalue of the predictor variance-covariance matrix in the second summand according to \cref{lma:Q_zz}\ref{lma:Q_zz_1}, and (iii) by the Cauchy-Schwarz inequality for vector spaces in the third summand. 

Isolating $1 / N \sum_{i=1}^{N} \| \hat{\bs{a}}_i \|_2^2$ and rearranging yields
\begin{equation*}
    \label{eq:T31_2_3}
        \frac{M}{\underbar{c}_{\tilde{z} \tilde{z}}} \left[ 2 \left[\frac{1}{N} \sum_{i=1}^{N} \| \hat{\bs{q}}_{i, \tilde{z} \tilde{u}} \|_2^2 \right]^{1/2} + \frac{2}{\sqrt{N}} \lambda ( \max_{i \in G_k^0, j \notin G_k^0} \dot{\omega}_{ij} )  \right] \left[ \frac{1}{N} \sum_{i = 1}^{N} \| \hat{\bs{a}}_i \|_2^2 \right]^{1/2}  \geq \frac{1}{N} \sum_{i=1}^{N} \| \hat{\bs{a}}_i \|_2^2
\end{equation*}
Again, recognize that $\lambda ( \max_{i \in G_k^0, j \notin G_k^0} \dot{\omega}_{ij} ) = O_p(T^{-1/2} + M^{-\theta - 1/2})$ by \cref{lma:adaptive_weight}\ref{lma:adaptive_weight_2} and Assumption \superref{line:A2}{2(iii)}. Furthermore, $N^{-1} \sum_{i = 1}^{N} \| \hat{\bs{q}}_{i, \tilde{z} \tilde{u}} \|_2^2 = O_p(T^{-1} + M^{-2\theta - 1})$ by \cref{lma:comp_error}\ref{lma:comp_error_2}. Putting these pieces together and collecting terms,
\begin{equation*}
    \label{eq:T31_2_2}
    \begin{split}
               \frac{2M}{\underbar{c}_{\tilde{z} \tilde{z}}} \left[ \left[ O_p(T^{-1} + M^{-2\theta - 1}) \right]^{1/2} + \frac{1}{\sqrt{N}} O_p(T^{-1/2} + M^{-\theta - 1/2} ) \right] \left[ \frac{1}{N} \sum_{i = 1}^{N} \| \hat{\bs{a}}_i \|_2^2 \right]^{1/2}  \geq \frac{1}{N} \sum_{i=1}^{N} \| \hat{\bs{a}}_i \|_2^2 \\ 
               \frac{2}{\underbar{c}_{\tilde{z} \tilde{z}}} \left[ O_p(M^2 T^{-1} + M^{-2\theta +1}) \frac{1}{N} \sum_{i=1}^{N} \| \hat{\bs{a}}_i \|_2^2 \right]^{1/2} \geq \frac{1}{N} \sum_{i=1}^{N} \| \hat{\bs{a}}_i \|_2^2,
    \end{split}
\end{equation*}
where by the same argument as in the \superref{sec:3.1}{proof} of \cref{sec:Theo_1}\ref{sec:Theo_1_1}, $N^{-1} \sum_{i=1}^{N} \| \hat{\bs{a}}_i \|_2^2 = N^{-1} \sum_{i=1}^{N} \| \hat{\bs{\pi}}_i - \bs{\pi}_i^0 \|_2^2 = O_p(M^2 T^{-1} + M^{-2 \theta + 1})$ holds to warrant $\mathcal{F}_{NT,i}(\hat{\bs{\pi}}_i, \lambda) \leq \mathcal{F}_{NT,i}(\bs{\pi}_i^0, \lambda)$.
\end{proof}


\begin{proof}[{\textbf{Proof of \Cref{sec:Coro_1}}}]
\label{sec:3.2}

 Following \citet[Corollary 4.1]{su2019sieve}, recall that $\hat{\bs{\beta}}_i(v) = \hat{\bs{\Pi}}_i^\prime \bs{b}(v) = (\hat{\bs{\Pi}}_i - \bs{\Pi}_i^0)^\prime \bs{b}(v) + \bs{\Pi}_i^{0 \prime} \bs{b}(v)$. Using this decomposition,
\begin{equation}
    \label{eq:beta_diff}
    \begin{split}
        \left\| \hat{\bs{\beta}}_i(v) - \bs{\beta}^0_i(v) \right\|_2 & = \left\| (\hat{\bs{\Pi}}_i - \bs{\Pi}_i^0)^\prime \bs{b}(v) + \bs{\Pi}_i^{0 \prime} \bs{b}(v) - \bs{\beta}^0_i(v) \right\|_2 \\
        & \leq \left\| (\hat{\bs{\Pi}}_i - \bs{\Pi}_i^0)^\prime \bs{b}(v) \right\|_2 + \left\| \bs{\Pi}_i^{0 \prime} \bs{b}(v) - \bs{\beta}^0_i(v) \right\|_2 = \left\| \bs{d}_{1i} \right\|_2 + \left\| \bs{d}_{2i} \right\|_2,
    \end{split}
\end{equation}
where the inequality holds because of the triangle property. 

\paragraph{\textnormal{\textit{Part (i):}}} Focusing on the first summand in \eqref{eq:beta_diff} and taking the supremum over the unit interval results in
\begin{equation*}
    \label{eq:A_1}
    \sup_{v \in [0,1]} \| \bs{d}_{1i} \|_2 = \sup_{v \in [0,1]} \left\| (\hat{\bs{\Pi}}_i - \bs{\Pi}_i^0)^\prime \bs{b}(v) \right\|_2 \leq \left\| \hat{\bs{\Pi}}_i - \bs{\Pi}_i^0 \right\|_F \sup_{v \in [0,1]} \left\| \bs{b}(v) \right\|_2 \leq \left\| \hat{\bs{\Pi}}_i - \bs{\Pi}_i^0 \right\|_F,
\end{equation*}
where the first inequality holds because of the Cauchy-Schwarz inequality and the second inequality since $\| \bs{b}(v) \|_2 \leq 1$ as shown in \cref{lma:spline_properties}\ref{lma:spline_properties_1}. Moreover, $ \left\| \hat{\bs{\Pi}}_i - \bs{\Pi}_i^0 \right\|_F = \left\| \hat{\bs{\pi}}_i - \bs{\pi}_i^0 \right\|_2 = O_p(MT^{-1/2} + M^{-\theta + 1/2})$ by \cref{sec:Theo_1}\ref{sec:Theo_1_1} and $\sup_{v \in [0,1]} \| \bs{d}_{2i} \|_2 = \sup_{v \in [0,1]} \left\| \bs{\Pi}_i^{0 \prime} \bs{b}(v) - \bs{\beta}^0_i(v) \right\|_2 = O(M^{-\theta})$ according to \cref{lma:spline_properties}\ref{lma:spline_properties_4}. As a consequence, $\sup_{v \in [0,1]} \left\| \hat{\bs{\beta}}_i(v) - \bs{\beta}^0_i(v) \right\|_2 = O_p(MT^{-1/2} + M^{-\theta + 1/2}) + O(M^{-\theta}) = O_p(MT^{-1/2} + M^{-\theta + 1/2})$.

\paragraph{\textnormal{\textit{Part (ii):}}} We employ the same decomposition as in \eqref{eq:beta_diff}, $\int_{0}^{1} \| \hat{\bs{\beta}}_i(v) - \bs{\beta}_i^0(v) \|_2^2 \,dv = \int_{0}^{1} \| \bs{d}_{1i} + \bs{d}_{2i} \|_2^2 \,dv$. By the parallelogram law $\int_{0}^{1} \| \bs{d}_{1i} + \bs{d}_{2i} \|_2^2 \,dv \leq 2 \int_{0}^{1} \| \bs{d}_{1i} \|_2^2 \,dv + 2 \int_{0}^{1} \| \bs{d}_{2i} \|_2^2 \,dv$.

Again, analyzing both summands on their own,
\begin{equation*}
    \label{eq:A_1_l2}
    \begin{split}
        \int_{0}^{1} \| \bs{d}_{1i} \|_2^2 \,dv & = \int_{0}^{1} \left\| (\hat{\bs{\Pi}}_i - \bs{\Pi}_i^0)^\prime \bs{b}(v) \right\|_2^2 \,dv = \text{tr} \left[(\hat{\bs{\Pi}}_i - \bs{\Pi}_i^0)^\prime \int_{0}^{1} \bs{b}(v) \bs{b}(v)^\prime \,dv (\hat{\bs{\Pi}}_i - \bs{\Pi}_i^0) \right] \\
                                                & = O(M^{-1}) \left\| \hat{\bs{\Pi}}_i - \bs{\Pi}_i^0 \right\|^2_F = O(M^{-1}) O_p(M^2 T^{-1} + M^{-2 \theta + 1}) = O_p(M T^{-1} + M^{-2 \theta}),
    \end{split}
\end{equation*}
since $\int_{0}^{1} \bs{b}(v) \bs{b}(v)^\prime \,dv = O(M^{-1})$ by \cref{lma:spline_properties}\ref{lma:spline_properties_3} and $\| \hat{\bs{\pi}}_i - \bs{\pi}_i^0 \|_2 = O_p(MT^{-1/2} + M^{-\theta + 1/2})$ by \cref{sec:Theo_1}\ref{sec:Theo_1_1}.

Following \cref{lma:spline_properties}\ref{lma:spline_properties_4}, $\int_{0}^{1} \| \bs{d}_{2i} \|_2^2 \,dv \leq \int_{0}^{1} \sup_{v \in [0,1]} \left\| \bs{\Pi}_i^{0 \prime} \bs{b}(v) - \bs{\beta}^0_i(v) \right\|_2^2 \,dv = O(M^{-2\theta})$. Subsequently, $\int_{0}^{1} \left\| \hat{\bs{\beta}}_i(v) - \bs{\beta}_i^0(v) \right\|_2^2 \,dv \leq O_p(M T^{-1} + M^{-2 \theta}) + O(M^{-2\theta}) = O_p(M T^{-1} + M^{-2 \theta})$.
\end{proof}


\begin{proof}[{\textbf{Proof of \Cref{sec:Theo_2}}}]\label{sec:3.3}
    Following \citet[Theorem 3.2]{mehrabani2023}, define the $Mp \times 1$ vector $\bs{a}_{ij} = {\bs{\pi}}_i - {\bs{\pi}}_j = (a_{ij,1}, \dots, a_{ij,Mp})^{\prime} $ and suppose there exists an $i \in G_k^0$ such that $\| {a}_{ij} \|_2 \neq 0$ for any $j \in G_k^0$. Let $|a_{ij,r}| = \max_{l = 1, \dots, Mp} | a_{ij,l} |$ and reorder $\bs{a}_{ij}$ such that $r = Mp$. As a consequence, each $| a_{ij,l} | \in [0, | a_{ij,Mp} |)$, $l = 1, \dots, Mp$. Hence $\| \bs{a}_{ij} \|_2 = \left[\sum_{l = 1}^{Mp} a_{ij,l}^2 \right]^{1/2} \leq \left[Mp a_{ij,Mp}^2 \right]^{1/2} = (Mp)^{1/2} |a_{ij,Mp}|$ and $(Mp)^{-1/2} \leq |a_{ij,Mp}| / \| \bs{a}_{ij} \|_2 \leq 1$.

Recall the criterion function
\begin{equation}
    \label{eq:T32_objective}
    \begin{split}
        \mathcal{F}_{NT,i}(\bs{\pi}_i, \lambda) & = \frac{1}{T} \sum_{t = 1}^T \left(\tilde{y}_{it} - \bs{\pi}_i^\prime \tilde{\bs{z}}_{it} \right)^2 + \frac{\lambda}{N} \sum_{j = 1, j\neq i}^N \dot{\omega}_{ij} \| \bs{\pi}_i - \bs{\pi}_j \|_2 \\
                                                & = \frac{1}{T} \sum_{t = 1}^T \left(\tilde{y}_{it}^2 - 2 \tilde{y}_{it} \bs{\pi}_i^\prime \tilde{\bs{z}}_{it} + \tilde{\bs{z}}_{it}^\prime \bs{\pi}_i \bs{\pi}_i^\prime \tilde{\bs{z}}_{it} \right) \\
                                                & \qquad \qquad \qquad \qquad \qquad \qquad + \frac{\lambda}{N} \sum_{j = 1, j\neq i}^N \dot{\omega}_{ij} \left[ \sum_{l=1}^{Mp} (\pi_{il}^2 - 2 \pi_{il} \pi_{jl} + \pi_{jl}^2)\right]^{1/2}.
    \end{split}
\end{equation}

Differentiating \eqref{eq:T32_objective} with respect to the scalar $\pi_{i,Mp}$ yields the first order condition (FOC)
\begin{equation}
    \label{eq:T32_FOC}
    \begin{split}
        \frac{\partial\mathcal{F}_{NT,i}}{\partial \pi_{i,Mp}} & \overset{!}{=} 0                                                                                                                                                                                                                                                      \\
                                                               & = -\frac{2}{T} \sum_{t=1}^{T} \tilde{y}_{it} \tilde{z}_{it,Mp} + \frac{2}{T} \sum_{t=1}^{T} \hat{\bs{\pi}}_i^\prime \tilde{\bs{z}}_{it} \tilde{z}_{it,Mp}                                                                                                             \\
                                                               & \qquad \qquad + \frac{\lambda}{N} \sum_{j = 1, j\neq i}^N \dot{\omega}_{ij} \left\{ \frac{1}{2} \left[ \sum_{l = 1}^{Mp} (\hat{\pi}^2_{il} - 2 \hat{\pi}_{il} \hat{\pi}_{jl} + \hat{\pi}^2_{jl})^2 \right]^{- 1/2} (2 \hat{\pi}_{i,Mp} - 2 \hat{\pi}_{j,Mp}) \right\} \\
                                                               & = -\frac{2}{T} \sum_{t=1}^{T} \tilde{z}_{it,Mp} (\tilde{y}_{it} - \hat{\bs{\pi}}_i^\prime \tilde{\bs{z}}_{it}) + \frac{\lambda}{N} \sum_{j = 1, j\neq i}^N \dot{\omega}_{ij} \frac{\hat{\pi}_{i,Mp} - \hat{\pi}_{j,Mp}}{\| \bs{\pi}_i - \bs{\pi}_j \|_2}.
    \end{split}
\end{equation}

Now, (i) expand \eqref{eq:T32_FOC} with $(T^{-1/2} + M^{-\theta - 1/2})^{-1}$, (ii) recognize $\hat{\bs{\pi}}_i = \bs{\pi}_i^0 + (\hat{\bs{\pi}}_i - \bs{\pi}_i^0)$, (iii) rewrite the penalty term to discriminate between $j \in G_k^0$ and $j \notin G_k^0$, and (iv) define $e_{ij,Mp} = (\hat{\pi}_{i,Mp} - \hat{\pi}_{j,Mp})/\| \hat{\bs{\pi}}_i - \hat{\bs{\pi}}_j \|_2$. This results in
\begin{equation}
    \label{eq:T32_FOC2}
    \begin{split}
        0 & = -2 (T^{-1/2} + M^{-\theta - 1/2})^{-1} \frac{1}{T} \sum_{t=1}^{T} \tilde{z}_{it,Mp} \left( \tilde{y}_{it} - \bs{\pi}_i^{0\prime} \tilde{\bs{z}}_{it} - (\hat{\bs{\pi}}_i - \bs{\pi}_i^{0})^\prime \tilde{\bs{z}}_{it} \right)                           \\
          & \qquad + (T^{-1/2} + M^{-\theta - 1/2})^{-1} \frac{\lambda}{N} \sum_{j \in G_k^0, j \neq i} \dot{\omega}_{ij} e_{ij,Mp} + (T^{-1/2} + M^{-\theta - 1/2})^{-1} \frac{\lambda}{N} \sum_{j \notin G_k^0, j \neq i} \dot{\omega}_{ij} e_{ij,Mp}              \\
          & = -2 (T^{-1/2} + M^{-\theta - 1/2})^{-1} \frac{1}{T} \sum_{t=1}^{T} \tilde{z}_{it,Mp} \tilde{u}_{it} + 2 (T^{-1/2} + M^{-\theta - 1/2})^{-1} \frac{1}{T} \sum_{t=1}^{T} \tilde{z}_{it,Mp} (\hat{\bs{\pi}}_i - \bs{\pi}_i^{0})^\prime \tilde{\bs{z}}_{it} \\
          & \qquad + (T^{-1/2} + M^{-\theta - 1/2})^{-1} \frac{\lambda}{N} \sum_{j \in G_k^0, j \neq i} \dot{\omega}_{ij} e_{ij,Mp} + (T^{-1/2} + M^{-\theta - 1/2})^{-1} \frac{\lambda}{N}  \sum_{j \notin G_k^0} \dot{\omega}_{ij} e_{ij,Mp},
    \end{split}
\end{equation}

\eqref{eq:T32_FOC2} is made up of four terms, say $0 = d_{1i} + d_{2i} + d_{3i} + d_{4i}$. Studying each summand in isolation,
\begin{itemize}
    \item $d_{1i} = -2 (T^{-1/2} + M^{-\theta - 1/2})^{-1} T^{-1} \sum_{t=1}^{T} \tilde{z}_{it,Mp} \tilde{u}_{it}$. Note that $\| \hat{\bs{q}}_{i, \tilde{z} \tilde{u}} \|_2 = O_p(T^{-1/2} + M^{-\theta - 1/2})$ by \cref{lma:Q_zz}\ref{lma:Q_zz_1}. Therefore, it is straightforward to see that $d_{1i}$ is $O_p(1)$.
    \item $d_{2i} = 2 (T^{-1/2} + M^{-\theta - 1/2})^{-1} T^{-1} \sum_{t=1}^{T} \tilde{z}_{it,Mp} (\hat{\bs{\pi}}_i - \bs{\pi}_i^{0})^\prime \tilde{\bs{z}}_{it}$. Since $\| \hat{\bs{q}}_{i,\tilde{z}_{Mp} \tilde{z}} (\hat{\bs{\pi}}_i - \bs{\pi}_i^{0}) \|_2  \leq \| \hat{\bs{q}}_{i,\tilde{z}_{Mp} \tilde{z}} \|_2 \| \hat{\bs{\pi}}_i - \bs{\pi}_i^{0} \|_2 = O_p(M^{-1}) O_p(M T^{-1/2} + M^{-\theta + 1/2}) = O_p(T^{-1/2} + M^{-\theta - 1/2})$ by \cref{lma:comp_error}\ref{lma:comp_error_1} and \cref{sec:Theo_1}\ref{sec:Theo_1_1}, where $T^{-1} \sum_{t=1}^{T} \tilde{z}_{it,Mp} \tilde{\bs{z}}_{it}^\prime = \hat{\bs{q}}_{i,\tilde{z}_{Mp} \tilde{z}}$, the rate of $d_{2i}$ follows as $O_p(1)$.
    \item $d_{3i} = (T^{-1/2} + M^{-\theta - 1/2})^{-1} \lambda N^{-1} \sum_{j \in G_k^0, j \neq i} \dot{\omega}_{ij} e_{ij,Mp}$. The lower bound $(Mp)^{-1/2} \leq |e_{ij,Mp}|$ gives the inequality $|d_{3i}| \geq  (T^{-1/2} + M^{-\theta - 1/2})^{-1} \lambda N^{-1} (Mp)^{-1/2} \sum_{j \in G_k^0, j \neq i} \dot{\omega}_{ij}$. Moreover, recall that \cref{lma:adaptive_weight}\ref{lma:adaptive_weight_1} states $\min_{i,j \in G_k^0} \dot{\omega}_{ij} = O_p((MT^{-1/2} + M^{-\theta + 1/2})^{-\kappa})$ and that $\sum_{j \in G_k^0, j \neq i}$ sums over $N_k - 1$ elements. As a consequence, $\sum_{j \in G_k^0, j \neq i} \dot{\omega}_{ij} \geq (N_k - 1) \min_{i,j \in G_k^0} \dot{\omega}_{ij} = (N_k - 1) O_p((MT^{-1/2} + M^{-\theta + 1/2})^{-\kappa})$. Combining all these elements yields $p^{1/2} |d_{3i}| \geq (T^{-1/2} + M^{-\theta - 1/2})^{-1} \lambda N^{-1} M^{-1/2} (N_k-1) O_p((MT^{-1/2} + M^{-\theta + 1/2})^{-\kappa})$. Note that by Assumption \superref{line:A1}{1(v)}, $(N_k - 1)N^{-1} = \tau_k - 1/N \geq 0$, where $\tau_k = N_k / N$. Rearranging and using the bound on $\tau_k$ yields $p^{1/2} |d_{3i}| \geq O_p((MT^{-1/2} + M^{-\theta + 1/2})^{-\kappa-1} M^{1/2} \lambda \tau_k)$, which, under Assumption \superref{line:A2}{2(iv)}, diverges to infinity in the limit.
    \item $d_{4i} = (T^{-1/2} + M^{-\theta - 1/2})^{-1} \lambda N^{-1}  \sum_{j \notin G_k^0} \dot{\omega}_{ij} e_{ij,Mp}$. Recall that by \cref{lma:adaptive_weight}\ref{lma:adaptive_weight_2}, $ \max_{i \in G_k^0} \dot{\omega}_{ij} = O_p(J^{-\kappa}_{\min})$. Furthermore, as shown above, $1 \geq | a_{ij,Mp} | / \| \bs{a}_{ij} \|_2 = |e_{ij,p}|$ and subsequently $\sum_{j \notin G_k^0} | e_{ij,Mp} | \leq N - N_k$ since $\sum_{j \notin G_k^0}$ sums over $N - N_k$ elements. Lastly, $| d_{4i} | \leq (T^{-1/2} + M^{-\theta - 1/2})^{-1} \lambda O_p(J_{min}^{-\kappa}) N^{-1} (N - N_k) = O_p(1)$ by Assumptions \superref{line:A2}{1(iv)} and \superref{line:A2}{2(iii)}.
\end{itemize}

Since $|d_{3i}| \gg |d_{1i} + d_{2i} + d_{3i}|$ cannot hold for large $(N,T)$, $\| \hat{\bs{\pi}}_i - \bs{\pi}_i^0 \|_2$ must not be differentiable w.p.a.1 for all $i,j \in G_k^0$ for $k = 1, \dots, K_0$ and $(T^{-1/2} + M^{-\theta - 1/2})^{-1} \lambda \dot{\omega}_{ij} e_{ij} = O_p(1)$ in \eqref{eq:T32_FOC2}. This translates to $\Pr(\| \hat{\bs{\pi}}_i - \hat{\bs{\pi}}_j \|_2 = 0) \to 1 \, \forall \, i,j \in G_k^0$ as $(N,T) \to \infty$.
\end{proof}

\begin{proof}[{\textbf{Proof of \Cref{sec:Coro_2}}}]
\item[] 
\paragraph{\textnormal{\textit{Part (i):}}} 
Theorem \hyperref[sec:3.3]{3.3} gives $\Pr(\| \hat{\bs{\pi}}_i - \bs{\pi}_i^0 \|_2 = 0) \to 1 \, \forall i,j \in G_k^0$ as $(N,T) \to \infty$. As a consequence, it also holds that $\lim_{(N,T) \to \infty} \Pr(\hat{K} = K_0) = 1$ since $\hat{K}$ is the number of unique subvectors of $\hat{\bs{\pi}} = (\hat{\bs{\pi}}_1^\prime, \dots, \hat{\bs{\pi}}_N^\prime)^\prime$.

\paragraph{\textnormal{\textit{Part (ii):}}} Suppose there exist two heterogeneous units  $i \in G_k^0, j \notin G_k^0$ that nonetheless exhibit $\| \hat{\bs{\pi}}_i - \hat{\bs{\pi}}_j \|_2 = 0$. Expanding $\hat{\bs{\pi}}_i - \hat{\bs{\pi}}_j$ by $\pm \bs{\pi}_i^0$ and recognizing $\| \hat{\bs{\pi}}_i - \bs{\pi}_i^0\|_2 = O_p(MT^{-1/2} + M^{-\theta+1/2})$ following Theorem \hyperref[sec:3.1.1]{3.1(i)}, we obtain
\begin{equation}
    \label{eq:Cor33}
    0 = \hat{\bs{\pi}}_i - \hat{\bs{\pi}}_j = \left( \bs{\pi}_i^0 + (\hat{\bs{\pi}}_i - \bs{\pi}_i^0) \right) - \left( \bs{\pi}_j^0 + (\hat{\bs{\pi}}_j - \bs{\pi}_j^0) \right) = \bs{\pi}_i^0 - \bs{\pi}_j^0 + O_p(MT^{-1/2} + M^{-\theta+1/2}).
\end{equation}
\eqref{eq:Cor33} makes it easy to see that if $\| \hat{\bs{\pi}}_i - \hat{\bs{\pi}}_j \|_2 = 0$, then $\| \bs{\pi}_i^0 - \bs{\pi}_j^0 \|_2 = O_p(MT^{-1/2} + M^{-\theta+1/2})$. Recall that $\| \bs{\pi}_i^0 - \bs{\pi}_j^0 \|_2 \geq J_{\min}$ by construction. However, Assumption \hyperref[line:A2]{2(ii)} states that $(MT^{-1/2} + M^{-\theta+1/2})^{-1} J_{\min} \to \infty$, in which case \eqref{eq:Cor33} cannot hold. In consequence, $\| \hat{\bs{\pi}}_i - \hat{\bs{\pi}}_j \|_2 \neq 0$ for all $i \in G_k^0, j \notin G_k^0$ and $\lim_{(N,T) \to \infty} \Pr(\hat{\mathcal{G}}_{K_0} = \mathcal{G}^0_{K_0}) = 1$.
\end{proof}

\begin{proof}[{\textbf{Proof of \Cref{sec:Theo_3}}}]
\label{sec:3.5}
\label{sec:3.5.1}
\label{sec:3.5.2}
\item[] 
\paragraph{\textnormal{\textit{Part (i):}}} Note the decomposition $\| \hat{\bs{\alpha}}_k(v) - \hat{\bs{\alpha}}_{\hat{G}_k}^p(v) \|_2 = \left\| (\hat{\bs{\Xi}}_k - \hat{\bs{\Xi}}_{\hat{G}_k}^p)^\prime \bs{b}(v) \right\|_2 \leq \left\| \hat{\bs{\Xi}}_k - \hat{\bs{\Xi}}_{\hat{G}_k}^p \right\|_F \| \bs{b}(v) \|_2$. Since $\| \bs{b}(v) \|_2 = 1$ by \cref{lma:spline_properties}\ref{lma:spline_properties_1}, it suffices to show that $(N_k T)^{1/2} M^{- 1/2} \left\| \hat{\bs{\Xi}}_k - \hat{\bs{\Xi}}_{\hat{G}_k}^p \right\|_F = o_p(1)$ in order to prove that the \textit{PSE} and \textit{post-Lasso} feature an identical limiting distribution.\footnote{The $\sqrt{N_k T /M}$ scaling is required for the \textit{post-Lasso} estimator to converge in distribution. See the \superref{sec:3.5.2}{proof} of \cref{sec:Theo_3}\ref{sec:Theo_3_2}.}

Recall the FOC from the \superref{sec:3.3}{proof} of \cref{sec:Theo_2}, \eqref{eq:T32_FOC}
\begin{equation*}
    \label{eq:T35_FOC}
    \frac{\partial \mathcal{F}_{NT,i}}{\partial \bs{\pi}_{i}} = 0 = -\frac{2}{T} \sum_{t=1}^{T} \tilde{\bs{z}}_{it} (\tilde{y}_{it} - \hat{\bs{\pi}}_i^\prime \tilde{\bs{z}}_{it}) + \frac{\lambda}{N} \sum_{j = 1, j\neq i}^N \dot{\omega}_{ij} (\hat{\bs{\pi}}_{i} - \hat{\bs{\pi}}_{j}) \| \hat{\bs{\pi}}_i - \hat{\bs{\pi}}_j \|_2^{-1}.
\end{equation*}

Write $\hat{\bs{\pi}}_i = \hat{\bs{\xi}}_k$ for any $i \in \hat{G}_k$ and sum over all $i \in \hat{G}_k$ to obtain
\begin{equation}
    \label{eq:T35_FOC_2}
    \begin{split}
        0 & = - \frac{2}{T} \sum_{i \in \hat{G}_k} \sum_{t=1}^{T} \tilde{\bs{z}}_{it} \left[\tilde{y}_{it} - \hat{\bs{\xi}}_k^\prime \tilde{\bs{z}}_{it} \right] + \frac{\lambda}{N} \sum_{i \in \hat{G}_k} \sum_{j = 1, j\neq i}^N \dot{\omega}_{ij} (\hat{\bs{\pi}}_{i} - \hat{\bs{\pi}}_{j}) \| \hat{\bs{\pi}}_i - \hat{\bs{\pi}}_j \|_2^{-1} \\
          & = - \frac{2}{T} \sum_{i \in \hat{G}_k} \sum_{t=1}^{T} \left[\tilde{\bs{z}}_{it} \tilde{y}_{it} - \tilde{\bs{z}}_{it} \hat{\bs{\xi}}_k^\prime \tilde{\bs{z}}_{it} \right] + \hat{\bs{r}}_{\hat{G}_k}                                                                                                                                  \\
          & = -2 \left[\hat{\bs{q}}_{\hat{G}_k, \tilde{z} \tilde{y}} - \hat{\bs{\mathcal{Q}}}_{\hat{G}_k, \tilde{z} \tilde{z}} \hat{\bs{\xi}}_k \right] + \hat{\bs{r}}_{\hat{G}_k},
    \end{split}
\end{equation}
where $\hat{\bs{r}}_{\hat{G}_k} = \lambda N^{-1} \sum_{i \in \hat{G}_k} \sum_{j = 1, j\neq i}^N \dot{\omega}_{ij} \hat{\bs{e}}_{ij}$, $\hat{\bs{e}}_{ij} = (\hat{\bs{\pi}}_{i} - \hat{\bs{\pi}}_{j}) \| \hat{\bs{\pi}}_i - \hat{\bs{\pi}}_j \|_2^{-1}$, $\hat{\bs{q}}_{\hat{G}_k,\tilde{z} \tilde{y}} = \sum_{i \in \hat{G}_k} T^{-1} \sum_{t=1}^{T} \tilde{\bs{z}}_{it} \tilde{y}_{it}$, and $\hat{\bs{\mathcal{Q}}}_{\hat{G}_k, \tilde{z} \tilde{z}} = \sum_{i \in \hat{G}_k} T^{-1} \sum_{t=1}^{T} \tilde{\bs{z}}_{it} \tilde{\bs{z}}_{it}^\prime$.

Solving \eqref{eq:T35_FOC_2} for $\hat{\bs{\xi}}_k$ gives $$
    \hat{\bs{\xi}}_k = \hat{\bs{\mathcal{Q}}}_{\hat{G}_k, \tilde{z} \tilde{z}}^{-1} \hat{\bs{q}}_{\hat{G}_k, \tilde{z} \tilde{y}} - 1/2 \hat{\bs{\mathcal{Q}}}_{\hat{G}_k, \tilde{z} \tilde{z}}^{-1} \hat{\bs{r}}_{\hat{G}_k} = \hat{\bs{\xi}}_{\hat{G}_k}^p - 1/2 \hat{\bs{\mathcal{Q}}}_{\hat{G}_k, \tilde{z} \tilde{z}}^{-1} \hat{\bs{r}}_{\hat{G}_k}.
$$ It is straightforward to see that the \textit{PSE} estimator $\hat{\bs{\xi}}_k$ and the \textit{post-Lasso} $\hat{\bs{\xi}}_{\hat{G}_k}^p$ are asymptotically equivalent if $\hat{\bs{\mathcal{Q}}}_{\hat{G}_k, \tilde{z} \tilde{z}}^{-1} \hat{\bs{r}}_{\hat{G}_k} = o_p(1)$. We employ \cref{sec:Coro_2} and make use of the oracle property to demonstrate this result. Subsequently, one can infer the limiting behavior of $\hat{\bs{\mathcal{Q}}}_{\hat{G}_k, \tilde{z} \tilde{z}}^{-1} \hat{\bs{r}}_{\hat{G}_k}$ by studying $\hat{\bs{\mathcal{Q}}}_{G_k^0, \tilde{z} \tilde{z}}^{-1} \hat{\bs{r}}_{G^0_k}$. Scaling by $\sqrt{N_k T / M}$ and taking the squared $L_2$ norm yields
\begin{equation}
    \label{eq:limit_resid}
    \begin{split}
         & \left\| \sqrt{\frac{N_k T}{M}} \hat{\bs{\mathcal{Q}}}_{G_k^0, \tilde{z} \tilde{z}}^{-1} \hat{\bs{r}}_{G^0_k} \right\|_2^2 = \left\| \left( \frac{M}{N_k} \hat{\bs{\mathcal{Q}}}_{G_k^0, \tilde{z} \tilde{z}} \right)^{-1} \sqrt{\frac{T M}{N_k}} \hat{\bs{r}}_{G^0_k} \right\|_2^2 \\
         & \leq \left[ \mu_{\min} \left(\frac{M}{N_k} \hat{\bs{\mathcal{Q}}}_{G_k^0, \tilde{z} \tilde{z}} \right) \right]^{-2} \left\| \sqrt{\frac{T M}{N_k}} \hat{\bs{r}}_{G^0_k} \right\|_2^2,
    \end{split}
\end{equation}
where $\underbar{c}_{\tilde{z} \tilde{z}} \leq \mu_{\min} \left( M N_k^{-1} \hat{\bs{\mathcal{Q}}}_{G_k^0, \tilde{z} \tilde{z}} \right)$ by \cref{lma:Q_zz}\ref{lma:Q_zz_2}. Focusing on the norm in \eqref{eq:limit_resid},
\begin{equation}
    \label{eq:resid_2}
    \begin{split}
        \left\| \sqrt{\frac{T M}{N_k}} \hat{\bs{r}}_{G^0_k} \right\|_2^2 & = \frac{TM}{N_k} \frac{\lambda^2}{N^2} \left\| \sum_{i \in G_k^0} \sum_{j \notin G_k^0} \dot{\omega}_{ij} \hat{\bs{e}}_{ij} \right\|_2^2                                                               \\
                                                                         & \leq \frac{TM}{N_k} \frac{\lambda^2}{N^2} \left( \sum_{i \in G_k^0} \sum_{j \notin G_k^0} | \dot{\omega}_{ij} | \| \hat{\bs{e}}_{ij} \|_2 \right)^2                                                    \\
                                                                         & \leq \frac{TM}{N_k} \frac{\lambda^2}{N^2} \left( \max_{i \in G_k^0, j \notin G_k^0} | \dot{\omega}_{ij} |\right)^2 \left( \sum_{i \in G_k^0} \sum_{j \notin G_k^0} \| \hat{\bs{e}}_{ij} \|_2 \right)^2 \\
                                                                         & \leq TM \lambda^2 \left(\max_{i \in G_k^0, j \notin G_k^0} | \dot{\omega}_{ij} |\right)^2 \frac{N_k^2 (N - N_k)^2}{N^2 N_k}                                                                            \\
                                                                         & \leq TM \lambda^2 \left(\max_{i \in G_k^0, j \notin G_k^0} | \dot{\omega}_{ij} |\right)^2 N_k,
    \end{split}
\end{equation}
where the first inequality by the Cauchy-Schwarz inequality; the second inequality holds by taking the maximum adaptive weight; the third inequality holds by recognizing that $\| \hat{\bs{e}}_{ij} \|_2 \leq 1$ by the Triangle inequality; the fourth inequality holds by taking $N^{-2} N_k (N - N_k)^2 = N_k (1 - 2 \tau_k + \tau_k^2) \leq N_k$ using Assumption \superref{line:A1}{1(v)}, where $\tau_k = N_k / N$. Furthermore, note that $\max_{i \in G_k^0, j \in G_k^0} \left(| \dot{\omega}_{ij} |^2\right) = O_p(J_{\min}^{-2\kappa})$ by \cref{lma:adaptive_weight}\ref{lma:adaptive_weight_2}. Plugging these pieces into \eqref{eq:resid_2}, it becomes apparent that
\begin{equation*}
    \label{eq:resid_3}
    \left\| \sqrt{\frac{TM}{N_k}} \hat{\bs{r}}_{G^0_k} \right\|_2^2 \leq TM \lambda^2 O_p(J_{\min}^{-2\kappa}) N_k = O_p(\sqrt{N_k T M} \lambda J_{\min}^{-\kappa})^2.
\end{equation*}
Assumption \superref{line:A3}{3} gives $O_p(\sqrt{N_k T M} \lambda J_{\min}^{-\kappa}) = o_p(1)$. In consequence, the whole term \eqref{eq:limit_resid} becomes negligible in the limit and $\| \hat{\bs{\alpha}}_k(v) - \hat{\bs{\alpha}}_{\hat{G}_k}^p(v) \|_2 = o_p(1)$.\\

\paragraph{\textnormal{\textit{Part (ii):}}} We make use of the decomposition
\begin{equation}
    \label{eq:alpha_decomp}
    \begin{split}
        \sqrt{\frac{N_k T}{M}} \left[\hat{\bs{\alpha}}_{\hat{G}_k}^p(v) - \bs{\alpha}^0_k(v) \right] & = \sqrt{\frac{N_k T}{M}} (\hat{\bs{\Xi}}_{\hat{G}_k}^p - \bs{\Xi}^0_k)^\prime \bs{b}(v) + \sqrt{\frac{N_k T}{M}} (\bs{\Xi}_k^{0\prime} \bs{b}(v) - \bs{\alpha}_k^0(v)) \\
                                                                                                     & = \bs{d}_{1,\hat{G}_k} + \bs{d}_{2,\hat{G}_k},
    \end{split}
\end{equation}
where $\bs{\alpha}_k^0(v) = \bs{\Xi}_k^{0 \prime} \bs{b}(v) + (\bs{\alpha}_k^0(v) - \bs{\Xi}_k^{0 \prime} \bs{b}(v))$.

It is easy to show that $\bs{d}_{2,\hat{G}_k} / s_{c, \hat{G}_k} = \sqrt{N_k T / M} O( M^{-\theta}) O_p(1) = o_p(1)$, since $\|\bs{\alpha}_k^0(v) - \bs{\Xi}_k^{0 \prime} \bs{b}(v) \|_2 = O(M^{-\theta})$ by \cref{lma:spline_properties}\ref{lma:spline_properties_4}, $\sqrt{N_k T / M^{2\theta}} = o_p(1)$ according to Assumption \superref{line:A4}{4(ii)}, and $s_{c, \hat{G}_k} = O_p(1)$ by \cref{lma:dist}\ref{lma:dist_2}.\footnote{$s_{c,G_k^0}$ is a scaling factor that is introduced at a later stage.} As a result, the second summand is negligible for the remainder of the proof and it is sufficient to study $\bs{d}_{1,\hat{G}_k}$ only.

Regarding $\bs{d}_{1,\hat{G}_k}$, consider $\bs{d}_{1,\hat{G}_k}^\prime \bs{c}$ for some nonrandom $p \times 1$ vector $ \bs{c}$ with $\| \bs{c} \|_2=1$. Define $ \bs{b}_{c} =  \bs{c} \otimes \bs{b}(v)$. Recognize that $\text{vec} \left( \bs{b}^\prime(v) (\hat{\bs{\Xi}}_{\hat{G}_k}^p - \bs{\Xi}^0_k) \bs{c} \right)=(\bs{c}^\prime \otimes \bs{b}(v)) \text{vec} \left(\hat{\bs{\Xi}}_{\hat{G}_k}^p - \bs{\Xi}^0_k \right)$ \citep[see][p. 249]{bernstein2009matrix}. In addition, let $\hat{\bs{q}}_{\hat{G}_k,\tilde{z} a} = \sum_{i \in \hat{G}_k} T^{-1} \sum_{t=1}^{T} \tilde{ \bs{z}}_{it} a_{it}$ for $a = \{\tilde{y}, \tilde{u}, \tilde{\epsilon}\}$. Recall that the error term $\tilde{u}_{it}$ can be decomposed into an idiosyncratic and a sieve component $\tilde{u}_{it} = \tilde{\varrho}_{it} + \tilde{\epsilon}_{it}$ where $\varrho_{it} = \bs{\eta}_{it}^\prime \bs{x}_{it}$ (see, e.g., \eqref{eq:DGP_spline}). Then,
\begin{equation}
    \label{eq:B1_1}
    \begin{split}
        \bs{d}_{1,\hat{G}_k}^\prime \bs{c} & = \sqrt{\frac{N_k T}{M}} \bs{b}^{\prime}(v) (\hat{\bs{\Xi}}_{\hat{G}_k}^p - \bs{\Xi}^0_k) \bs{c}  = \sqrt{\frac{N_k T}{M}} \bs{b}_{c}^\prime (\hat{\bs{\xi}}_{\hat{G}_k}^p - \bs{\xi}^0_k) = \sqrt{\frac{N_k T}{M}} \bs{b}_{c}^\prime \left[ \hat{\bs{\mathcal{Q}}}_{\hat{G}_k,\tilde{z} \tilde{z}}^{-1} \hat{\bs{q}}_{\hat{G}_k, \tilde{z} \tilde{y}} - \bs{\xi}^0_k \right] \\
                                           & = \sqrt{\frac{N_k T}{M}} \bs{b}_{c}^\prime \hat{\bs{\mathcal{Q}}}_{\hat{G}_k,\tilde{z} \tilde{z}}^{-1} \left[ \hat{\bs{q}}_{\hat{G}_k, \tilde{z} \tilde{y}} - \hat{\bs{\mathcal{Q}}}_{\hat{G}_k,\tilde{z} \tilde{z}} \bs{\xi}^0_k \right]                                                                                                                                     \\
                                           & = \sqrt{\frac{N_k T}{M}} \bs{b}_{c}^\prime \hat{\bs{\mathcal{Q}}}_{\hat{G}_k,\tilde{z} \tilde{z}}^{-1} \left[ \hat{\bs{q}}_{\hat{G}_k,\tilde{z} \tilde{u}} + \sum_{i \in \hat{G}_k} \frac{1}{T} \sum_{t = 1}^{T} \bs{z}_{it} \bs{z}_{it}^\prime (\bs{\pi}_i^0 - \bs{\xi}_k^0) \right]                                                                                         \\
                                           & = \sqrt{\frac{N_k T}{M}} \bs{b}_{c}^\prime \hat{\bs{\mathcal{Q}}}_{\hat{G}_k,\tilde{z} \tilde{z}}^{-1} \hat{\bs{q}}_{\hat{G}_k,\tilde{z} \tilde{\epsilon}} + \sqrt{\frac{N_k T}{M}} \bs{b}_{c}^\prime \hat{\bs{\mathcal{Q}}}_{\hat{G}_k,\tilde{z} \tilde{z}}^{-1} \hat{\bs{q}}_{\hat{G}_k,\tilde{z} \tilde{\varrho}}                                                             \\
                                           & \qquad \qquad \qquad \qquad \qquad \qquad \qquad + \sqrt{\frac{N_k}{T M}} \bs{b}_{c}^\prime \hat{\bs{\mathcal{Q}}}_{\hat{G}_k,\tilde{z} \tilde{z}}^{-1} \sum_{i \in \hat{G}_k} \sum_{t = 1}^{T} \bs{z}_{it} \bs{z}_{it}^\prime (\bs{\pi}_i^0 - \bs{\xi}_k^0).
    \end{split}
\end{equation}

Exploiting the oracle property by \cref{sec:Coro_2}, we study $\bs{d}_{1,\hat{G}_k}$ by analyzing $\bs{d}_{1,G_k^0}$. Subsequently,
\begin{equation}
    \label{eq:B1_G0}
    \begin{split}
        \bs{d}_{1,G_k^0}^\prime \bs{c} & = \sqrt{\frac{N_k T}{M}} \bs{b}_{c}^\prime \hat{\bs{\mathcal{Q}}}_{G_k^0,\tilde{z} \tilde{z}}^{-1} \hat{\bs{q}}_{G_k^0,\tilde{z} \tilde{\epsilon}} + \sqrt{\frac{N_k T}{M}} \bs{b}_{c}^\prime \hat{\bs{\mathcal{Q}}}_{G_k^0,\tilde{z} \tilde{z}}^{-1} \hat{\bs{q}}_{\hat{G}_k,\tilde{z} \tilde{\varrho}} \\
                                       & = d_{11,G_k^0} + d_{12,G_k^0},
    \end{split}
\end{equation}
where the last summand in \eqref{eq:B1_1} equals zero since $\bs{\pi}_i^0 = \bs{\xi}_k^0$ for $i \in G_k^0$.

We show that $d_{12,G_k^0}$ vanishes in the limit before studying the asymptotic behavior of $d_{11,G_k^0}$. But first, note two preliminary results that are used later.
\begin{equation}
    \label{eq:eta_limit}
    \begin{split}
        \frac{1}{N_k T} \sum_{i \in G_k^0} \sum_{t = 1}^{T} \tilde{\varrho}_{it}^2 & = \frac{1}{N_k T} \sum_{i \in G_k^0} \sum_{t = 1}^{T} \left(\left[ \bs{\beta}_i^0 \left(\frac{t}{T}\right) - \bs{\Pi}_i^{0 \prime} \bs{b} \left(\frac{t}{T}\right)\right]^\prime \tilde{\bs{x}}_{it} \right)^2      \\
                                                                                & \leq \left[\sup_{v \in [0,1]} \left\| \bs{\alpha}_k^0 (v) - \bs{\Xi}_k^{0 \prime} \bs{b}(v) \right\|_2 \right]^2 \frac{1}{N_k T} \sum_{i \in G_k^0} \sum_{t = 1}^{T} \tilde{\bs{x}}_{it} \tilde{\bs{x}}_{it}^\prime \\
                                                                                & \leq O_p(M^{-\theta})^2 \bar{c}_{xx} (1 + o_p(1)) = O_p(M^{-2\theta}),
    \end{split}
\end{equation}
where the rates are given by \cref{lma:spline_properties}\ref{lma:spline_properties_4} and Assumption \superref{line:A1}{1(iv)}. In addition,
\begin{equation}
    \label{eq:z_limit}
    \begin{split}
         & \frac{M}{N_k T} \sum_{i \in G_k^0} \sum_{t = 1}^{T} \left[ \bs{b}_{c}^\prime \left( M N_k^{-1} \hat{\bs{\mathcal{Q}}}_{G_k^0, \tilde{z} \tilde{z}} \right)^{-1} \tilde{\bs{z}}_{it} \right]^2                                                                                                                          \\
         & = \frac{M}{N_k T} \sum_{i \in G_k^0} \sum_{t = 1}^{T} \left[ \bs{b}_{c}^\prime \left( M N_k^{-1} \hat{\bs{\mathcal{Q}}}_{G_k^0, \tilde{z} \tilde{z}} \right)^{-1} \tilde{\bs{z}}_{it} \tilde{\bs{z}}_{it}^\prime \left( M N_k^{-1} \hat{\bs{\mathcal{Q}}}_{G_k^0, \tilde{z} \tilde{z}} \right)^{-1} \bs{b}_{c} \right] \\
         & = \frac{M}{N_k T} \sum_{i \in G_k^0} \sum_{t = 1}^{T} \left[ \bs{\varpi}^{\prime} \tilde{\bs{z}}_{it} \tilde{\bs{z}}_{it}^\prime \bs{\varpi}\right]                                                                                                                                                                    \\
         & = \frac{M}{N_k T} \sum_{i \in G_k^0} \sum_{t = 1}^{T} \left[ \bs{g}_{\varpi}^\prime \left(\frac{t}{T}\right) \tilde{\bs{x}}_{it} \tilde{\bs{x}}_{it}^\prime \bs{g}_{\varpi} \left(\frac{t}{T}\right) \right],
    \end{split}
\end{equation}
where $\bs{g}_{\varpi}(v)$ is a $p \times 1$ vector of spline functions $\bs{g}_{\varpi}(v) = (g_1(v, \bs{\varpi}_1), \dots, g_p(v, \bs{\varpi}_p))^\prime$, with $g_l(v, \varpi_l) = \bs{\varpi}_l^{\prime} \bs{b}(v)$, and $\bs{\varpi}_l = (\varpi_{l1}, \dots, \varpi_{lM})^\prime$. The $M \times p$ matrix $\bs{W}$ is defined as $\bs{\varpi} = \text{vec}(\bs{W}) = \bs{b}_c^{\prime} \left( M N_k^{-1} \hat{\bs{\mathcal{Q}}}_{G_k^0, \tilde{z} \tilde{z}} \right)^{-1}$ with $\bs{W}=(\bs{\varpi}_1, \dots, \bs{\varpi}_p)$. Additionally, $\bs{g}_{\varpi} (v) = \bs{W}^{\prime} \bs{b}(v)$. The last inequality in \eqref{eq:z_limit} holds due to the decomposition $\bs{\varpi}^\prime \tilde{\bs{z}}_{it} = \bs{\varpi}^\prime \left( \tilde{\bs{x}}_{it} \otimes \bs{b}(t/T) \right) = \bs{b}(t/T)^\prime \bs{W} \tilde{\bs{x}}_{it} = \bs{g}_{\varpi}^\prime (t/T) \tilde{\bs{x}}_{it} $.

We rewrite \eqref{eq:z_limit} as
\begin{equation}
    \label{eq:z_limit_2}
    \begin{split}
         & \frac{M}{N_k T} \sum_{i \in G_k^0} \sum_{t = 1}^{T} \bs{g}_{\varpi}^\prime \left(\frac{t}{T}\right) \tilde{\bs{x}}_{it} \tilde{\bs{x}}_{it}^\prime \bs{g}_{\varpi} \left(\frac{t}{T}\right)                                                                                                                                   \\
         & = \frac{M}{N_k T} \sum_{i \in G_k^0} \sum_{t = 1}^{T} \bs{g}_{\varpi}^\prime \left(\frac{t}{T}\right) E (\tilde{\bs{x}}_{it} \tilde{\bs{x}}_{it}^\prime) \bs{g}_{\varpi} \left(\frac{t}{T}\right) (1 + o_p(1))                                                                                                                \\
         & \leq \frac{M \bar{c}_{\tilde{x} \tilde{x}}}{T} \sum_{t = 1}^{T} \bs{g}_{\varpi}^\prime \left(\frac{t}{T}\right) \bs{g}_{\varpi} \left(\frac{t}{T}\right) = \frac{M \bar{c}_{\tilde{x} \tilde{x}}}{T} \sum_{t = 1}^{T} \bs{\varpi}^{\prime} \bs{b} \left(\frac{t}{T}\right) \bs{b} \left(\frac{t}{T}\right)^\prime \bs{\varpi} \\
         & = \bar{c}_{\tilde{x} \tilde{x}} \| \bs{\varpi} \|_2^2 \frac{1}{T} \sum_{t = 1}^{T} M \bs{b} \left(\frac{t}{T} \right) \bs{b} \left(\frac{t}{T}\right)^\prime.
    \end{split}
\end{equation}
The first equality holds since $T^{-1} \sum_{t = 1}^{T} \tilde{\bs{x}}_{it} \tilde{\bs{x}}_{it}^\prime = T^{-1} \sum_{t = 1}^{T} E(\tilde{\bs{x}}_{it} \tilde{\bs{x}}_{it}^\prime) (1+o_p(1))$ by \citet[Lemma A.2]{huang2004polynomial}. The first inequality uses the fact that $E(\tilde{\bs{x}}_{it} \tilde{\bs{x}}_{it}^\prime) = E(\bs{x}_{it} \bs{x}_{it}^\prime) - E(\bs{x}_{it}) E(\bs{x}_{it}^\prime) \leq E(\bs{x}_{it} \bs{x}_{it}^\prime) \leq \bar{c}_{\tilde{x} \tilde{x}}$ by Assumption \superref{line:A1}{1(iv)} and consequently $(N_k T)^{-1} \sum_{i \in G_k^0} \sum_{t = 1}^{T} E(\tilde{\bs{x}}_{it} \tilde{\bs{x}}_{it}^\prime) \leq \bar{c}_{\tilde{x} \tilde{x}}$. Recall that by property of the Riemann sum and \cref{lma:spline_properties}\ref{lma:spline_properties_3}, $\lim_{T \to \infty} T^{-1} \sum_{t=1}^{T} M \bs{b}(t/T) \bs{b}(t/T)^\prime = \int_{0}^{1} M \bs{b}(v) \bs{b}(v)^\prime \,dv = O(1)$. Moreover, $\| \bs{W} \|_F = O_p(1)$ since $\text{vec}(\bs{W}) = \bs{b}_{c}^\prime \left( M N_k^{-1} \hat{\bs{\mathcal{Q}}}_{G_k^0, \tilde{z} \tilde{z} } \right)^{-1}$, with $\| M N_k^{-1} \hat{\bs{\mathcal{Q}}}_{G_k^0, \tilde{z} \tilde{z} } \|_F = O_p(1)$ following \cref{lma:Q_zz}\ref{lma:Q_zz_2} and $\| \bs{b}_{c} \|_2 = O_p(1)$ following \cref{lma:dist}\ref{lma:dist_1}. Plugging these results into \eqref{eq:z_limit_2}, the rate of \eqref{eq:z_limit_2} and therefore of \eqref{eq:z_limit} becomes apparent as $\bar{c}_{\tilde{x} \tilde{x}} \text{vec}(\bs{\varpi})^\prime \text{vec}(\bs{\varpi}) T^{-1} \sum_{t = 1}^{T} M \bs{b}(t/T) \bs{b}(t/T)^\prime = O_p(1)$.

Turning to $d_{12,G_k^0}$,
\begin{equation}
    \label{eq:B_2}
    \begin{split}
        | d_{12,G_k^0} | & = \left| \sqrt{\frac{N_k T}{M}} \bs{b}_{c}^\prime \hat{\bs{\mathcal{Q}}}_{G_k^0,\tilde{z} \tilde{z}}^{-1} \hat{\bs{q}}_{\hat{G}_k,\tilde{z} \tilde{\varrho}} \right| = \left| (N_k T)^{-1/2} M^{1/2} \frac{1}{T} \sum_{i \in G_k^0} \sum_{t = 1}^{T} \bs{b}_{c}^\prime (M N_k^{-1} \hat{\bs{\mathcal{Q}}}_{G_k^0,\tilde{z} \tilde{z}})^{-1} \tilde{\bs{z}}_{it} \tilde{\varrho}_{it} \right| \\
                         & \leq (N_k T)^{-1/2} M^{1/2} \frac{1}{T} \sum_{i \in G_k^0} \sum_{t = 1}^{T} \left| \bs{b}_{c}^\prime (M N_k^{-1} \hat{\bs{\mathcal{Q}}}_{G_k^0,\tilde{z} \tilde{z}})^{-1} \tilde{\bs{z}}_{it} \right| \left| \tilde{\varrho}_{it} \right|                                                                                                                                                 \\
                         & \leq (N_k T)^{1/2} \left[ \frac{M}{N_k T} \sum_{i \in G_k^0} \sum_{t = 1}^{T} \left| \bs{b}_{c}^\prime (M N_k^{-1} \hat{\bs{\mathcal{Q}}}_{G_k^0,\tilde{z} \tilde{z}})^{-1} \tilde{\bs{z}}_{it} \right|^2 \right]^{1/2} \left[ \frac{1}{N_k T} \sum_{i \in G_k^0} \sum_{t = 1}^{T} | \tilde{\varrho}_{it} |^2 \right]^{1/2},
    \end{split}
\end{equation}
where the first inequality holds by the triangle property and the second because of the Cauchy-Schwarz inequality. \eqref{eq:z_limit}-\eqref{eq:z_limit_2} provide the rate of the first term in squared brackets and \eqref{eq:eta_limit} gives the rate of the second term in squared brackets of \eqref{eq:B_2}. Consequently, it is apparent that $| d_{12,G_k^0} | = (N_k T)^{1/2} O_p(1) O_p(M^{-\theta})$ and $O_p((N_k T)^{1/2} M^{-\theta}) = o_p(1)$ under Assumption \superref{line:A4}{4(ii)}. Moreover, as $|s_{c,G_k^0}| = O_p(1)$ by \cref{lma:dist}\ref{lma:dist_2}, $| d_{12,G_k^0} | / s_{c,G_k^0} = o_p(1)$ and $d_{12,G_k^0}$ is negligible in the limit.

As a result, it suffices to study the behavior of $d_{11,G_k^0}$ in \eqref{eq:B1_G0} to derive the limiting distribution of the \textit{post-Lasso} estimator. To this end,
\begin{equation*}
    \label{eq:B_1_1}
    \begin{split}
        d_{11,G_k^0} & = \sqrt{\frac{N_k T}{M}} \bs{b}_{c}^\prime \hat{\bs{\mathcal{Q}}}_{G_k^0,\tilde{z} \tilde{z}}^{-1} \hat{\bs{q}}_{G_k^0,\tilde{z} \tilde{\epsilon}} = \bs{b}_{c}^\prime \left(\frac{M}{N_k} \hat{\bs{\mathcal{Q}}}_{G_k^0,\tilde{z} \tilde{z}} \right)^{-1} \sqrt{\frac{M}{N_k T}} \sum_{i \in G_k^0} \sum_{t = 1}^{T} \tilde{\bs{z}}_{it} \tilde{\epsilon}_{it}              \\
                     & = \left[ \bs{b}_{c}^\prime \left(\frac{M}{N_k} \hat{\bs{\mathcal{Q}}}_{G_k^0,\tilde{z} \tilde{z}} \right)^{-1} \frac{M}{N_k T} \sum_{i \in G_k^0} \left( \tilde{\bs{Z}}_{i}^\prime E(\bs{\epsilon}_i \bs{\epsilon}_i^\prime) \tilde{\bs{Z}}_{i} \right) \left(\frac{M}{N_k} \hat{\bs{\mathcal{Q}}}_{G_k^0,\tilde{z} \tilde{z}} \right)^{-1} \bs{b}_{c} \right]^{1/2} \zeta_i \\
                     & = \sum_{i \in G_k^0} a_i \zeta_i,
    \end{split}
\end{equation*}
where $\bs{\epsilon}_i = (\epsilon_{i1}, \dots, \epsilon_{iT})^\prime$, $\tilde{\bs{Z}}_{i} = (\tilde{\bs{z}}_{i1}, \dots, \tilde{\bs{z}}_{it})^\prime$, and 
$$
a_i = \left[ \bs{b}_{c}^\prime \left(\frac{M}{N_k} \hat{\bs{\mathcal{Q}}}_{G_k^0,\tilde{z} \tilde{z}} \right)^{-1} \frac{M}{N_k T} \left( \tilde{\bs{Z}}_{i}^\prime E(\bs{\epsilon}_i \bs{\epsilon}_i^\prime) \tilde{\bs{Z}}_{i} \right) \left(\frac{M}{N_k} \hat{\bs{\mathcal{Q}}}_{G_k^0,\tilde{z} \tilde{z}} \right)^{-1} \bs{b}_{c} \right]^{1/2}.
$$ 
$\zeta_{i}$ is an independent random variable with variance one and mean $\bs{c}^{\prime} \bs{q}_{G_k^0} = \bs{c}^{\prime} \sqrt{M / (N_k T)} \sum_{i \in G_k^0} \sum_{t = 1}^{T} E(\tilde{\bs{z}}_{it} \tilde{\epsilon}_{it})$, conditional on $\{\bs{z}_i\}_{i = 1}^N$.

Subsequently, we prove
$$
    \frac{d_{11,G_k^0}}{\sqrt{\sum_{i \in G_k^0} a_i^2}} - \frac{\bs{c}^\prime \bs{q}_{G_k^0}}{\sqrt{M / (N_k T) \bs{c}^\prime \sum_{i \in G_k^0} \left( \tilde{\bs{Z}}_{i}^\prime E(\bs{\epsilon}_i \bs{\epsilon}_i^\prime) \tilde{\bs{Z}}_{i} \right) \bs{c}}} \overset{D}{\to} N(0, 1),
$$
by verifying the Lindeberg condition
$$
    \frac{\max_{i \in G_k^0} a_i^2}{\sum_{i \in G_k^0} a_i^2} = o_p(1).
$$

Considering $\max_{i \in G_k^0} a_i^2$ first,
\begin{equation*}
    \label{eq:Lind_num}
    \begin{split}
        \max_{i \in G_k^0} a_i^2 & = \max_{i \in G_k^0} \left[ \bs{b}_{c}^\prime \left(\frac{M}{N_k} \hat{\bs{\mathcal{Q}}}_{G_k^0,\tilde{z} \tilde{z}} \right)^{-1} \frac{M}{N_k T} \left( \tilde{\bs{Z}}_i^\prime E(\bs{\epsilon}_i \bs{\epsilon}_i^\prime) \tilde{\bs{Z}}_i \right) \left(\frac{M}{N_k} \hat{\bs{\mathcal{Q}}}_{G_k^0,\tilde{z} \tilde{z}} \right)^{-1} \bs{b}_{c}^\prime \right]          \\
                                 & \leq \max_{i \in G_k^0} \left[ \frac{M}{N_k T} \mu_{\max} \left(\tilde{\bs{Z}}_i^\prime E(\bs{\epsilon}_i \bs{\epsilon}_i^\prime) \bs{Z}_i \right) \bs{b}_{c}^\prime \left(\frac{M}{N_k} \hat{\bs{\mathcal{Q}}}_{G_k^0,\tilde{z} \tilde{z}} \right)^{-1}\left(\frac{M}{N_k} \hat{\bs{\mathcal{Q}}}_{G_k^0,\tilde{z} \tilde{z}} \right)^{-1} \bs{b}_{c} \right]             \\
                                 & \leq \frac{1}{N_k} \max_{i \in G_k^0} \left[\mu_{\max} \left( E(\bs{\epsilon}_i \bs{\epsilon}_i^\prime) \right) \right] \max_{i \in G_k^0} \left[\mu_{\max} \left( \frac{M}{T} \tilde{\bs{Z}}_i^\prime \tilde{\bs{Z}}_i \right)\right] \left[ \mu_{\min} \left( \frac{M}{N_k} \hat{\bs{\mathcal{Q}}}_{G_k^0,\tilde{z} \tilde{z}} \right) \right]^{-2} \| \bs{b}_{c} \|_2^2 \\
                                 & = \frac{1}{N_k} \bar{c}_{\epsilon \epsilon} O_p(1) O_p(1) O_p(1) = o_p(1)
    \end{split}
\end{equation*}
by Assumptions \superref{line:A1}{1(iv)} and \superref{line:A4}{4(i)}. Note that the inequalities hold by the property of a p.s.d. symmetric matrix $A$ and a conformable vector $a$, $a^\prime A a \leq \mu_{\max}(A) a^\prime a$. Furthermore, $\mu_{\max} \left( E(\bs{\epsilon}_i \bs{\epsilon}_i^\prime) \right)$ is bounded from above by Assumption \superref{line:A4}{4(i)}, while $\mu_{\max} \left( M T^{-1} \tilde{\bs{Z}}_i^\prime \tilde{\bs{Z}}_i \right) = \mu_{\max} \left( M T^{-1} \sum_{t = 1}^T \tilde{\bs{z}}_{it} \tilde{\bs{z}}_{it}^\prime \right)$ and $\mu_{\min} \left( M N_k^{-1} \hat{\bs{\mathcal{Q}}}_{G_k^0,\tilde{z} \tilde{z}} \right)$ are bounded by \cref{lma:Q_zz}. In addition, $\|\bs{b}_{c}\|_2 = 1$ following \cref{lma:dist}\ref{lma:dist_1}.

In conclusion, it is straightforward to see that the whole term in \eqref{eq:alpha_decomp} converges in distribution
$$
    \frac{\sqrt{N_k T / M} \bs{c}^\prime [\hat{\bs{\alpha}}_{\hat{G}_k}^p(v) - \bs{\alpha}_k^0(v)]}{\sqrt{\sum_{i \in G_k^0} a_i^2}} - \frac{\bs{c}^\prime \bs{q}_{G_k^0}}{\sqrt{M / (N_k T) \bs{c}^\prime \sum_{i \in G_k^0} \left( \tilde{\bs{Z}}_{i}^\prime E(\bs{\epsilon}_i \bs{\epsilon}_i^\prime) \tilde{\bs{Z}}_{i} \right) \bs{c}}} \overset{D}{\to} N(0, 1),
$$
for all $k = 1, \dots, K_0$. By defining $\sum_{i \in G_k^0} a_i^2 = \bs{c}^{\prime} \hat{\bs{\Omega}}_{G_k^0} \bs{c}$  and making use of the Cramér-Wold Theorem, this generalizes to
\begin{equation*}
    \label{eq:limit_dist}
    \sqrt{\frac{N_k T}{M}} \hat{\bs{\Omega}}_{G_k^0}^{-1/2} \left[\hat{\bs{\alpha}}_{\hat{G}_k}^p(v) - \bs{\alpha}_k^0(v) \right] - \hat{\bs{\mathcal{E}}}_{G_k^0}^{-1/2} \bs{q}_{G_k^0} \overset{D}{\to} N \left(0, \bs{I}_p \right), \text{ for } k = 1, \dots, K_0,
\end{equation*}
where $\hat{\bs{\Omega}}_{G_k^0}^{-1/2}$is the symmetric square root of the inverse of $\hat{\bs{\Omega}}_{G_k^0} = \hat{\bs{\nu}}_{G_k^0}^{\prime} \hat{\bs{\mathcal{E}}}_{G_k^0} \hat{\bs{\nu}}_{G_k^0}$, with
$$
    \hat{\bs{\mathcal{E}}}_{G_k^0} = \frac{M}{N_k T} \sum_{i \in G_k^0} \tilde{\bs{Z}}_{i}^\prime E(\bs{\epsilon}_i \bs{\epsilon}_i^\prime) \tilde{\bs{Z}}_{i} \text{ and } \hat{\bs{\nu}}_{G_k^0} = \left( \frac{M}{N_k} \hat{\bs{\mathcal{Q}}}_{G_k^0, \tilde{z} \tilde{z}} \right)^{-1} \left( \bs{I}_p \otimes \bs{b} (v) \right).
$$
See \citet[Theorem 3.5]{qian2016shrinkage}, \citet[Theorem 3.5]{mehrabani2023}, and \citet[Theorem 4.5]{su2019sieve} for similar derivations.
\end{proof}
\begin{proof}[{\textbf{Proof of \Cref{sec:Theo_4}}}]
\label{sec:3.6}
First, recall that $IC(\lambda) = \ln( \sigma^2_{\hat{\mathcal{G}}_{\hat{K}, \lambda}}) + \rho_{NT} p M \hat{K}_\lambda$, where $\rho_{NT}$ is a tuning parameter and $\sigma^2_{\hat{\mathcal{G}}_{\hat{K}, \lambda}}$ reflects the \textit{MSE} $\sigma^2_{\hat{\mathcal{G}}_{\hat{K}, \lambda}} = (NT)^{-1} \sum_{k = 1}^{\hat{K}_\lambda} \sum_{i \in \hat{G}_{k, \lambda}} \sum_{t=1}^{T} \left( \tilde{y}_{it} - \hat{\bs{\alpha}}_{\hat{G}_{k, \lambda}}^{p \prime} (t/T) \tilde{\bs{x}}_{it}\right)^2$.

Let $\Lambda = [0, \lambda_{\max}]$ be an interval in $\mathbb{R}^+$. Define three subsets
$\Lambda_0$, $\Lambda_-$ and $\Lambda_+$, such that
\begin{equation*}
    \label{eq:T36_Lambda}
    \Lambda_0 = \{\lambda \in \Lambda: \hat{K}_{\lambda} = K_0 \}, \; \Lambda_- = \{\lambda \in \Lambda: \hat{K}_\lambda < K_0 \}, \; \Lambda_+ = \{\lambda \in \Lambda: \hat{K}_\lambda > K_0 \}.
\end{equation*}
Moreover, all $\lambda \in \Lambda_0$ comply with the regularity conditions in Assumption \superref{line:A2}{2}, which allows the use of the oracle property by \cref{sec:Coro_2}. In addition, since the \textit{post-Lasso} estimator converges to the true estimator by \cref{sec:Theo_3}, $\sigma^2_{\hat{\mathcal{G}}_{\hat{K}, \lambda}} = \sigma^2_{\mathcal{G}^0}$ w.p.a.1. Therefore, let  $\inf_{\lambda \in \Lambda_{0,NT}} IC(\lambda) = IC(\lambda_0)$ and
\begin{equation*}
    \label{eq:T36_lambda_0}
    IC(\lambda_0) = \ln \left( \sigma^2_{\mathcal{G}^0} \right) + \rho_{NT} p M \hat{K}_\lambda \overset{p}{\to} \ln \left( \sigma^2_0\right),
\end{equation*}
where $\sigma^2_0$ is the irreducible \textit{MSE} $\sigma^2_0 = \underset{(N,T)\to \infty}{\text{plim}}(NT)^{-1} \sum_{i = 1}^{N} \sum_{t = 1}^{T} \tilde{\epsilon}_{it}^2$ and Assumption \superref{line:A6}{6(i)} gives $\rho_{NT} M K_0 \to 0$ as $(N,T) \to \infty$.

The proof of \cref{sec:Theo_4} works by showing that $IC(\lambda)$ converges in probability to some value strictly larger than $IC(\lambda_0)$, $\forall \lambda \in \Lambda_- \cup \Lambda_+$.\\

\textit{Case 1 Underfitting}: $\lambda \in \Lambda_-$ such that $\hat{K}_\lambda < K_0$. Let $\mathbb{G}$ denote the set of all possible partitions over $N$. Note that
\begin{equation*}
    \label{eq:T36_sigma}
    \begin{split}
        \sigma^2_{\hat{\mathcal{G}}_{\hat{K}, \lambda}} & = (NT)^{-1} \sum_{k = 1}^{\hat{K}_\lambda} \sum_{i \in \hat{G}_{k, \lambda}} \sum_{t=1}^{T} \left( \tilde{y}_{it} - \hat{\bs{\xi}}_{\hat{G}_{k, \lambda}}^{p \prime} \tilde{\bs{z}}_{it}\right)^2             \\
                                                       & \geq \min_{1 \leq K < K_0} \inf_{\mathcal{G}_K \in \mathbb{G}} \sum_{k = 1}^{K} \sum_{i \in {G}_k} \sum_{t=1}^{T} \left(\tilde{y}_{it} - \hat{\bs{\xi}}_{G_k}^{p \prime} \tilde{\bs{z}}_{it}\right)^2 \\
                                                       & = \min_{1 \leq K < K_0} \inf_{\mathcal{G}_K \in \mathbb{G}} \hat{\sigma}_{\mathcal{G}_{(K)}}^2.
    \end{split}
\end{equation*}

Assumption \superref{line:A5}{5} states, $\lim_{(N,T) \to \infty} \min_{1 \leq K < K_0} \inf_{\mathcal{G}_K \in \mathbb{G}} \hat{\sigma}_{\mathcal{G}_K}^2 \overset{p}{\to} \underline{\sigma}^2 > \sigma^2_0$. Therefore, applying Slutsky,
\begin{equation*}
    \label{eq:T36_Lambda_min}
    \inf_{\lambda \in \Lambda_-} IC(\lambda) \geq \min_{1 \leq K < K_0} \inf_{\mathcal{G}_K \in \mathbb{G}} \ln (\sigma_{\mathcal{G}_K}^2) + \rho_{NT} p M K \overset{p}{\to} \ln (\underline{\sigma}^2) > \ln (\sigma^2_0)
\end{equation*}
and in consequence $\Pr \left( \inf_{\lambda \in \Lambda_-} IC(\lambda) \geq IC(\lambda_0) \right) \to 1$.
\\

\textit{Case 2 Overfitting: $\lambda \in \Lambda_+$ such that $\hat{K}_\lambda > K_0$.}
Define the set $\bar{\mathbb{G}}_K = \{\bar{\mathcal{G}}_K = \{\bar{G}_k\}_{k = 1}^K: \nexists \, i,j \in \bar{G}_K \; \text{where} \; i \in G_l^0, \, j \notin G_l^0, \, 1 \leq l \leq K_0\}$ with $K > K_0$. That is, each $\bar{\mathcal{G}}_K \in \bar{\mathbb{G}}_K$ denotes a partition of $K>K_0$ over $N$, where no heterogeneous cross-sectional units are pooled together in any of the $K$ groups. Recognize that such an over-fitted model yields a weakly lower \textit{MSE} than when using the true group structure $\hat{\sigma}^2_{\bar{\mathcal{G}}_{K}} \leq \hat{\sigma}^2_{\mathcal{G}^0}$.

Let
\begin{equation*}
    \label{eq:T_36_Lambda_plus}
    \begin{split}
         & \Pr \left(\inf_{\lambda \in \Lambda_+} IC(\lambda) > IC(\lambda_0) \right)                                                                                                                                                                                       \\
         & \geq \Pr \left( \min_{K_0 < K \leq N} \inf_{\bar{\mathcal{G}}_K \in \bar{\mathbb{G}}_K} \left( \ln(\hat{\sigma}^2_{\bar{\mathcal{G}}_K}) + \rho_{NT} p M K \right) > \ln (\sigma_{\mathcal{G}^0}^2) + \rho_{NT} p M K_0 \right)                          \\
         & \geq \Pr \left( \min_{K_0 < K \leq N} \inf_{\bar{\mathcal{G}}_K \in \bar{\mathbb{G}}_K} \left( \ln (\hat{\sigma}^2_{\bar{\mathcal{G}}_{K}}) + \rho_{NT} p M K \right) > \ln (\sigma_{\mathcal{G}^0}^2) + \rho_{NT} p M K_0 \right)                       \\
         & = \Pr \left( \min_{K_0 < K \leq N} \inf_{\bar{\mathcal{G}}_K \in \bar{\mathbb{G}}_K} \left[ \ln( \hat{\sigma}^2_{\bar{\mathcal{G}}_{K}} / \sigma_{\mathcal{G}^0}^2) + \rho_{NT} p M (K - K_0) \right] > 0 \right)                                        \\
         & = \Pr \left( \min_{K_0 < K \leq N} \inf_{\bar{\mathcal{G}}_K \in \bar{\mathbb{G}}_K} \left[ (\hat{\sigma}^2_{\bar{\mathcal{G}}_{K}} - \sigma_{\mathcal{G}^0}^2) / \sigma_{\mathcal{G}^0}^2 + o_p(1) + \rho_{NT} p M (K - K_0) \right] > 0 \right).
    \end{split}
\end{equation*}
Employing \cref{lma:mse}, $T M^{-1} (\hat{\sigma}^2_{\bar{\mathcal{G}}_{K}} - \sigma_{\mathcal{G}^0}^2) = O_p(1)$ and $\sigma_{\mathcal{G}^0}^2 \to \sigma_0^2$ as shown above. Therefore, after expanding by $T M^{-1}$ and using the fact that $T \rho_{NT} \to \infty$ by Assumption \superref{line:A6}{6(ii)},
\begin{equation*}
    \label{eq:T_36_fin}
    \begin{split}
        \Pr \left( \min_{K_0 < K \leq N} \inf_{\bar{\mathcal{G}}_K \in \bar{\mathbb{G}}_K} \left[O_p(1) + o_p(1) + T \rho_{NT} p (K - K_0) \right] > 0 \right) \to 1,
    \end{split}
\end{equation*}
as $(N,T) \to \infty$.    
\end{proof}

\label{sec:Lemmas}
\begin{proof}[{\textbf{Proof of \Cref{lma:spline_properties}}}]
\item[] 
 \paragraph{\textnormal{\textit{Part (i):}}} $b_m(v) > 0$, $m = -d, \dots, M^*$, is a basis function of degree $d$ (order $d+1$) and defined on the interval $[v_m, v_{m+d+1})$ \citep[ch. 9, eq. 21]{de2001practical}. Moreover, by property of B-slines, all spline basis functions sum up to one for each $v$: $\sum_{m = -d}^{M^*} b_m(v) = 1$ \citep[ch. 9, eq. 37]{de2001practical}. From here it follows that $\| \bs{b}(v)\|_2 = \left[ \sum_{j = -d}^{M^*} b_m(v)^2 \right]^{1/2} \leq \left[ \sum_{j = -d}^{M^*} b_m(v) \right]^{1/2} = 1$.
\paragraph{\textnormal{\textit{Part (ii):}}} As stated above, $b_m(v)$ is uniformly bounded on the unit interval and $\sum_{m = -d}^{M^*} b_m(v) = 1$. As a consequence, $\int_{0}^{1} b_m(v) \, dv = O(M^{-1})$ and
$$
    \int_{0}^{1} \| \bs{b}(v) \|_2 \, dv = \int_{0}^{1} \left[ \sum_{m = -d}^{M^*} b_m(v)^2 \right]^2 \, dv \leq \int_{0}^{1} \left[ \sum_{m = -d}^{M^*} b_m(v) \right]^2 \, dv = 1.
$$
\paragraph{\textnormal{\textit{Part (iii):}}} As shown in \citet[ch. 11, eq. 8]{de2001practical}, there exist two constants $0 < \underbar{c}_{b} \leq \bar{c}_{b} < \infty$ such that $\underbar{c}_{b} \| \bs{c} \|_2^2 \leq M \int \left( \bs{c}^\prime \bs{b}(v)\right)^2 \,dv \leq \bar{c}_{b} \| \bs{c} \|_2^2$ for all nonrandom $\bs{c} \in \mathbb{R}^M $. Using the Cramér-World device, $M \int_{0}^{1} \bs{b}(v) \bs{b}(v)^\prime \,dv = O(1)$ and the minimum and maximum eigenvalues are bounded $\underbar{c}_{b} \leq \mu_{\min} \left(M \int_{0}^{1} \bs{b}(v) \bs{b}(v)^\prime \,dv \right) \leq \mu_{\max} \left(M \int_{0}^{1} \bs{b}(v) \bs{b}(v)^\prime \,dv \right) \leq \bar{c}_b $.
\paragraph{\textnormal{\textit{Part (iv):}}} The proof is analogous to \citet[ch. 12, Theorem 6]{de2001practical}.
\end{proof}
\begin{proof}[{\textbf{Proof of \Cref{lma:Q_zz}}}]
\item[] 
\paragraph{\textnormal{\textit{Part (i):}}} Consistent with \citet[Lemma A.3]{su2019sieve}, the proof exploits the inherent property that B-splines are bounded. Recall that $\hat{\bs{Q}}_{i, \tilde{z} \tilde{z}} = T^{-1} \sum_{t = 1}^T \tilde{\bs{z}}_{it} \tilde{\bs{z}}_{it}^\prime$ and $\hat{\bs{\mathcal{Q}}}_{i, \tilde{z} \tilde{z}} = \sum_{i \in G_k^0} T^{-1} \sum_{t = 1}^T \tilde{\bs{z}}_{it} \tilde{\bs{z}}_{it}^\prime$. Define the $M \times p$ matrix $\bs{W} = (\bs{\varpi}_1, \dots, \bs{\varpi}_p)$ with $\bs{\varpi} = \vect{\bs{W}}$, $\bs{\varpi}_l = (\varpi_{l,1}, \dots, \varpi_{l,M})^\prime$, and $\| \bs{\varpi} \|_2 \leq c_\varpi < \infty$ for some positive constant $c_\varpi$. Let $\mathbb{B}_{\mathbb{V}, M}$ denote a linear space as defined in Subsection \ref{sec:sieve}. The vector $\bs{g}_{\varpi}(v) = \bs{W}^\prime \bs{b}(v)$ collects spline basis functions $\bs{g}_{\varpi}(v) = (g_1(v, \varpi_1), \dots, g_p(v, \varpi_p))^\prime$, where $g_l(v, \varpi_l) = \bs{\varpi}_{l}^\prime \bs{b}(v) \in \mathbb{B}_{\mathbb{V}, M}$ for $l = 1, \dots, p$ and $\bs{g}_{\varpi}(v) \in \mathbb{B}^{\otimes p}_{\mathbb{V}, M}$. Recognize that $\hat{\bs{Q}}_{i, \tilde{z} \tilde{z}} = T^{-1} \sum_{t = 1}^T \bs{z}_{it} \bs{z}_{it}^\prime - T^{-1} \sum_{t = 1}^{T} \bs{z}_{it} T^{-1} \sum_{t = 1}^{T} \bs{z}_{it}^\prime = \bs{A}_{1i} - \bs{A}_{2i}$.\\

\textit{Case 1: $\bs{\bs{x}_{it}}$ does not contain an intercept}. 
Consider $\bs{\varpi}^\prime \bs{A}_{1i} \bs{\varpi}$,
\begin{equation}
    \label{eq:Q_zz_w}
    \begin{split}
         & \bs{\varpi}^\prime \frac{1}{T} \sum_{t=1}^{T} \bs{z}_{it} \bs{z}_{it}^\prime \bs{\varpi} = \bs{\varpi}^\prime \frac{1}{T} \sum_{t=1}^{T} \left[\bs{x}_{it} \otimes \bs{b} \left(\frac{t}{T}\right) \right] \left[ \bs{x}_{it} \otimes \bs{b} \left(\frac{t}{T}\right) \right]^\prime \bs{\varpi}                                                                                \\
         & = \frac{1}{T} \sum_{t=1}^{T} \left[ \bs{b} \left( \frac{t}{T} \right)^\prime \bs{W} \bs{x}_{it} \right]^2 = \frac{1}{T} \sum_{t=1}^{T} \left[ \bs{g}_{\varpi} \left( \frac{t}{T} \right)^\prime \bs{x}_{it} \right]^2 = \frac{1}{T} \sum_{t=1}^{T} \bs{g}_{\varpi} \left( \frac{t}{T} \right)^\prime \bs{x}_{it} \bs{x}_{it}^\prime \bs{g}_{\varpi} \left( \frac{t}{T} \right),
    \end{split}
\end{equation}
since $(\bs{x}_{it} \otimes \bs{b}(v)) \text{vec} \left( \bs{W} \right)^{\prime} = \text{vec} \left( \bs{b}^\prime(v) \bs{W} \bs{x}_{it} \right)$ by property of the Kronecker product \citep[see][p. 249]{bernstein2009matrix}.

Note that the maximum eigenvalue of $E(\bs{x}_{it} \bs{x}_{it}^\prime)$ is bounded away from infinity by Assumption \superref{line:A1}{1(iv)}. Furthermore, as shown in Lemma \citet[A.2]{huang2004polynomial},
\begin{equation}
    \label{eq:E_xx}
    \frac{1}{T} \sum_{t=1}^{T} \bs{g}_{\varpi} \left( \frac{t}{T} \right)^\prime \bs{x}_{it} \bs{x}_{it}^\prime \bs{g}_{\varpi} \left( \frac{t}{T} \right) = \frac{1}{T} \sum_{t=1}^{T} \bs{g}_{\varpi} \left( \frac{t}{T} \right)^\prime E (\bs{x}_{it} \bs{x}_{it}^\prime) \bs{g}_{\varpi} \left( \frac{t}{T} \right) (1 + o_p(1))
\end{equation}
for strong mixing processes and
\begin{equation}
    \label{eq:Q_zz_w_2}
    \frac{1}{T} \sum_{t=1}^{T} \bs{g}_{\varpi} \left( \frac{t}{T} \right)^\prime E (\bs{x}_{it} \bs{x}_{it}^\prime) \bs{g}_{\varpi} \left( \frac{t}{T} \right) (1 + o_p(1)) \leq \bar{c}_{xx} \frac{1}{T} \sum_{t=1}^{T} \bs{g}_{\varpi} \left( \frac{t}{T} \right)^\prime \bs{g}_{\varpi} \left( \frac{t}{T} \right) (1 + o_p(1)),
\end{equation}
where $T^{-1} \sum_{t=1}^{T} \bs{g}_{\varpi} \left( t/T \right)^\prime \bs{g}_{\varpi} \left( t/T \right) = \int_{0}^{1} \bs{g}_{\varpi} (v)^\prime \bs{g}_{\varpi} (v) \, dv \left[ 1 + O(T^{-1}) \right]$ by property of the Riemann sum and $\int_{0}^{1} \bs{g}_{\varpi} (v)^\prime \bs{g}_{\varpi} (v) \, dv = \sum_{l = 1}^{p} \varpi_l^\prime \int_{0}^{1} \bs{b}(v) \bs{b}(v)^\prime \, dv \varpi_l = O_p(M^{-1})$ by \cref{lma:spline_properties}\ref{lma:spline_properties_3}, since $\| \bs{\varpi} \|_2^2 < \infty$. As a consequence, after plugging into \eqref{eq:Q_zz_w_2} it is apparent that the whole term in \eqref{eq:Q_zz_w} $\bs{\varpi}^\prime \bs{A}_{1i} \bs{\varpi} \leq O(M^{-1}) \left[ 1 + O(T^{-1}) \right] (1 + o(1)) = O(M^{-1})$. Since $\mu_{\max} (\bs{B} - \bs{C}) \leq \mu_{\max}(\bs{B}) + \mu_{\max}(-\bs{C}) = \mu_{\max}(\bs{B}) - \mu_{\min}(\bs{C}) \leq \mu_{\max}(\bs{B})$ for two generic real matrices $\bs{B}, \bs{C}$, $\mu_{\max}(\hat{\bs{Q}}_{i, \tilde{z} \tilde{z}}) \leq \mu_{\max} (\bs{A}_{1i})$ and $\mu_{\max} \left( M \hat{\bs{Q}}_{i, \tilde{z} \tilde{z}} \right)$ is bounded away from infinity in probability, uniformly in $\bs{\varpi}$. This result also applies uniformly across the cross-section to $\max_{i} \mu_{\max} \left( M \hat{\bs{Q}}_{i, \tilde{z} \tilde{z}} \right)$ with probability $1 - o(N^{-1})$.

To study the behavior of the minimum eigenvalue of $\hat{\bs{Q}}_{i, \tilde{z} \tilde{z}}$, one must also consider $\bs{A}_{2i}$. For this purpose, take
\begin{equation}
    \label{eq:Q_zz_w_3}
    \begin{split}
         & \bs{\varpi}^\prime \bs{A}_{2i} \bs{\varpi} = \bs{\varpi}^\prime \frac{1}{T} \sum_{t = 1}^{T} \bs{z}_{it} \frac{1}{T} \sum_{t = 1}^{T} \bs{z}_{it}^\prime \bs{\varpi}                                                                        \\
         & = \bs{\varpi}^\prime \frac{1}{T} \sum_{t = 1}^{T} \left[ \bs{x}_{it} \otimes \bs{b} \left( \frac{t}{T} \right) \right] \frac{1}{T} \sum_{t = 1}^{T} \left[ \bs{x}_{it} \otimes \bs{b} \left( \frac{t}{T} \right) \right]^\prime \bs{\varpi} \\
         & = \left[ \frac{1}{T} \sum_{t=1}^{T} \bs{b} \left( \frac{t}{T} \right)^\prime \bs{\varpi} \bs{x}_{it} \right]^2 = \left[ \frac{1}{T} \sum_{t=1}^{T} \bs{g}_{\varpi} \left( \frac{t}{T} \right)^\prime \bs{x}_{it} \right]^2                  \\
         & = E \left[ \frac{1}{T} \sum_{t=1}^{T} \bs{g}_{\varpi} \left( \frac{t}{T} \right)^\prime \bs{x}_{it} \right]^2 (1 + o_p(1)),
    \end{split}
\end{equation}
where the last equality holds again by Lemma \citet[A.2]{huang2004polynomial}. Using \eqref{eq:Q_zz_w_3} and the result in \eqref{eq:E_xx},
\begin{equation}
    \label{eq:Q_zz_w_4}
    \begin{split}
         & \bs{\varpi}^\prime \left(\bs{A}_{1i} - \bs{A}_{2i}\right)\bs{\varpi} = \bs{\varpi}^\prime \left(\hat{\bs{Q}}_{i, zz} - T^{-1} \sum_{t = 1}^{T} \bs{z}_{it} T^{-1} \sum_{t = 1}^{T} \bs{z}_{it}^\prime\right)\bs{\varpi}                             \\
         & = \frac{1}{T} \sum_{t=1}^{T} E \left[ \left( \bs{g}_{\varpi} \left(\frac{t}{T}\right)^\prime \bs{x}_{it} \right)^2 \right] - E \left[ \frac{1}{T} \sum_{t=1}^{T} \bs{g}_{\varpi} \left(\frac{t}{T}\right)^\prime \bs{x}_{it} \right]^2 + o(N^{-1}),
    \end{split}
\end{equation}
uniformly in $\bs{\varpi}$ and $i$.

In order to show that the minimum eigenvalue of $M \hat{\bs{Q}}_{i, \tilde{z} \tilde{z}}$ is bounded away from 0, define
$$\bs{\mu}_i(v)=\left\{\begin{array}{ll}\mathbf{0}_{p \times 1} & \text { if } v=0 \\ E\left(\bs{x}_{it}\right) & \text { if } v \in (\frac{t-1}{T}, \frac{t}{T}] \end{array} \quad\right. \text{ and } \bs{\Sigma}_i(v)= \begin{cases}\mathbf{0}_{p \times p} & \text { if } v=0 \\ E\left(\bs{x}_{it} \bs{x}_{it}^{\prime}\right) & \text { if } v \in (\frac{t-1}{T}, \frac{t}{T} ], \end{cases}$$
for $v \in [0,1]$. Note that $\bar{\bs{\Sigma}}_i(t/T) = \bs{\Sigma}_i (t/T) - \bs{\mu}_i (t/T) \bs{\mu}_i (t/T)^\prime = \text{Var}(\bs{x}_{it})$. Rewrite $T^{-1} \sum_{t=1}^{T} E \left[ \left( \bs{g}_{\varpi} (t/T)^\prime \bs{x}_{it} \right)^2 \right] - E \left[ T^{-1} \sum_{t=1}^{T} \bs{g}_{\varpi} (t/T)^\prime \bs{x}_{it} \right]^2$ in \eqref{eq:Q_zz_w_4} to
\begin{equation*}
    \label{eq:Q_zz_w_5}
    \begin{split}
         & \frac{1}{T} \sum_{t=1}^{T} \left[ \bs{g}_{\varpi} \left(\frac{t}{T}\right)^\prime \bs{\Sigma}_i \left(\frac{t}{T}\right) \bs{g}_{\varpi} \left(\frac{t}{T}\right) \right] - \left[ \frac{1}{T} \sum_{t=1}^{T} \bs{g}_{\varpi} \left(\frac{t}{T}\right)^\prime \bs{\mu}_i \left(\frac{t}{T}\right) \right]^2 \\
         & = \int_{0}^{1} \bs{g}_{\varpi} (v)^\prime \bs{\Sigma}_i (v) \bs{g}_{\varpi} (v) \, dv \left[ 1 + O(T^{-1}) \right] - \left[ \int_{0}^{1} \bs{g}_{\varpi} (v)^\prime \bs{\mu}_i (v) \, dv \left[ 1 + O(T^{-1}) \right] \right]^2                                                                             \\
         & \geq \int_{0}^{1} \bs{g}_{\varpi} (v)^\prime \bs{\Sigma}_i (v) \bs{g}_{\varpi} (v) \, dv \left[ 1 + O(T^{-1}) \right] - \int_{0}^{1} \bs{g}_{\varpi} (v)^\prime \bs{\mu}_i (v) \bs{\mu}_i (v)^\prime \bs{g}_{\varpi} (v) \, dv \left[ 1 + O(T^{-1}) \right]                                                 \\
         & = \left[ \int_{0}^{1} \bs{g}_{\varpi} (v)^\prime \bs{\Sigma}_i (v) \bs{g}_{\varpi} (v) \, dv - \int_{0}^{1} \bs{g}_{\varpi} (v)^\prime \bs{\mu}_i (v) \bs{\mu}_i (v)^\prime \bs{g}_{\varpi} (v) \, dv \right]                                                                                               \\
         & \qquad \qquad - O(T^{-1}) \left[ \int_{0}^{1} \bs{g}_{\varpi} (v)^\prime \bs{\Sigma}_i (v) \bs{g}_{\varpi} (v) \, dv - \int_{0}^{1} \bs{g}_{\varpi} (v)^\prime \bs{\mu}_i (v) \bs{\mu}_i (v)^\prime \bs{g}_{\varpi} (v) \, dv \right]                                                                       \\
         & = d_{1i} - d_{2i},
    \end{split}
\end{equation*}
where the inequality holds because of Jensen's inequality. Studying $d_{1i}$ first,
\begin{equation*}
    \label{eq:A_1i}
    \begin{split}
        d_{1i} & = \int_{0}^{1} \bs{g}_{\varpi} (v)^\prime \bs{\Sigma}_i (v) \bs{g}_{\varpi} (v) \, dv - \int_{0}^{1} \bs{g}_{\varpi} (v)^\prime \bs{\mu}_i (v) \bs{\mu}_i (v)^\prime \bs{g}_{\varpi} (v) \, dv \\
               & = \int_{0}^{1} \bs{g}_{\varpi} (v)^\prime \bar{\bs{\Sigma}}_i (v) \bs{g}_{\varpi} (v) \, dv \geq \underbar{c}_{zz} \sum_{l = 1}^{p} \int_{0}^{1} g_l(v,\varpi_l)^\prime g_l(v,\varpi_l) \, dv  \\
               & = \underbar{c}_{zz} \sum_{l = 1}^{p} \varpi_{il}^\prime \int_{0}^{1} \bs{b}(v) \bs{b}(v)^\prime \, dv \varpi_{il} = \| \bs{\varpi} \|_2^2 O(M^{-1}) = O(M^{-1}),
    \end{split}
\end{equation*}
where $\underbar{c}_{zz} \leq \text{Var}(\bs{x}_{it}) = \bar{\bs{\Sigma}}_i(t/T)$ by Assumption \superref{line:A1}{1(iv)}.

Since $D_{2i} = O(T^{-1}) D_{1i} = O((MT)^{-1})$, the minimum eigenvalue $\min_{i} \mu_{\min} (M \hat{\bs{Q}}_{i, \tilde{z} \tilde{z}})$ is bounded away from 0 uniformly in $\bs{W}$ and $i$ with probability $1 - o(N^{-1})$.
\\

\textit{Case 2: $\bs{x}_{it} = 1$}. Then \eqref{eq:Q_zz_w} collapses to
\begin{equation}
    \label{eq:D_1}
    \bs{\varpi}^\prime \frac{1}{T} \sum_{t=1}^{T} \bs{z}_{it} \bs{z}_{it}^\prime \bs{\varpi} = \bs{\varpi}^\prime \frac{1}{T} \sum_{t=1}^{T} \bs{b}\left(\frac{t}{T}\right) \bs{b}\left(\frac{t}{T}\right)^\prime \bs{\varpi}
\end{equation}
and \eqref{eq:Q_zz_w_4} becomes
\begin{equation}
    \label{eq:D_2}
    \begin{split}
        \bs{\varpi}^\prime & \left(\hat{\bs{Q}}_{i, zz} - \frac{1}{T} \sum_{t = 1}^{T} \bs{z}_{it} \frac{1}{T} \sum_{t = 1}^{T} \bs{z}_{it}^\prime\right)\bs{\varpi}                                                                                                                                        \\
                           & = \bs{\varpi}^\prime \left(\frac{1}{T} \sum_{t=1}^{T} \bs{b}\left(\frac{t}{T}\right) \bs{b}\left(\frac{t}{T}\right)^\prime - \frac{1}{T} \sum_{t = 1}^{T} \bs{b}\left(\frac{t}{T}\right) \frac{1}{T} \sum_{t = 1}^{T} \bs{b}\left(\frac{t}{T}\right)^\prime\right)\bs{\varpi}.
    \end{split}
\end{equation}
It suffices to apply the basic properties of B-splines in order to study the minimum and maximum eigenvalues of $\hat{\bs{Q}}_{i, zz}$. To this end, \cref{lma:spline_properties}\ref{lma:spline_properties_3} shows that $M \int_{0}^{1} \bs{b}(t/T) \bs{b}(t/T)^\prime dv$ has bounded minimum and maximum eigenvalues, where $T^{-1} \sum_{t=1}^{T} \bs{b}(t/T) \bs{b}(t/T)^\prime = \int_{0}^{1} \bs{b}(t/T) \bs{b}(t/T)^\prime dv \left[1 + O(T^{-1})\right]$ by the Riemann sum and $\| \bs{W} \|_F < \infty$ in \eqref{eq:D_1}. Furthermore, $T^{-1} \sum_{t=1}^{T} \bs{b}(t/T) \bs{b}(t/T)^\prime - T^{-1} \sum_{t = 1}^{T} \bs{b}(t/T) T^{-1} \sum_{t = 1}^{T} \bs{b}(t/T)^\prime$ in \eqref{eq:D_2} can be considered as a vector product of standardized spline basis functions. Subsequently, the properties of B-splines, including the ones stated in \cref{lma:spline_properties}\ref{lma:spline_properties_3} carry over. As a consequence, the results obtained under Case 1 hold by \cref{lma:spline_properties}\ref{lma:spline_properties_3}.
\\

\textit{Case 3: $\bs{x}_{it}$ contains an intercept and stochastic regressors}. Reorder $\bs{x}_{it}$ so that $\bs{x}_{it} = (1, \bs{x}_{it}^{(2)^\prime})^\prime$ and $\bs{x}_{it}^{(2)}$ is a $(p - 1) \times 1$ vector of random variables. This instance presents a mix of the two previously considered cases, where Case 1 applies to $\bs{x}_{it}^{(2)}$ and Case 2 concerns the intercept in $\bs{x}_{it}$. Therefore, the results with respect to the minimum and maximum eigenvalues of $\hat{\bs{Q}}_{i, \tilde{z} \tilde{z}}$ also hold for $\bs{x}_{it} = (1, \bs{x}_{it}^{(2)^\prime})$, 
see   \citet[Lemma A.3(i)]{su2019sieve} for more details.

\paragraph{\textnormal{\textit{Part (ii):}}} Consider $\hat{\bs{\mathcal{Q}}}_{i, \tilde{z} \tilde{z}} = \sum_{i \in G_k^0} \hat{\bs{Q}}_{i, \tilde{z} \tilde{z}}$. Since cross-sectional individuals are weakly dependent as given in Assumption \superref{line:A1}{1(i)}, it holds that
\begin{equation*}
    \Pr \left( \underbar{c}_{\tilde{z} \tilde{z}} \leq \mu_{min} \left(\frac{M}{N_k} \hat{\bs{\mathcal{Q}}}_{i, \tilde{z} \tilde{z}} \right) < \mu_{max} \left( \frac{M}{N_k} \hat{\bs{\mathcal{Q}}}_{i, \tilde{z} \tilde{z}} \right) \leq \bar{c}_{\tilde{z} \tilde{z}} \right) = 1 - o(N^{-1})
\end{equation*}
for $i = 1, \dots, N$ and two constants $0 < \underbar{c}_{\tilde{z} \tilde{z}} \leq \bar{c}_{\tilde{z} \tilde{z}} < \infty$, as shown in \cref{lma:Q_zz}\ref{lma:Q_zz_1}.
\end{proof}
\begin{proof}[{\textbf{Proof of \Cref{lma:comp_error}}}]
    Recall $\hat{\bs{q}}_{i, \tilde{z} \tilde{u}} = T^{-1} \sum_{t = 1}^{T} \tilde{\bs{z}}_{it} \tilde{u}_{it}$, where $\tilde{u}_{it} = \tilde{\epsilon}_{it} + \tilde{\varrho}_{it}$, $\varrho_{it} = \bs{\eta}_{it}^\prime \bs{x}_{it}$. Following \citet[Lemma A.4]{su2019sieve}, the proof of \cref{lma:comp_error} is constructed by decomposing the error $u_{it}$ into the idiosyncratic and the sieve elements and analyzing both separately.

\paragraph{\textnormal{\textit{Part (i):}}} As defined in Section \ref{sec:sieve}, the sieve-bias equals $\bs{\eta}_{it} = \bs{\beta}_i^0(t/T) - \bs{\Pi}_i^{0 \prime} \bs{b}(t/T)$ and $\varrho_{it} =\bs{\eta}_{it}^\prime \bs{x}_{it}$. By the identity $u_{it} = \epsilon_{it} + \left[ \bs{\beta}_i^{0 \prime}(t/T) \bs{x}_{it} - \bs{\pi}_{i}^{0\prime} \bs{z}_{it} \right] = \epsilon_{it} + \left[ \bs{\beta}_i^0(t/T) - \bs{\Pi}_i^{0 \prime} \bs{b}(t/T) \right]^\prime \bs{x}_{it} = \epsilon_{it} + \varrho_{it}$ we have $\tilde{u}_{it} = \tilde{\varrho}_{it} + \tilde{\epsilon}_{it}$, $\hat{\bs{q}}_{i, \tilde{z} \tilde{u}} = \hat{\bs{q}}_{i, \tilde{z} \tilde{\varrho}} + \hat{\bs{q}}_{i, \tilde{z} \tilde{\epsilon}}$, and $\| \hat{\bs{q}}_{i, \tilde{z} \tilde{u}} \|_2 \leq \| \hat{\bs{q}}_{i, \tilde{z} \tilde{\varrho}} \|_2 + \| \hat{\bs{q}}_{i, \tilde{z} \tilde{\epsilon}} \|_2$ by the Triangle inequality, where $\hat{\bs{q}}_{i, \tilde{z} a} = T^{-1} \sum_{t = 1}^{T} \bs{z}_{it} a_{it}$ for $a_{it} = \{\tilde{\varrho}_{it}, \tilde{\epsilon}_{it} \}$.  First we derive the rate of $\| \hat{\bs{q}}_{i, \tilde{z} \tilde{\varrho}} \|_2$, then the rate of $\| \hat{\bs{q}}_{i, \tilde{z} \tilde{\epsilon}} \|_2$.

By the triangle inequality and the sub-multiplicative property
\begin{equation}
    \label{eq:Q_z_eta_sq}
    \| \hat{\bs{q}}_{i, \tilde{z} \tilde{\varrho}} \|_2 = \left\| T^{-1} \sum_{t = 1}^{T} \bs{z}_{it} \varrho_{it} - T^{-1} \sum_{t = 1}^{T} \bs{z}_{it} T^{-1} \sum_{t = 1}^{T} \varrho_{it} \right\|_2 \leq \left\| T^{-1} \sum_{t = 1}^{T} \bs{z}_{it} \varrho_{it} \right\|_2 + \left\| T^{-1} \sum_{t = 1}^{T} \bs{z}_{it} \right\|_2 \left| T^{-1} \sum_{t = 1}^{T} \varrho_{it} \right|.
\end{equation}
Studying all these elements in isolation, consider
\begin{equation}
    \label{eq:Eta}
    \begin{split}
        \left| \frac{1}{T} \sum_{t=1}^{T} \varrho_{it} \right|^2 & = \left| \frac{1}{T} \sum_{t=1}^{T} \bs{x}_{it}^\prime \bs{\eta}_{it} \right|^2 \leq \left\| \frac{1}{T} \sum_{t=1}^{T} \bs{x}_{it} \bs{x}_{it}^\prime \right\|_F \left\| \frac{1}{T} \sum_{t=1}^{T} \bs{\eta}_{it} \bs{\eta}_{it}^\prime \right\|_F,
    \end{split}
\end{equation}
where the inequality holds because of the Chauchy-Schwarz property. Assumption \superref{line:A1}{1(iv)} ensures that \newline $\left\| T^{-1} \sum_{t=1}^{T} \bs{x}_{it} \bs{x}_{it}^\prime \right\|_2 \leq \bar{c}_{xx}$ in probability. Furthermore, \cref{lma:spline_properties}\ref{lma:spline_properties_4} gives $\sup_{v \in [0,1]} \| \bs{\beta}_i^0(t/T) - \bs{\Pi}_i^{0 \prime} \bs{b}(t/T) \|_2 = O(M^{-\theta})$. As a consequence,
\begin{equation}
    \label{eq:Eta_2}
    \begin{split}
        \left\| \frac{1}{T} \sum_{t=1}^{T} \bs{\eta}_{it} \bs{\eta}_{it}^\prime \right\|_F & = \left\| \left[\bs{\beta}_i^0 \left( \frac{t}{T} \right) - \bs{\Pi}_i^{0 \prime} \bs{b} \left( \frac{t}{T} \right) \right] \left[\bs{\beta}_i^0 \left( \frac{t}{T} \right) - \bs{\Pi}_i^{0 \prime} \bs{b} \left( \frac{t}{T} \right) \right]^\prime \right\|_F \\
                                                                           & \leq \left[ \sup_{v \in [0,1]} \left\| \bs{\beta}_i^0 (v) - \bs{\Pi}_i^{0 \prime} \bs{b}(v) \right\|_2  \right]^2 = O(M^{-2\theta}),
    \end{split}
\end{equation}
and $\left| T^{-1} \sum_{t=1}^{T} \varrho_{it} \right|^2 = \bar{c}_{xx} O_p(M^{-2\theta})$, from which follows $\left| T^{-1} \sum_{t=1}^{T} \varrho_{it} \right| = O_p(M^{-\theta})$.

Define the real $M \times p$ matrix $\bs{W}$ with $\vect{\bs{W}} = \bs{\varpi}$ and $\| \bs{\varpi} \|_2 < \infty$. Then
\begin{equation*}
    \label{eq:Z_sum}
    \begin{split}
         & \bs{\varpi}^\prime \left\| \frac{1}{T} \sum_{t=1}^{T} \bs{z}_{it} \right\|_2^2 \bs{\varpi} = \left| \bs{\varpi}^\prime \frac{1}{T} \sum_{t=1}^{T} \left[\bs{x}_{it} \otimes \bs{b} \left(\frac{t}{T}\right) \right] \right|^2                                                                                                        \\
         & = \left| \frac{1}{T} \sum_{t=1}^{T} \bs{b} \left(\frac{t}{T}\right)^\prime \bs{W} \bs{x}_{it} \right|^2 \leq \left\| \frac{1}{T} \sum_{t=1}^{T} \bs{W}^\prime \bs{b} \left(\frac{t}{T}\right) \bs{b} \left(\frac{t}{T}\right)^\prime \bs{W} \right\|_F \left\| \frac{1}{T} \sum_{t=1}^{T} \bs{x}_{it} \bs{x}_{it}^\prime \right\|_F,
    \end{split}
\end{equation*}
where by Assumption \superref{line:A1}{1(iv)} $\left\| T^{-1} \sum_{t=1}^{T} \bs{x}_{it} \bs{x}_{it}^\prime \right\|_F \leq \bar{c}_{xx}$ in probability. In addition, using the B-spline property in \cref{lma:spline_properties}\ref{lma:spline_properties_3}
\begin{equation}
    \label{eq:B_norm}
    \begin{split}
         & \left\| \frac{1}{T} \sum_{t=1}^{T} \bs{W}^\prime \bs{b} \left(\frac{t}{T}\right) \bs{b} \left(\frac{t}{T}\right)^\prime \bs{W} \right\|_F = \left\| \bs{W}^\prime \int_{0}^{1} \bs{b} \left(v\right) \bs{b} \left(v\right)^\prime \, dv \bs{W} (1 + O(T^{-1})) \right\|_F \\
         & \leq \|\bs{W} \|^2_F O(M^{-1}) (1 + O(T^{-1})) = O(M^{-1}),
    \end{split}
\end{equation}
since $\|\bs{W} \|^2_F$ is bounded. Subsequently, $\left\| T^{-1} \sum_{t = 1}^{T} \bs{z}_{it} \right\|_2 = O_p(M^{-1/2})$.

Lastly, making use of the Cauchy-Schwarz inequality,
\begin{equation}
    \label{eq:Q_z_eta}
    \| \hat{\bs{q}}_{i, z \varrho} \|_2^2 = \left\| \frac{1}{T} \sum_{t=1}^{T} \bs{z}_{it} \varrho_{it} \right\|_2^2 \leq \left\| \frac{1}{T} \sum_{t=1}^{T} \bs{z}_{it} \bs{z}_{it}^\prime \right\|_F \left| \frac{1}{T} \sum_{t=1}^{T} \varrho_{it}^2 \right|,
\end{equation}
where $\left\| T^{-1} \sum_{t=1}^{T} \bs{z}_{it} \bs{z}_{it}^\prime \right\|_F = O_p(M^{-1})$ by \cref{lma:Q_zz}\ref{lma:Q_zz_1} and
\begin{equation}
    \label{eq:eta_limit_1}
    \begin{split}
        E \left| \frac{1}{T} \sum_{t=1}^{T} \varrho_{it}^2 \right| & = E \left| \frac{1}{T} \sum_{t=1}^{T} (\bs{x}_{it}^\prime \bs{\eta}_{it})^2 \right| = \left[\sup_{v \in [0,1]} \left\| \alpha_k^0(v) - \bs{\Pi}_i^{0 \prime} \bs{b}(v) \right\|_2 \right]^2 \frac{1}{T} \sum_{t=1}^{T} E \| \bs{x}_{it} \|_2^2 \\
                                                                & \leq O(M^{-2\theta}) \bar{c}_{x}^{2/q} = O(M^{-2\theta}),
    \end{split}
\end{equation}
where $\sup_{v \in [0,1]} \left\| \bs{\alpha}_k^0(v) - \bs{\Pi}_i^{0 \prime} \bs{b}(v) \right\|_2 = O(M^{-\theta})$ by \cref{lma:spline_properties}\ref{lma:spline_properties_4} and $E \| \bs{x}_{it} \|_2^2 \leq \bar{c}_{x}^{2/q}$ by Assumption \superref{line:A1}{1(iii)}. As a consequence, $\left| \frac{1}{T} \sum_{t=1}^{T} \varrho_{it}^2 \right| = O_p(M^{-2\theta})$ by Chebyshev's inequality and $\| \hat{\bs{q}}_{i, z \varrho} \|_2^2 = O_p(M^{-1}) O_p(M^{-2\theta})$ with $\| \hat{\bs{q}}_{i, z \varrho} \|_2 = O_p(M^{-\theta - 1/2})$.

Plugging the results of \eqref{eq:Eta_2}, \eqref{eq:B_norm}, \eqref{eq:Q_z_eta}, and \eqref{eq:eta_limit_1} into \eqref{eq:Q_z_eta_sq} yields
\begin{equation}
    \label{eq:Q_z_eta_2}
    \left\| \hat{\bs{q}}_{i, \tilde{z} \tilde{\varrho}} \right\|_2 \leq \left\|\hat{\bs{q}}_{i, z \varrho} \right\|_2 + \left\| T^{-1} \sum_{t = 1}^{T} \bs{z}_{it} \right\|_2 \left| T^{-1} \sum_{t = 1}^{T} \varrho_{it} \right| = O_p(M^{-\theta - 1/2}) + O_p(M^{-1/2}) O_p(M^{-\theta}) = O_p(M^{-\theta - 1/2}).
\end{equation}

Turning to $\hat{\bs{q}}_{i, \tilde{z} \tilde{\epsilon}}$, note that $\hat{\bs{q}}_{i, \tilde{z} \tilde{\epsilon}} = \hat{\bs{q}}_{i, z \epsilon} - T^{-1} \sum_{t = 1}^{T} \bs{z}_{it} T^{-1} \sum_{t = 1}^{T} \epsilon_{it} $. Studying $\hat{\bs{q}}_{i, z \epsilon}$ first,
\begin{equation*}
    \label{eq:Q_t_u}
    \begin{split}
        E \| \hat{\bs{q}}_{i, z \epsilon} \|_2^2 & = E \left\| \frac{1}{T} \sum_{t=1}^{T} \bs{z}_{it} {\epsilon}_{it} \right\|_2^2 = E \left\| \frac{1}{T} \sum_{t=1}^{T} \left[ \bs{x}_{it} \otimes \bs{b} \left(\frac{t}{T}\right) \right] {\epsilon}_{it} \right\|_2^2 = E \left\| \frac{1}{T} \sum_{t=1}^{T} \left( \bs{x}_{it} {\epsilon}_{it} \right) \otimes \bs{b} \left(\frac{t}{T}\right) \right\|_2^2 \\
                                                 & = \frac{1}{T^2} \sum_{t = 1}^{T} \sum_{s = 1}^{T} E(\bs{x}_{it}^\prime \bs{x}_{is} \epsilon_{it} \epsilon_{is}) \bs{b} \left(\frac{t}{T}\right)^\prime \bs{b} \left(\frac{s}{T}\right)                                                                                                                                                                        \\
                                                 & = \frac{1}{T^2} \sum_{t = 1}^{T} E(\bs{x}_{it}^\prime \bs{x}_{it} \epsilon_{it}^2) \left\| \bs{b} \left(\frac{t}{T}\right) \right\|_2^2                                                                                                                                                                                                                       \\
                                                 & \qquad \qquad + \frac{2}{T^2} \sum_{t = 1}^{T-1} \sum_{s = 1}^{T} E(\bs{x}_{it}^\prime \bs{x}_{is} \epsilon_{it} \epsilon_{is}) \bs{b} \left(\frac{t}{T}\right)^\prime \bs{b} \left(\frac{s}{T}\right) = d_{1i} + d_{2i}.
    \end{split}
\end{equation*}

Employing Assumption \superref{line:A3}{1(iii)} gives $E(\bs{x}_{it}^\prime \bs{x}_{it} \epsilon_{it}^2) \leq E( \| \bs{x}_{it} \|_2^2 \| \epsilon_{it} \|_2^2 ) \leq (\bar{c}_x \bar{c}_\epsilon)^{2/q}$. In addition, $T^{-1} \sum_{t = 1}^{T} \| \bs{b}(t/T) \|_2^2 = \int_{0}^{1} \| \bs{b}(v) \|_2^2 \, dv (1 + O(T^{-1})) \leq (1 + O(T^{-1}))$ by \cref{lma:spline_properties}\ref{lma:spline_properties_2} and the Riemann sum. Therefore,
$$d_{1i} \leq T^{-1} (\bar{c}_x \bar{c}_\epsilon)^{2/q} (1 + O(T^{-1})) = O(T^{-1}).$$
Similarly,
\begin{equation*}
    \label{eq:Q_t_u_A2}
    \begin{split}
        | d_{2i} | & = \left| \frac{2}{T^2} \sum_{t = 1}^{T-1} \sum_{s = 1}^{T} E(\bs{x}_{it}^\prime \bs{x}_{is} \epsilon_{it} \epsilon_{is}) \bs{b} \left(\frac{t}{T}\right)^\prime \bs{b} \left(\frac{s}{T}\right) \right| \\
                   & \leq \frac{2}{T^2} \sum_{l = 1}^{p} \sum_{t = 1}^{T-1} \sum_{s = t+1}^{T} \left| E(x_{it,l} \epsilon_{it} x_{is,l} \epsilon_{is}) \bs{b} \left(\frac{t}{T}\right) \right|                               \\
                   & \leq \frac{16}{T^2} \sum_{l = 1}^{p} \max_{i,t} \left\{ E \left| x_{it,l} \epsilon_{it} \right|^{2/q} \right\}^{4/q} \sum_{t = 1}^{T-1} \sum_{j = 1}^{\infty} \phi(j)^{q-4/q},
    \end{split}
\end{equation*}
where the last inequality holds by the Davydov inequality for strong mixing processes in combination with the moment conditions in Assumption \superref{line:A1}{1(iii)}, qualified by the strong mixing condition on $\left\{ (\bs{x}_{it}^{(2)}, \epsilon_{it}, t = 1, \dots, T) \right\}$ in Assumption \superref{line:A1}{1(ii)}. Subsequently,
\begin{equation*}
    \label{eq:Davydov}
    \begin{split}
         & \frac{16}{T^2} \sum_{l = 1}^{p} \max_{i,t} \left\{ E \left| x_{it,l} \epsilon_{it} \right|^{2/q} \right\}^{4/q} \sum_{t = 1}^{T-1} \sum_{j = 1}^{\infty} \phi(j)^{q-4/q} \\
         & \leq \frac{16}{T} \sum_{l = 1}^{p} \max_{i,t} \left\{ E \left| x_{it,l} \epsilon_{it} \right|\right\}^{6/q} c_\phi = O_p(T^{-1}).
    \end{split}
\end{equation*}
As a result, $E \| \hat{\bs{q}}_{i, z \epsilon} \|_2^2 = O(T^{-1})$ and, using Chebyshev's inequality, $ \| \hat{\bs{q}}_{i, z \epsilon} \|_2 = O_p(T^{-1/2})$. Furthermore, $\|T^{-1} \sum_{t = 1}^{T} \bs{z}_{it} \|_2 = O_p(M^{-1/2})$ as shown in \eqref{eq:B_norm} and $| T^{-1} \sum_{t = 1}^{T} \epsilon_{it} | = O_p(T^{-1/2})$ by Assumption \superref{line:A1}{1(ii)}. Using these intermediary steps, the triangle inequality, and the sub-multiplicative property,
\begin{equation}
    \label{eq:Q_z_epsilon}
    \| \hat{\bs{q}}_{i, \tilde{z} \tilde{\epsilon}} \|_2 \leq \| \hat{\bs{q}}_{i, z \epsilon} \|_2 + \| T^{-1} \sum_{t = 1}^{T} \bs{z}_{it} \|_2 | T^{-1} \sum_{t = 1}^{T} \epsilon_{it} | = O_p(T^{-1/2}) + O_p(M^{-1/2}) O_p(T^{-1/2}) = O_p(T^{-1/2}).
\end{equation}

Combining the results in \eqref{eq:Q_z_eta_2} and \eqref{eq:Q_z_epsilon} yields
\begin{equation*}
    \label{eq:LA3_fin}
    \| \hat{\bs{q}}_{i, \tilde{z} \tilde{u}} \|_2 = \| \hat{\bs{q}}_{i, \tilde{z} \tilde{\epsilon}} + \hat{\bs{q}}_{i, \tilde{z} \tilde{\varrho}} \|_2 \leq \| \hat{\bs{q}}_{i, \tilde{z} \tilde{\epsilon}} \|_2 + \| \hat{\bs{q}}_{i, \tilde{z} \tilde{\varrho}} \|_2 =  O_p(T^{-1/2}) + O_p (M^{-\theta - 1/2} ) = O_p(T^{-1/2} + M^{-\theta - 1/2})
\end{equation*}
by the triangle inequality.

\paragraph{\textnormal{\textit{Part (ii):}}} Since the cross-sectional dependence is bounded according to Assumption \superref{line:A1}{1(i)}, it follows readily that
\begin{equation*}
    \label{eq:Q_zu_MS}
    \frac{1}{N} \sum_{i = 1}^{N} \| \hat{\bs{q}}_{i, \tilde{z} \tilde{\epsilon}} \|_2^2 = O_p(T^{-1} + M^{-2\theta - 1}),
\end{equation*}
using \cref{lma:comp_error}\ref{lma:comp_error_1}.
\end{proof}
\begin{proof}[{\textbf{Proof of \Cref{lma:prelim_estimate}}}]
    Consider the un-penalized criterion function $\mathcal{F}^*_{NT,i}(\bs{\pi}_i) = T^{-1}\sum_{t = 1}^T \left[ \tilde{y}_{it} - \tilde{\bs{z}}_{it}^\prime \bs{\pi}_i \right]^2$. Define $\bs{a}_i = \bs{\pi}_i - \bs{\pi}^0_i$ and recognize that $\mathcal{F}^*_{NT,i}(\bs{\pi}_i) - \mathcal{F}^*_{NT,i}(\bs{\pi}_i^0) = -2 \bs{a}_i^\prime \hat{\bs{q}}_{i, \tilde{z}\tilde{u}} + \bs{a}_i^\prime \hat{\bs{Q}}_{i, \tilde{z}\tilde{z}} \bs{a}_i$ as shown in \eqref{eq:Q_diff}. The inequality $\mathcal{F}^*_{NT,i}(\dot{\bs{\pi}}_i, \lambda) \leq \mathcal{F}^*_{NT,i}(\bs{\pi}_i^0, \lambda)$ holds trivially since $\dot{\bs{\pi}}_i = \arg \min_{\bs{\pi}_i} \mathcal{F}^*_{NT,i}(\bs{\pi}_i, \lambda)$. Plugging the decomposition into this inequality gives
\begin{equation*}
    \label{eq:Q_diff_ineq}
    \begin{split}
        0 & \geq \mathcal{F}^*_{NT,i}(\dot{\bs{\pi}}_i, \lambda) - \mathcal{F}^*_{NT,i}(\bs{\pi}_i^0, \lambda) = -2 \dot{\bs{a}}_i^\prime \hat{\bs{q}}_{i, \tilde{z}\tilde{u}} + \dot{\bs{a}}_i^\prime \hat{\bs{Q}}_{i, \tilde{z}\tilde{z}} \dot{\bs{a}}_i                                      \\
          & \geq -2 \| \dot{\bs{a}}_i \|_2 \| \hat{\bs{q}}_{i, \tilde{z}\tilde{u}} \|_2 + \| \dot{\bs{a}}_i \|_2^2 M^{-1} \underbar{c}_{\tilde{z} \tilde{z}} = 2 \| \dot{\bs{a}}_i \|_2 O_p(T^{-1/2} + M^{-\theta - 1/2}) + \| \dot{\bs{a}}_i \|_2^2 M^{-1} \underbar{c}_{\tilde{z} \tilde{z}},
    \end{split}
\end{equation*}
where \cref{lma:comp_error}\ref{lma:comp_error_1} provides the rate of $\| \hat{\bs{q}}_{i, \tilde{z}\tilde{u}} \|_2$ and the predictor variance-covariance matrix is substituted with its lower bound $\underbar{c}_{\tilde{z} \tilde{z}}$ according to \cref{lma:Q_zz}\ref{lma:Q_zz_1}, similar to the argument in the \superref{sec:3.1}{proof} of \cref{sec:Theo_1}.

Averaging over all $i = 1, \dots, N$ and rearranging yields
\begin{equation}
    \label{eq:Q_diff_fin_2}
    \frac{2}{\underbar{c}_{\tilde{z} \tilde{z}}} O_p(MT^{-1/2} + M^{-\theta + 1/2}) \frac{1}{N} \sum_{i=1}^{N} \| \dot{\bs{a}}_i \|_2 \geq \sum_{i=1}^{N} \| \dot{\bs{a}}_i \|_2^2.
\end{equation}

As in Theorem \hyperref[sec:3.1]{3.1}, $\| \dot{\bs{a}}_i \|_2 = \| \dot{\bs{\pi}}_i - \bs{\pi}_i^0 \|_2 = O_p(T^{-1/2} M + M^{-\theta + 1/2})$ to ensure that the inequality in \eqref{eq:Q_diff_fin_2} holds for an arbitrarily large $\| \dot{\bs{a}}_i \|_2$. Assumption \superref{line:A2}{2(i)} ensures that $T^{-1/2} M$ is not explosive and $-\theta + 1/2 < 0$ by Assumption \hyperref[line:A1]{1(vi)}.    
\end{proof}
\begin{proof}[{\textbf{Proof of \Cref{lma:adaptive_weight}}}]
    Recall that $\dot{\bs{\pi}} = (\dot{\bs{\pi}}_1^{\prime}, \dots, \dot{\bs{\pi}}_N^{\prime})^\prime$ represents an initial least squares estimate, which, given \cref{lma:prelim_estimate}, is $(MT^{-1/2} + M^{-\theta + 1/2})^{-1}$ consistent. Define $\dot{\bs{\pi}}_i - \bs{\pi}_i^0 = (MT^{-1/2} + M^{-\theta + 1/2}) \dot{\bs{v}}_i$ and recognize that $\| \dot{\bs{v}}_i \|_2 = O_p(1)$. Along the lines of \citet[Lemma B.2]{qian2016shrinkage}, rewrite
\begin{equation*}
    \label{eq:LA5_w}
    \begin{split}
        \dot{\omega}_{ij} & = \| \dot{\bs{\pi}}_i -\dot{\bs{\pi}}_j \|_2^{-\kappa} = \left\| \left[ (MT^{-1/2} + M^{-\theta + 1/2}) \dot{\bs{v}}_i + \bs{\pi}_i^0 \right] - \left[(MT^{-1/2} + M^{-\theta + 1/2}) \dot{\bs{v}}_j + \bs{\pi}_j^0 \right] \right\|_2^{-\kappa} \\
                          & = \left\| \bs{\pi}_i^0 - \bs{\pi}_j^0 + (MT^{-1/2} + M^{-\theta + 1/2}) (\dot{\bs{v}}_i - \dot{\bs{v}}_j) \right\|_2^{-\kappa} = \left\| \bs{\pi}_i^0 - \bs{\pi}_j^0 + (MT^{-1/2} + M^{-\theta + 1/2}) \right\|_2^{-\kappa}.
    \end{split}
\end{equation*}

\paragraph{\textnormal{\textit{Part (i):}}} Let $i,j \in G_k^0$ and subsequently $\bs{\pi}_i^0 = \bs{\pi}_j^0 = \bs{\xi}_k^0$. Considering the minimum adaptive weight,
\begin{equation*}
    \label{eq:LA5_Case1}
    \begin{split}
        \min_{i,j \in G_k^0} \dot{\omega}_{ij} & = \min_{i,j \in G_k^0} \left\| \bs{\pi}_i^0 - \bs{\pi}_j^0 + O_p (MT^{-1/2} + M^{-\theta + 1/2}) \right\|_2^{-\kappa}                                  \\
                                               & = \min_{i,j \in G_k^0} \left\| O_p (MT^{-1/2} + M^{-\theta + 1/2} ) \right\|_2^{-\kappa} = O_p((MT^{-1/2} + M^{-\theta + 1/2})^{-\kappa}).
    \end{split}
\end{equation*}

\paragraph{\textnormal{\textit{Part (ii):}}} Let $i \in G_k^0, j \notin G_k^0$ and as a result $\bs{\pi}_i^0 \neq \bs{\pi}_j^0$. Now, analyzing the maximum weight,
\begin{equation*}
    \label{eq:LA5_Case2}
    \begin{split}
        \max_{i \in G_k^0, j \notin G_k^0} \dot{\omega}_{ij} & = \max_{i \in G_k^0, j \notin G_k^0} \left\| \bs{\pi}_i^0 - \bs{\pi}_j^0 + O_p(MT^{-1/2} + M^{-\theta + 1/2}) \right\|_2^{-\kappa}                 \\
                                                             & = \left\| \min_{i \in G_k^0, j \notin G_k^0} \left( \bs{\pi}_i^0 - \bs{\pi}_j^0 \right) + O_p (MT^{-1/2} + M^{-\theta + 1/2}) \right\|_2^{-\kappa} \\
                                                             & = O_p(J_{\min}^{-\kappa}),
    \end{split}
\end{equation*}
where $J_{\min} = \min_{i \in G_k^0, j \notin G_k^0} \| \bs{\pi}_i^0 - \bs{\pi}_j^0 \|_2$ and Assumption \superref{line:A2}{2(ii)} ensures that $J_{\min}$ dominates an $O_p(MT^{-1/2} + M^{-\theta + 1/2})$ term in the limit.
\end{proof}

\begin{proof}[\textbf{Proof of \Cref{lma:dist}}]
\item[] 
\paragraph{\textnormal{\textit{Part (i):}}} 
Recognize that $\| \bs{b}_c \|_2 = \| \bs{c} \otimes \bs{b}(v)\|_2 = \| \bs{c} \|_2 \| \bs{b}(v) \|_2$ by property of the Kronecker product and a nonrandom $p \times 1$ vector $\bs{c}$ with $\| \bs{c} \|_2= 1$ (see the \superref{sec:3.5.2}{proof} of \cref{sec:Theo_3}\ref{sec:Theo_3_2}) and, following \cref{lma:spline_properties}\ref{lma:spline_properties_1}, $\| \bs{b}(v) \|_2 = 1$. Subsequently, $\| \bs{b}_c \|_2 = O_p(1)$.

\paragraph{\textnormal{\textit{Part (ii):}}} Consider
\begin{equation*}
    \label{eq:LA6_2}
    \begin{split}
        s_{c, \hat{G}_k}^2 & = \frac{M}{N_k T} \bs{b}_c^\prime \left(M N_k^{-1} \hat{\bs{\mathcal{Q}}}_{G_k^0, \tilde{z} \tilde{z}} \right)^{-1} \sum_{i \in G_k^0} \left[ {\tilde{\bs{Z}}_i^\prime} E \left(\bs{\epsilon}_i \bs{\epsilon}_i^{\prime} \right) {\tilde{\bs{Z}}_i} \right] \left(M N_k^{-1} \hat{\bs{\mathcal{Q}}}_{G_k^0, \tilde{z} \tilde{z}} \right)^{-1} \bs{b}_c \\
                           & \leq \| \bs{b}_c \|_2^2 \left[\mu_{\min}\left(M N_k^{-1} \hat{\bs{\mathcal{Q}}}_{G_k^0, \tilde{z} \tilde{z}} \right) \right]^{-2} \frac{M}{N_k T} \sum_{i \in G_k^0} \left[ {\tilde{\bs{Z}}_i^\prime} E \left(\bs{\epsilon}_i \bs{\epsilon}_i^{\prime} \right) {\tilde{\bs{Z}}_i} \right]                                                              \\
                           & \leq \| \bs{b}_c \|_2^2 \left[\mu_{\min}\left(M N_k^{-1} \hat{\bs{\mathcal{Q}}}_{G_k^0, \tilde{z} \tilde{z}} \right) \right]^{-2} \max_{i \in G_k^0} \left[ \mu_{\max} (E \left(\bs{\epsilon}_i \bs{\epsilon}_i^{\prime} \right)) \right] \mu_{max} \left( \frac{M}{N_k} \hat{\bs{\mathcal{Q}}}_{G_k^0, \tilde{z} \tilde{z}} \right)                   \\
                           & \leq \frac{\| \bs{b}_c \|_2^2  \bar{c}_{\epsilon \epsilon} \bar{c}_{\tilde{z} \tilde{z}}}{\underbar{c}_{\tilde{z} \tilde{z}}^2},
    \end{split}
\end{equation*}
by the Triangle property and $\bs{a}^\prime \bs{A} \bs{a} \geq \mu_{\min}(\bs{A}) \bs{a}^\prime \bs{a}$ for a conformable vector $\bs{a}$ and matrix $\bs{A}$. Furthermore, notice that $\|\bs{b}_c\|_2^2 = O_p(1)$ as shown in \cref{lma:dist}\ref{lma:dist_1}, $\max_{1 \leq i \leq N} \mu_{max} \left(M N_k^{-1} \hat{\bs{\mathcal{Q}}}_{G_k^0, \tilde{z} \tilde{z}} \right) \leq \bar{c}_{\tilde{z} \tilde{z}}$ and $\min_{1 \leq i \leq N} \mu_{min} \left( M N_k^{-1} \hat{\bs{\mathcal{Q}}}_{G_k^0, \tilde{z} \tilde{z}} \right) \geq \underbar{c}_{\tilde{z} \tilde{z}}$ by \cref{lma:Q_zz}\ref{lma:Q_zz_2}, and $\max_{i \in G_k^0} \mu_{\max} (E(\bs{\epsilon}_i \bs{\epsilon}_i^\prime)) \leq \bar{c}_{\epsilon \epsilon}$ by Assumption \superref{line:A4}{4(i)}. In consequence,  $s_{c, \hat{G}_k}^2 = O_p(1)$ \citep[see][Lemma A.8]{su2019sieve}.    
\end{proof}

\begin{proof}[\textbf{Proof of \Cref{lma:mse}}]
    The proof of \eqref{eq:LA6_claim} works by demonstrating that the \textit{MSE} of an over-fitted model and the \textit{MSE} of a model with the true grouping structure both converge towards the non-reducible residual error variance in the limit. The different converge rates give the result in \eqref{eq:LA6_claim}.

\begin{equation}
    \label{eq:LA6_truegroups}
    \begin{split}
        \hat{\sigma}^2_{\mathcal{G}^0} & = \frac{1}{NT} \sum_{k = 1}^{K_0} \sum_{i \in G^0_k} \sum_{t = 1}^T \hat{\tilde{u}}_{it}^2 =  \frac{1}{NT} \sum_{k = 1}^{K_0} \sum_{i \in G^0_k} \sum_{t = 1}^T \left[ \tilde{y}_{it} - \hat{\bs{\alpha}}_{G_k^0}^{p \prime} \left(\frac{t}{T}\right) \tilde{\bs{x}}_{it} \right]^2                                              \\
                                       & = \frac{1}{NT} \sum_{k = 1}^{K_0} \sum_{i \in G^0_k} \sum_{t = 1}^T \left[ \tilde{\epsilon}_{it} - \left( \hat{\bs{\alpha}}_{G_k^0}^{p} \left(\frac{t}{T}\right) - \bs{\alpha}_{G_k^0}^0 \left(\frac{t}{T}\right) \right)^\prime \tilde{\bs{x}}_{it} \right]^2                                                                   \\
                                       & = \frac{1}{NT} \sum_{i = 1}^{N} \sum_{t = 1}^T \tilde{\epsilon}_{it}^2 - \frac{1}{NT} \sum_{k = 1}^{K_0} \sum_{i \in G^0_k} \sum_{t = 1}^T \left[ \left(\hat{\bs{\alpha}}_{G_k^0}^{p} \left(\frac{t}{T}\right) - \bs{\alpha}_{G_k^0}^0 \left(\frac{t}{T}\right) \right)^\prime \tilde{\bs{x}}_{it} \tilde{\epsilon}_{it} \right] \\
                                       & \qquad + \frac{1}{NT} \sum_{k = 1}^{K_0} \sum_{i \in G^0_k} \sum_{t = 1}^T \left[ \left(\hat{\bs{\alpha}}_{G_k^0}^{p} \left(\frac{t}{T}\right) - \bs{\alpha}_{G_k^0}^0 \left(\frac{t}{T}\right) \right)^\prime \tilde{\bs{x}}_{it} \right]^2
    \end{split}
\end{equation}

Denote the irreducible sample \textit{MSE} as $\bar{\sigma}_{NT}^2 = (NT)^{-1} \sum_{i = 1}^N \sum_{t = 1}^T \tilde{\epsilon}_{it}^2$. Moreover, given \cref{sec:Theo_3}\ref{sec:Theo_3_2}, $\hat{\bs{\alpha}}_{\hat{G}_k}^p (t/T)$ is $((N_{k, \min} T)^{1/2} M^{1/2})$-consistent, where $N_{k,\min} = \min_{k=1,\dots,K_0} N_k$. Due to Assumption \superref{line:A1}{1(ii)}, $\| T^{-1} \sum_{t = 1}^T \tilde{\bs{x}}_{it} \tilde{\epsilon}_{it} \|_2 = O_p(T^{1/2})$. In addition, note that $ N^{-1} \sum_{k = 1}^{K_0} \sum_{i \in G_k^0} \| \hat{\bs{q}}_{i, \tilde{x} \tilde{\epsilon}} \|_2 = N^{1/2} O_p(T^{1/2})$ as a function of weak cross-sectional dependence (see Assumption \superref{line:A1}{1(i)}). Plugging these results into \eqref{eq:LA6_truegroups} yields the rates
\begin{equation}
    \label{eq:LA6_truegroups2}
    \begin{split}
        \hat{\sigma}^2_{\mathcal{G}^0} & = \bar{\sigma}_{NT}^2 + O_p\left( (N T)^{1/2} M^{1/2}\right) O_p((NT)^{1/2}) + O_p\left( (N T)^{-1} M \right) O_p(1) \\
                                       & = \bar{\sigma}_{NT}^2 + O_p\left( (T N)^{-1} M \right).
    \end{split}
\end{equation}

Define the set $\bar{\mathbb{G}}_K = \{\bar{\mathcal{G}}_K = \{\bar{G}_k\}_{k = 1}^K: \nexists \, i,j \in \bar{G}_K \; \text{where} \; i \in G_l^0, \, j \notin G_l^0, \, 1 \leq l \leq K_0\}$ with $K_0 < K \leq K_{\max} = N$. That is, each $\bar{\mathcal{G}}_K \in \bar{\mathbb{G}}_K$ denotes a partition of $K>K_0$ over $N$ where no heterogeneous cross-sectional units are pooled together in any group. Hence, it is trivial that $\hat{\sigma}^2_{\bar{\mathcal{G}}_K} \leq \hat{\sigma}^2_{\mathcal{G}^0}$. Given the result in \eqref{eq:LA6_truegroups2}, we can expand this inequality to
\begin{equation}
    \label{eq:LA6_ineq}
    0 \leq \hat{\sigma}^2_{\mathcal{G}^0} - \hat{\sigma}^2_{\bar{\mathcal{G}}_K} = \bar{\sigma}_{NT}^2 - \hat{\sigma}^2_{\bar{\mathcal{G}}_K} + O_p((N T)^{-1} M).
\end{equation}

Let $\mathcal{J}_{NT}$ be the largest distance between the estimated and irreducible mean squared error for any of the $K_0 < K \leq K_{\max}$ groups
\begin{equation*}
    \label{eq:LA6_M}
    \mathcal{J}_{NT} = \max_{K_0 < k \leq K} \inf_{\alpha} \left\{ \frac{1}{N_k T} \sum_{i \in \bar{G}_k} \sum_{t = 1}^T \left[\tilde{\epsilon}_{it}^2 - \left( \tilde{y}_{it} - \tilde{\bs{x}}_{it}^\prime \bs{\alpha}_k \left(\frac{t}{T}\right) \right)^2 \right] \right\},
\end{equation*}
such that $\bar{\sigma}_{NT}^2 - \hat{\sigma}^2_{\bar{\mathcal{G}}_K} \leq K \mathcal{J}_{NT}$. Then
\begin{equation*}
    \label{eq:LA6_J2}
    \breve{\alpha}_k \left(\frac{t}{T}\right) = \breve{\bs{\Xi}}_k^\prime B \left(\frac{t}{T}\right) = \arg \min_{\xi_k} \frac{1}{N_k T} \sum_{i \in \bar{G}_k} \sum_{t = 1}^T \left[ \tilde{\epsilon}_{it}^2 - \left( \tilde{y}_{it} - \tilde{\bs{z}}_{it}^\prime \bs{\xi}_k \left(\frac{t}{T}\right) \right)^2 \right],
\end{equation*}
with $\breve{\bs{\xi}}_k = (\sum_{i \in \bar{G}_k} \sum_{t = 1}^{T} \tilde{\bs{z}}_{it} \tilde{\bs{z}}_{it}^\prime)^{-1} \sum_{i \in \bar{G}_k} \sum_{t = 1}^{T} \tilde{\bs{z}}_{it} \tilde{y}_{it}$. Since $K_0 < K \leq K_{\max} = N$, the minimum group cardinality is 1 and $\breve{\alpha}_k (t/T)$ is, in line with \cref{sec:3.5}, only $\sqrt{M/T}$-consistent. As a consequence, $\mathcal{J}_{NT} = O_p(T^{-1}M)$.

Plugging $\bar{\sigma}_{NT}^2 - \hat{\sigma}^2_{\bar{\mathcal{G}}_K} \leq K \mathcal{J}_{NT} = O_p(T^{-1}M)$ into \eqref{eq:LA6_ineq}, it becomes apparent that $0 \leq \hat{\sigma}^2_{\mathcal{G}^0} - \hat{\sigma}^2_{\bar{\mathcal{G}}_K} \leq O_p(T^{-1}M) + O_p((N T)^{-1} M) = O_p(T^{-1}M)$. Then, by construction of $\mathcal{J}_{NT}$,
\begin{equation}
    \label{eq:LA6_claim}
    \max_{K_0 < K\leq K_{\max}} \; \sup_{\bar{\mathcal{G}}_K \in \bar{\mathbb{G}}_K} |\hat{\sigma}^2_{\bar{\mathcal{G}}_K} - \hat{\sigma}^2_{\mathcal{G}^0} | = O_p ( T^{-1} M ).
\end{equation}
\end{proof}

\section{Numerical Implementation}
\label{sec:Appendix_Algo}

We minimize the criterion \eqref{eq:Obj_penalty} using an iterative \textit{ADMM} algorithm, adapted from \citet[sec. 3.1]{ma2017} and \citet[sec. 5.1]{mehrabani2023}. Let $\bs{a}_{ij} = \bs{\pi}_i - \bs{\pi}_j$. Then minimizing
\begin{equation}
    \label{eq:Obj_penalty_const}
    \mathcal{F}_{NT}(\bs{\pi}, \bs{a}, \lambda) = \frac{1}{T} \sum_{i = 1}^{N} \sum_{t=1}^{T} \left(\tilde{y}_{it} - \bs{\pi}_i^\prime \tilde{\bs{z}}_{it} \right)^2 + \frac{\lambda}{N} \sum_{i = 1}^{N - 1} \sum_{j=i+1}^{N} \dot{\omega}_{ij} \left\| \bs{a}_{ij} \right\|_2 \text{ subject to } \bs{a}_{ij} = \bs{\pi}_i - \bs{\pi}_j,
\end{equation}
is equivalent to optimizing the objective in \eqref{eq:Obj_penalty}. We can rewrite this constraint optimization as an Augmented Lagrangian problem
\begin{equation}
    \label{eq:Obj_penalty_const_2}
    \begin{split}
        \mathcal{L}_{NT, \vartheta}(\bs{\pi}, \bs{a}, \lambda, \bs{\upsilon}) & = \frac{T}{2} \mathcal{F}_{NT}(\bs{\pi}, \bs{a}, \lambda) + \sum_{i = 1}^{N - 1} \sum_{j=i+1}^{N} \bs{\upsilon}_{ij}^\prime \left(\bs{\pi}_i - \bs{\pi}_j - \bs{a}_{ij} \right) + \frac{\vartheta}{2} \sum_{i = 1}^{N - 1} \sum_{j=i+1}^{N} \left\| \bs{\pi}_i - \bs{\pi}_j - \bs{a}_{ij} \right\|^2_2 \\
                                                                              & = \mathcal{F}_{NT}^*(\bs{\pi})  + \mathcal{P}_{NT}^*(\bs{a}, \lambda)  + \sum_{i = 1}^{N - 1} \sum_{j=i+1}^{N} \bs{\upsilon}_{ij}^\prime \left(\bs{\pi}_i - \bs{\pi}_j - \bs{a}_{ij} \right)                                                                                                           \\
                                                                              & \qquad \qquad \qquad \qquad \qquad \; \  + \frac{\vartheta}{2} \sum_{i = 1}^{N - 1} \sum_{j=i+1}^{N} \left\| \bs{\pi}_i - \bs{\pi}_j - \bs{a}_{ij} \right\|_2^2,
    \end{split}
\end{equation}
where $\mathcal{F}_{NT}^*$ reflects the goodness-of-fit term and $\mathcal{P}_{NT}^*(\bs{a}, \lambda)$ the \textit{PAGFL} penalty in \eqref{eq:Obj_penalty_const}, $\bs{\upsilon}_{ij}$ denotes the dual variable and $\vartheta > 0$ the \textit{ADMM} penalty parameter controlling the trade-off between feasibility and optimization.

The \textit{ADMM} algorithm \ref{alg:ADMM} iteratively minimizes \eqref{eq:Obj_penalty_const_2} until convergence or until a maximum number of iterations $l_{\max}$ is reached. We set the $l_{\max} = 50,000$ and the convergence tolerance $\varepsilon_{\text{tol}}^{\textit{ADMM}}$ to $\varepsilon_{\text{tol}}^{\textit{ADMM}} = 1 \times 10^{-10}$ for the simulation study and empirical illustration. Define $\tilde{\bs{y}} = (\tilde{\bs{y}}_1^\prime, \dots, \tilde{\bs{y}}_N^\prime)^\prime$, $\tilde{\bs{y}}_i = (\tilde{y}_{i1}, \dots, \tilde{y}_{iT})^\prime$; the $NT \times NMp$ regressor block-matrix $\tilde{\bs{Z}} = \text{diag}(\tilde{\bs{Z}}_1, \dots, \tilde{\bs{Z}}_N)$ with $\tilde{\bs{Z}}_i = (\tilde{\bs{z}}_{i1}, \dots, \tilde{\bs{z}}_{iT})^\prime$; the $NMp \times 1$ coefficient vector $\bs{\pi} = (\bs{\pi}_1^\prime, \dots, \bs{\pi}_N^\prime)^\prime$; the $MpN(N-1)/2 \times NMp$ differencing matrix $\bs{\Delta} = \bs{\varsigma} \otimes \bs{I}_{Mp}$ with $\bs{\varsigma} = \left\{\bs{\zeta}_i - \bs{\zeta}_j, 1 \leq i < j \leq N\right\}^\prime$, and the $N \times 1$ indicator vector $\bs{\zeta}_i$ with a one as its $i\textsuperscript{th}$ element and zeros elsewhere; the $MpN(N-1)/2 \times 1$ vectors $\bs{a} = \left\{\bs{a}_{ij}^\prime,1 \leq i < j \leq N \right\}^\prime$ and $\bs{\upsilon} = \left\{\bs{\upsilon}_{ij}^\prime, 1 \leq i < j \leq N\right\}^\prime$.
\RestyleAlgo{ruled}
\begin{algorithm}
    \caption{\textit{ADMM} algorithm to minimize the \textit{FUSE-TIME} criterion \eqref{eq:Obj_penalty}.}
    \label{alg:ADMM}
    \KwResult{$\hat{\bs{\pi}}_i = \bs{\pi}_i^{(l)}$}
    $l \gets 0$\;
    $\bs{\pi}^{(l)} \gets \dot{\bs{\pi}}$\;
    $\bs{a}^{(l)} \gets \bs{\Delta} \bs{\pi}^{(l)}$\;
    $\bs{\upsilon}^{(l)} \gets 0$\;
    \While{$\| \bs{\Delta} \bs{\pi}^{(l)} - \bs{a}^{(l)} \|_2 > \varepsilon_{\text{tol}}^{\textit{ADMM}} \; \& \; l < l_{\max}$}{
    $l \gets l + 1$\;
    \textbf{Primal step 1}: $\bs{\pi}^{(l)} \gets \left( \tilde{\bs{Z}}^\prime \tilde{\bs{Z}} + \vartheta \bs{\Delta}^\prime \bs{\Delta} \right)^{-1} \left( \tilde{\bs{Z}}^\prime \tilde{\bs{y}} + \vartheta \bs{\Delta}^\prime \bs{a}^{(l-1)} -  \bs{\Delta}^\prime \bs{\upsilon}^{(l-1)} \right)$\;
    \textbf{Primal step 2}: $\bs{\psi}^{(l)}_{ij} \gets \bs{\pi}_i^{(l)} - \bs{\pi}_j^{(l)} - \vartheta^{-1} \bs{\upsilon}_{ij}^{(l-1)}$, \\ 
    \qquad \qquad \qquad \, \, \, \ $\bs{a}^{(l)}_{ij} \gets \max \left\{ 1 - \dot{\bs{\omega}}_{ij} T \lambda / \left( 2 N \vartheta \| \bs{\psi}^{(l)}_{ij} \|_2 \right) , 0 \right\} \bs{\psi}^{(l)}_{ij}$ for each $1 \leq i < j \leq N$, \\ 
    \qquad \qquad \qquad \quad \, \ $\bs{a}^{(l)} \gets \left\{\bs{a}_{ij}^{(l) \prime}, 1 \leq i < j \leq N \right\}^\prime$\;
    \textbf{Dual step}: $\bs{\upsilon}^{(l)} \gets \bs{\upsilon}^{(l-1)} + \vartheta \left( \bs{\Delta} \bs{\pi}^{(l)} - \bs{a}^{(l)} \right)$\;
    }
\end{algorithm}
The primal steps of the \textit{ADMM} are proximal updates of $\hat{\bs{\pi}}$ and $\hat{\bs{a}}$ and take the form
\begin{equation*}
    \label{eq:ADMM_updates}
    \begin{split}
         & \bs{\pi}^{(l+1)} = \arg \min_{\bs{\pi}} \left\{ \mathcal{F}_{NT}^* (\bs{\pi}) + \frac{\vartheta}{2} \left\| \bs{\Delta} \bs{\pi} - \bs{a}^{(l)} + \vartheta^{-1} \bs{\upsilon}^{(l)} \right\|^2_2 \right\}    \\
         & \bs{a}^{(l+1)} = \arg \min_{\bs{a}} \left\{ \mathcal{P}_{NT}^*(\bs{a}, \lambda) + \frac{\vartheta}{2} \left\| \bs{\Delta} \bs{\pi}^{(l)} - \bs{a} + \vartheta^{-1} \bs{\upsilon}^{(l)} \right\|^2_2 \right\}.
    \end{split}
\end{equation*}
In particular, we obtain $\bs{\pi}^{(l+1)}$ by solving
$$
    \frac{\partial}{\partial \bs{\pi}} \left[ \mathcal{F}_{NT}^*(\bs{\pi})  + \vartheta / 2 \left\| \bs{\Delta} \bs{\pi} - \bs{a}^{(l)} + \vartheta^{-1} \bs{\upsilon}^{(l)} \right\|^2_2 \right] = 0,
$$ 
where $\mathcal{F}_{NT}^*(\bs{\pi})  = 1/2 \| \tilde{\bs{y}} - \tilde{\bs{Z}} \bs{\pi} \|^2_2$. 

The \textit{PAGFL} penalty is applied to each pair of cross-sectional units separately. Subsequently, $\bs{a}^{(l+1)}$ must be derived individually for each $i,j$. Therefore, we take
$$
    \bs{a}_{ij}^{(l+1)} = \arg \min_{a_{ij}} \left\{\frac{T \lambda}{2 N} \sum_{i = 1}^{N - 1} \sum_{j=i+1}^{N} \dot{\omega}_{ij} \| \bs{a}_{ij} \|_2 + \frac{\vartheta}{2} \sum_{i = 1}^{N - 1} \sum_{j=i+1}^{N} \| \bs{\pi}_i^{(l)} - \bs{\pi}_j^{(l)} - \bs{a}_{ij} - \vartheta^{-1} \bs{\upsilon}_{ij} \|^2_2 \right\},
$$
the closed-form solution of which is the typical soft-thresholding rule
$$
    \bs{a}_{ij}^{(l+1)} = \max \left\{ 1 - \frac{T \lambda}{2N} \frac{\dot{\omega}_{ij} / \vartheta}{\| \bs{\pi}_i^{(l)} - \bs{\pi}_j^{(l)} - \bs{a}_{ij} - \vartheta^{-1} \bs{\upsilon}_{ij} \|_2} , 0 \right\} \left( \bs{\pi}_i^{(l)} - \bs{\pi}_j^{(l)} - \bs{a}_{ij} - \vartheta^{-1} \bs{\upsilon}_{ij} \right),
$$
which appears in Algorithm \ref{alg:ADMM}. The dual step of the \textit{ADMM} algorithm enforces feasibility by updating the Lagrange multipliers to drive the constraint $\bs{\pi}_i - \bs{\pi}_j = \bs{a}_{ij}$, moderated by the \textit{ADMM} penalty $\vartheta$.

Since the objective function of \textit{FUSE-TIME} \eqref{eq:Obj_penalty} is convex, Algorithm \ref{alg:ADMM} achieves convergence to the optimal point as $l \to \infty$. Furthermore, given that the primal residual $\bs{r}_1^{(l)} = \bs{\Delta} \bs{\pi}^{(l)} - \bs{a}^{(l)}$ and dual residual $\bs{r}_2^{(l)} = \vartheta \bs{\Delta} (\bs{\pi}^{(l)} - \bs{\pi}^{(l-1)}) $ tend to zero asymptotically 
$$
\lim_{l \to \infty} \| \bs{r}_d^{(l)} \|^2_2 = 0, \quad d = \{1,2\},
$$ the algorithm achieves both primal and dual feasibility. The proof of these properties is identical to \citet[Supplement, Appendix C]{mehrabani2023} and thus omitted.

In theory, two individuals $i,j$ are only fused together if $\| \hat{\bs{\pi}}_i - \hat{\bs{\pi}}_j\|_2 = 0$. However, employing the \textit{ADMM} algorithm, it is not always computationally feasible for normed coefficient vector differences to equal zero exactly. As a consequence, we relax this condition and group two individuals if $\| \hat{\bs{\pi}}_i - \hat{\bs{\pi}}_j\|_2 < \varepsilon_{\text{tol}}^{\mathcal{G}}$, where $\varepsilon_{\text{tol}}^{\mathcal{G}}$ is set to a machine inaccuracy value. The smaller $\varepsilon_{\text{tol}}^{\mathcal{G}}$, the more \textit{ADMM} algorithm iterations are required to obtain suitable results. As a consequence, there is an efficiency-accuracy trade-off when selecting $\varepsilon_{\text{tol}}^{\mathcal{G}}$. We find that $\varepsilon_{\text{tol}}^{\mathcal{G}} = 0.001$ yields good computational efficiency while still providing sharp between-group distinction. Similarly, it may occur that $\| \hat{\bs{\pi}}_i - \hat{\bs{\pi}}_j\|_2 < \varepsilon_{\text{tol}}^{\mathcal{G}}$ and $\| \hat{\bs{\pi}}_i - \hat{\bs{\pi}}_l\|_2 < \varepsilon_{\text{tol}}^{\mathcal{G}}$, but $\| \hat{\bs{\pi}}_j - \hat{\bs{\pi}}_l\|_2 \geq \varepsilon_{\text{tol}}^{\mathcal{G}}$. Since the group structure must be transitive, we nonetheless assign such triplets $(i,j,l)$ to the same group in our numerical implementation.

Furthermore, splinter groups with only one or a few members may emerge in empirical or simulation settings. This characteristic also has been previously documented for similar models \citep[see][]{su2016identifying,su2019sieve,mehrabani2023,dzemski2024confidence}. Such trivial groups can be precluded by either increasing $\varepsilon_{\text{tol}}^{\mathcal{G}}$ or by increasing the number of iterations of the \textit{ADMM} algorithm. However, another solution is to specify a floor on the group cardinalities, e.g., 5\% of $N$, and place individuals of groups that fall short of this threshold into remaining groups of sufficient size according to the lowest \textit{MSE}. We impose such a $N_k/N \geq 5\%$ rule for the simulation study in Section \ref{sec:Simulation} and the empirical illustration in Section \ref{sec:Empirical}. Note that in the subsequent simulation study, the preliminary coefficient vector differences $\| \dot{\bs{\pi}}_i - \dot{\bs{\pi}}_j \|_2$ exceed $\varepsilon_{\text{tol}}^{\mathcal{G}} = 0.001$ in all but a negligible amount of instances. Subsequently, this simple $MSE$-based classifier in combination with $\varepsilon_{\text{tol}}^{\mathcal{G}}$ leads to almost no groupings and the clustering procedure is driven by the penalization routine.

The function \texttt{fuse\_time()} of our companion R-package \texttt{PAGFL} \citep{haimerl2025pagfl} provides an user-friendly and efficient open-source software implementation of the numerical algorithm presented in this section, along with a plethora of other functions and extensions.

\section{Extensions}
\label{sec:Extensions}

In the following, we provide details on several extensions to broaden the scope of \textit{FUSE-TIME}.

\subsection{Panels with Coefficient-specific Groups or Breaks in the Group Structure}
\label{sec:Extension_coefgroups}

Consider a panel data model where each coefficient function follows a distinct group structure. This a setup reflects, for example, heterogeneous responses of economies to commodity price shocks: a specific country may be equally exposed to fluctuations in oil prices than one set of peers but form a group with a different set of countries regarding the effect of coal prices \citep{cashin2014differential}.

We extend the DGP in \eqref{eq:DGP} to allow each of the $p$ functional coefficients in $\bs{\beta}_i^0 (t/T)$ to follow a unique unobserved group pattern
\begin{equation}
    \label{eq:beta_coef_group}
    \beta_{il}^0 \left(\frac{t}{T}\right) = \sum_{k = 1}^{K_{0l}} \alpha_{kl}^0 \left(\frac{t}{T}\right) \bs{1} \{i \in G_{kl}^0\}, \quad l = 1, \dots, p.
\end{equation}
Note that both the group adherence $G_{kl}^0$ and the total number of groups $K_{0l}$ may vary for each of the $p$ coefficient functions. To identify the coefficient-specific groups in \eqref{eq:beta_coef_group}, we adjust the penalty in the \textit{FUSE-TIME} criterion \eqref{eq:Obj_penalty} such that the group-Lasso operates on the $M$ control points associated with each functional coefficient, as opposed to the entire $M \times p$ matrix $\bs{\Pi}_i$. Let $\bs{\pi}_{il}$ denote the $l \textsuperscript{th}$ column of $\bs{\Pi}_i$ and specify the criterion
\begin{equation*}
    \label{eq:Obj_penalty_coef_group}
    \mathcal{F}_{NT}(\pi, \lambda) = \frac{1}{T} \sum_{i = 1}^{N} \sum_{t=1}^{T} \left(\tilde{y}_{it} - \bs{\pi}_i^\prime \tilde{\bs{z}}_{it} \right)^2 + \frac{\lambda}{N} \sum_{l = 1}^{p} \sum_{i = 1}^{N - 1} \sum_{j=i+1}^{N} \dot{\omega}_{ijl} \left\| \pi_{il} - \pi_{jl} \right\|_2,
\end{equation*}
where the adaptive weight is defined as $\dot{\omega}_{ijl} = \| \dot{\bs{\pi}}_{il} - \dot{\bs{\pi}}_{jl} \|_2^{-\kappa}$ and $\dot{\bs{\pi}}_{i}$ is identical to the base-case in Section \ref{sec:pagfl}.

Having obtained the $p$ distinct group structures, the \textit{post-Lasso} estimator follows as
\begin{equation*}
    \label{eq:label}
    \hat{\bs{\xi}}^p = (\breve{\bs{Z}}^\prime \breve{\bs{Z}})^{-1} \breve{\bs{Z}}^\prime \tilde{y},
\end{equation*}
where the $M \times \sum_{l = 1}^{p} \hat{K}_l$ matrix $\hat{\bs{\Xi}}^p = (\hat{\bs{\Xi}}_1^p, \dots, \hat{\bs{\Xi}}_p^p)$, $\vect{\hat{\bs{\Xi}}^p} = \hat{\bs{\xi}}^p $, stacks coefficient-wise with $\hat{\bs{\Xi}}_l^p = (\hat{\bs{\xi}}_{l1}^p, \dots, \hat{\bs{\xi}}_{l \hat{K}_l}^p)$ and $\hat{\bs{\xi}}_{lk}^p$ holds the $M$ control points shaping the time-varying function $l$ of group $k$, with $k = 1, \dots, \hat{K}_l$ and $l = 1, \dots, p$. Moreover, let $\tilde{\bs{y}} = (\tilde{\bs{y}}_1^\prime, \dots, \tilde{\bs{y}}_N^\prime)^\prime$, $\tilde{\bs{y}}_i = (\tilde{y}_{i1}, \dots, \tilde{y}_{iT})^\prime$, and $\breve{\bs{Z}} = (\breve{\bs{Z}}_1, \dots, \breve{\bs{Z}}_p)$, where $\breve{\bs{Z}}_l$ is a $NT \times \hat{K}_l M$ regressor matrix $\breve{\bs{Z}}_l = (\bs{D}_{l1}, \dots, \bs{D}_{l \hat{K}_l}) (\bs{I}_{\hat{K}_l} \otimes \tilde{\bs{z}}_l)$. $\bs{D}_{kl}$ indicates a $NT \times NT$ matrix selecting observations pertaining to group $k$ of the $l\textsuperscript{th}$ coefficient function $\bs{D}_{kl} = \text{diag}(\bs{d}_{kl}) \otimes \bs{I}_T$, where $\bs{d}_{kl}$ is a vector of length $N$ with ones for all $\{ i \in \hat{G}_{kl}, 1 \leq i \leq N \}$ and zeros elsewhere. $\bs{\tilde{Z}}_l$ is a $NT \times M$ matrix that collects all basis functions corresponding to the $l\textsuperscript{th}$ coefficient function $\tilde{\bs{Z}}_l = (\tilde{\bs{z}}_{1l}^\prime, \dots, \tilde{\bs{z}}_{Nl}^\prime)^\prime$, $\tilde{\bs{z}}_{il} = (\tilde{z}_{i1l}, \dots, \tilde{z}_{iTl})^\prime$, with $\tilde{z}_{itl} = (\bs{r}_l \tilde{\bs{x}}_{it}) \otimes \bs{b}(t/T)$ and the $1 \times p$ selection vector $\bs{r}_l$ featuring a one at position $l$ and zeros for all remaining elements. The time-varying functional coefficients for individual $i$ are given by $\hat{\bs{\alpha}}_{i}^p (t/T) = \bs{W}_i \hat{\bs{\Xi}}^{p \prime} \bs{b}(t/T)$, where $\bs{W}_i$ represents a $p \times \sum_{l = 1}^{p} \hat{K}_l$ selection matrix $\bs{W}_i = \{ ( \bs{0}_{\bs{1} \{l \neq 1 \} \sum_{s = 1}^{l-1} \hat{K}_s}^\prime, \bs{w}_l^\prime, \bs{0}_{\bs{1} \{l \neq p \} \sum_{s = l+1}^{p} \hat{K}_s}^\prime )^\prime , 1 \leq l \leq p\}^\prime$ and $\bs{w}_l$ is a vector of length $\hat{K}_l$ with a one on position $k = \{k: i \in \hat{G}_{kl}, 1 \leq k \leq \hat{K}_l\}$ and zeros elsewhere.

Based on the derivations in Section \ref{sec:Asymptotics}, we conjecture that the preliminary convergence rates extend naturally to coefficient-wise groups. Moreover, the same applies to the limiting distribution of each coefficient function in Theorem \superref{sec:Theo_3}{3} after replacing the global $N_k$ with the coefficient-specific group size $N_{kl}$ and inserting the coefficient group equivalents of $\hat{\bs{\Omega}}_{G_k^0}$, $\hat{\bs{\mathcal{E}}}_{G_k^0}$, and $\bs{q}_{G_k^0}$.

Analogous to defining the group-Lasso over the columns of $\hat{\bs{\Pi}}_i$, it is also straightforward to penalize $\hat{\bs{\Pi}}_i$ row-wise. In that case, the groupings do not vary among the $p$ functional coefficients but among each of the $M$ basis functions in $\bs{b}(t/T)$. Since the basis functions vanish outside their respective interval $\{ V_m \}_{m = -d}^{M^*}$ (cf. \eqref{eq:b_spline_interval}), the groups change over time. Furthermore, as $d+1$ basis functions overlap at any point along the domain (see Figure \ref{fig:Splines}, left panel), such a specification produces up to $\prod_{j = 0}^{d} \hat{K}_{-d+j + \varphi}$ functional coefficient vectors unique to each interval $[v_\varphi, v_{\varphi+1})$ for $\varphi = 0, \dots, M^* - 1$ and $[v_{M^*}, v_{M^*+1}]$, where $\hat{K}_{-d + j + \varphi} = \hat{K}_{m}$ denotes the estimated number of groups of basis function $m$ and $v_\varphi$ indicates the respective knot (cf. Appendix \ref{sec:Appendix_sieve}). Each interior knot reflects a break-point at which the group adherence may change, allowing for a total of $M^* + 1$ distinct groupings with a potential switch every $T / (M^* + 1)$ time periods. This significantly complicates the interpretation of the group structure but allows the coefficient functions of different cross-sectional individuals to coincide only for specific periods.

In addition, models with a mix of global, group-specific, and individual coefficients have found ample empirical use, mirroring models with both common, group-level, and idiosyncratic factors \citep[see][among others]{diebold2008global,ando2016panel,freyaldenhoven2022factor}. Such a structure is nested within coefficient-specific groups as introduced above. This becomes evident by reordering $\bs{\beta}_i(t/T)$ such that $\bs{\beta}_i(t/T) = (\bs{\beta}_i^{(1)\prime}(t/T), \bs{\beta}_i^{(2)\prime}(t/T), \beta^{(3)\prime}(t/T))^\prime$, where only the $p^{(1)} \times 1$ functional vector $\bs{\beta}_i^{(1)}(t/T)$ follows a latent group structure as described in \eqref{eq:beta_group}, that is $\hat{\mathcal{G}}_l$ is identical for all $l = 1, \dots, p^{(1)}$. The $p^{(2)} \times 1$ vector $\bs{\beta}_i^{(2)}(t/T)$ is idiosyncratic, $K_l = N$ for all $l = p^{(1)} + 1, \dots, p^{(2)}$. The $p^{(3)} \times 1$ vector $\beta^{(3)}(t/T)$ is common across all cross-sectional units such that $K_l = 1$ for all $l = p^{(2)} + 1, \dots, p^{(3)}$. Estimation then follows readily from the derivations above, where $\dot \omega_{ijl}$ is set to zero for all $l \in (p^{(1)}, p^{(2)}]$---inducing groups of cardinality one---and to an arbitrarily large value for all $l \in (p^{(2)}, p^{(3)}]$---inducing groups of cardinality $N$. The limiting distributions in Theorem \hyperref[sec:Theo_3]{3.5} now hold for $\hat{\alpha}_{\hat{G}_k}^{(1)p}(t/T)$ when replacing $p$ with $p^{(1)}$ and for $\hat{\alpha}_{i}^{(2)p}(t/T)$ when setting $p = p^{(2)}$ and $N_k = 1$. The global coefficients are subject to quicker convergence rates since all $N$ individuals are pooled. Subsequently, $\hat{\bs{\pi}}_i^{(3)}$ converges pointwise with a rate of $O_p(M(NT)^{-\frac{1}{2}} + M^{-\theta + \frac{1}{2}})$ and in mean-squares with $O_p(M^2(NT)^{-1} + M^{-2\theta + 1})$. Likewise, to obtain the limiting distributions of the global coefficients in $\hat{\alpha}_{i}^{(3)p}(t/T)$, one only needs to replace $p$ with $p^{(3)}$ and substitute $N_k$ with $N$ in Theorem \hyperref[sec:Theo_3]{3.5}.

\subsection{Panels with Time-Varying and Time-Constant Coefficients}

Despite the frequent occurrence of time-variant functional relationships in applied settings, some coefficients may nonetheless remain constant. To accommodate this, reorder $\bs{\beta}_i(t/T)$ such that $\bs{\beta}_i(t/T) = (\bs{\beta}_i^{(1)\prime}(t/T), \bs{\beta}_i^{(2)\prime})^\prime$, where the $p^{(1)} \times 1$ vector $\bs{\beta}_i^{(1)}(t/T)$ varies smoothly over time and the $p^{(2)} \times 1$ vector $\bs{\beta}^{(2)}$ is constant. After partitioning $\bs{x}_{it} = (\bs{x}_{it}^{(1)\prime}, \bs{x}_{it}^{(2)\prime})^\prime$ to conform with $\bs{\beta}_i(t/T)$, extend the criterion \eqref{eq:Obj_penalty} to
\begin{equation*}
    \label{eq:Obj_penalty_constant}
    \begin{split}
        \mathcal{F}_{NT}(\bs{\pi}^{(1)}, \bs{\beta}^{(2)}, \lambda) = \frac{1}{T} \sum_{i = 1}^{N} \sum_{t=1}^{T} & \left(\tilde{y}_{it} - \bs{\pi}_i^{(1) \prime} \tilde{\bs{z}}_{it}^{(1)} - \bs{\beta}_i^{(2) \prime} \tilde{\bs{x}}_{it}^{(2)} \right)^2                  \\
                                                                                                                  & \qquad \qquad \qquad \qquad + \frac{\lambda}{N} \sum_{i = 1}^{N - 1} \sum_{j=i+1}^{N} \dot{\omega}_{ij} \left\| \bs{\varpi}_i - \bs{\varpi}_j \right\|_2,
    \end{split}
\end{equation*}
where $\bs{\beta}^{(2)} = (\bs{\beta}^{(2)\prime}_1, \dots, \bs{\beta}^{(2)\prime}_N)^\prime$ and $\bs{\varpi}_i = \left(\bs{\pi}_i^{(1) \prime}, \bs{\beta}_i^{(2)\prime} \right)^\prime$. Furthermore, we now estimate the adaptive penalty weight $\dot{\omega}_{ij} = \left\| \dot{\bs{\varpi}}_i - \dot{\bs{\varpi}}_j \right\|_2^{-\kappa}$ as
\begin{equation*}
    \label{eq:Ini_estim_const}
    \dot{\bs{\varpi}}_i = \arg \min_{\varpi_i} \frac{1}{T} \sum_{t=1}^{T} \left(\tilde{y}_{it} - \bs{\pi}_i^{(1) \prime} \tilde{\bs{z}}_{it}^{(1)} - \bs{\beta}_i^{(2)\prime} \tilde{\bs{x}}_{it}^{(2)} \right)^2.
\end{equation*}

The \textit{post-Lasso} estimator is defined as
\begin{equation*}
    \label{eq:post-Lasso_const}
    \hat{\bs{\varpi}}^p = \arg \min_{\varpi} \frac{1}{T} \sum_{t=1}^{T} \left( \tilde{y}_{t} - \text{diag}(\tilde{\bs{Z}}_{\hat{G}_1 t}^{(1)}, \dots, \tilde{\bs{Z}}_{\hat{G}_{\hat{K}} t}^{(1)})^{\prime} \bs{\xi}^{(1)} - \text{diag}(\tilde{\bs{X}}_{\hat{G}_1 t}^{(2)}, \dots, \tilde{\bs{X}}_{\hat{G}_{\hat{K}} t}^{(2)})^{\prime} \bs{\alpha}^{(2)} \right)^2,
\end{equation*}
where $\tilde{\bs{Z}}_{\hat{G}_k t}^{(1)} = \{\tilde{\bs{z}}_{it}^{(1)} : i \in \hat{G}_k, 1 \leq i \leq N \}$ and $\tilde{\bs{X}}^{(2)}_{\hat{G}_k t} = \{\tilde{\bs{x}}^{(2)}_{it} : i \in \hat{G}_k, 1 \leq i \leq N \}$, $\hat{\bs{\varpi}}^p = (\hat{\bs{\varpi}}_{\hat{G}_1}^{p \prime}, \dots, \hat{\bs{\varpi}}_{\hat{G}_{\hat{K}}}^{p \prime})^\prime$ are $M p^{(1)} \times \hat{K}$ and $M p^{(2)} \times \hat{K}$ matrices collecting all observations pertaining to group $k$, respectively. Furthermore, $\hat{\bs{\varpi}}_{\hat{G}_k}^p = \left( \hat{\bs{\xi}}_{\hat{G}_k}^{(1)p \prime}, \hat{\alpha}_{\hat{G}_k}^{(2)p\prime} \right)^\prime$, ${\hat{\bs{\xi}}}^{(1)p} = (\hat{\bs{\xi}}^{(1){p}}_{\hat{G}_1}, \dots, {\hat{\bs{\xi}}^{(1){p}}_{\hat{G}_{\hat{K}}}} )$, $\hat{\bs{\alpha}}^{(2)p}_{\hat{\mathcal{G}}_{\hat{K}}} = \left( \hat{\bs{\alpha}}^{(2)p\prime}_{\hat{G}_1}, \dots, \hat{\bs{\alpha}}^{(2)p\prime}_{\hat{G}_{\hat{K}}} \right)^{\prime}$, and $\hat{\bs{\alpha}}_{\hat{G}_k}^{(1)p}(t/T) = \hat{\bs{\Xi}}_{\hat{G}_k}^{(1){p\prime}} \bs{b}(t/T)$. The remaining terms are as introduced in Section \ref{sec:ModelandEstimation} and above.

It is straightforward to see that the asymptotic properties of $\bs{\pi}_i^{(1)}$ remain unchanged from Section \ref{sec:Asymptotics} when substituting $p$ with $p^{(1)}$. The same holds for $\hat{\bs{\beta}}_i^{(2)}$ when taking $M = 1$, $p = p^{(2)}$, and treating $M$ as constant. The limiting distributions in Theorem \ref{sec:Theo_3} apply, when replacing $p$ with $p^{(l)}, \, l=\{1,2\}$, respectively, and when fixing $M=1$ for the vector of time-invariant coefficients.

\subsection{Unbalanced Panels}

Panel datasets with varying numbers of observations among individuals are ubiquitous in empirical applications. We accommodate such unbalanced panels in the \textit{FUSE-TIME} framework by adjusting the DGP in \eqref{eq:DGP_spline} to
\begin{equation*}
    \label{eq:DGP_spline_unbalanced}
    y_{it} = \gamma_i + \bs{\pi}_i^{0 \prime} \bs{z}_{it} + u_{it}, \quad i = 1, \dots, N, \quad t = t_i, \dots, T_i,
\end{equation*}
where $t_i$ indicates the individual start of the observational period for each cross-sectional unit, $T_i$ the end, and $\mathcal{T}_i = T_i - t_i - 1$ the number of observed time periods per unit. It is straightforward to extend the objective function \eqref{eq:Obj_penalty} to this scenario by rewriting
\begin{equation*}
    \label{eq:Obj_penalty_unbalanced}
    \mathcal{F}_{NT}({\bs{\pi}}, \lambda) = \frac{1}{\mathcal{T}_i} \sum_{i = 1}^{N} \sum_{t=t_i}^{T_i} \left(\tilde{y}_{it} - \bs{\pi}_i^\prime \tilde{\bs{z}}_{it} \right)^2 + \frac{\lambda}{N} \sum_{i = 1}^{N - 1} \sum_{j=i+1}^{N} \dot{\omega}_{ij} \left\| \bs{\pi}_i - \bs{\pi}_j \right\|_2,
\end{equation*}
with $\tilde{a}_{it} = a_{it} - \mathcal{T}_i^{-1} \sum_{t=t_i}^{T_i} a_{it}$ for $a_{it} = \{y_{it}, \bs{z}_{it}\}$. The remaining notation is left unchanged.

Similarly, the \textit{post-Lasso} is given by
\begin{equation*}
    \label{eq:post-Lasso_unbalanced}
    \hat{\bs{\xi}}^p_{\hat{G}_k} = \left( \sum_{i \in \hat{G}_k} \sum_{t = t_i}^{T_i} \tilde{\bs{z}}_{it} \tilde{\bs{z}}_{it}^\prime \right)^{-1} \sum_{i \in \hat{G}_k} \sum_{t = t_i}^{T_i} \tilde{\bs{z}}_{it} \tilde{y}_{it}.
\end{equation*}

In order to study the asymptotic behavior, let $T_{\min} = \min_{i} \mathcal{T}_i$ and assume $T_{\min} \to \infty$. When substituting $T$ with $T_{\min}$, all proofs and assumptions carry seamlessly over to unbalanced panel datasets \citep[see][sec. 5.2]{su2019sieve}. Notice that this extension also applies to missing observations in the middle of the panel, i.e. $t = t_i, \dots, t_{i+j}, t_{i+j+l}, \dots, T_i$, for some integers $0 < j < T_i - 2$ and $1 < l < T_i - i - j$, since nonparametric splines implicitly interpolate missing values. Furthermore, the \textit{post-Lasso} pools homogeneous cross-sectional units. In consequence, remaining group members compensate for a missing observation in individual series. These two aspects make our methodology particularly powerful in empirical applications with unbalanced panel datasets.

\section{Additional Simulation Studies and Details}
\label{sec:Appendix_Sim}

Subsections \ref{sec:Appendix_corr_errors} and \ref{sec:Appendix_NA} present the simulation study results when the errors are serially correlated and when $30\%$ of observations are randomly discarded, respectively. Subsection \ref{sec:Appendix_C_lasso} provides details on the implementation of the \textit{time-varying C-Lasso} benchmark model.

Table \ref{tab:lambda_values} reports the tuning parameter candidate values employed in the Monte Carlo experiments.
\begin{table}[h]
    \centering
    \caption{$\lambda$ tuning parameter candidate values in the simulation studies}
    \begin{tabular}{cc|cc|cc|cc}
        \hline \hline
        \multicolumn{2}{c|}{DGP 1} & \multicolumn{2}{c|}{Section \ref{sec:Simulation_results}} & \multicolumn{2}{c|}{$AR(1)$ errors (\ref{sec:Appendix_corr_errors})} & \multicolumn{2}{c}{\shortstack{30\% of the sample                                                                                               \\ discarded (\ref{sec:Appendix_NA})}} \\
        N                          & T                                                         & $\underline{\lambda}$                                                & $\overline{\lambda}$                              & $\underline{\lambda}$ & $\overline{\lambda}$ & $\underline{\lambda}$ & $\overline{\lambda}$ \\
        \hline
        50                         & 50                                                        & 0.1                                                                  & 50                                                & 0.1                   & 20                   & 0.1                   & 10                   \\
        100                        & 50                                                        & 0.1                                                                  & 50                                                & 0.1                   & 20                   & 0.1                   & 10                   \\
        50                         & 100                                                       & 0.1                                                                  & 50                                                & 0.1                   & 20                   & 0.1                   & 10                   \\
        100                        & 100                                                       & 0.1                                                                  & 50                                                & 0.1                   & 20                   & 0.1                   & 10                   \\
        \hline
        \multicolumn{2}{c|}{DGP 2} & \multicolumn{2}{c|}{Section \ref{sec:Simulation_results}} & \multicolumn{2}{c|}{$AR(1)$ errors (\ref{sec:Appendix_corr_errors})} & \multicolumn{2}{c}{\shortstack{30\% of the sample                                                                                               \\ discarded (\ref{sec:Appendix_NA})}} \\
        N                          & T                                                         & $\underline{\lambda}$                                                & $\overline{\lambda}$                              & $\underline{\lambda}$ & $\overline{\lambda}$ & $\underline{\lambda}$ & $\overline{\lambda}$ \\
        \hline
        50                         & 50                                                        & 10                                                                   & 35                                                & 30                    & 75                   & 15                    & 47                   \\
        100                        & 50                                                        & 10                                                                   & 35                                                & 25                    & 65                   & 10                    & 30                   \\
        50                         & 100                                                       & 1                                                                    & 20                                                & 8                     & 25                   & 10                    & 60                   \\
        100                        & 100                                                       & 1                                                                    & 20                                                & 18                    & 37                   & 10                    & 60                   \\
        \hline
        \multicolumn{2}{c|}{DGP 3} & \multicolumn{2}{c|}{Section \ref{sec:Simulation_results}} & \multicolumn{2}{c|}{$AR(1)$ errors (\ref{sec:Appendix_corr_errors})} & \multicolumn{2}{c}{\shortstack{30\% of the sample                                                                                               \\ discarded (\ref{sec:Appendix_NA})}} \\
        N                          & T                                                         & $\underline{\lambda}$                                                & $\overline{\lambda}$                              & $\underline{\lambda}$ & $\overline{\lambda}$ & $\underline{\lambda}$ & $\overline{\lambda}$ \\
        \hline
        50                         & 50                                                        & 0.01                                                                 & 15                                                & 0.1                   & 20                   & 0.1                   & 20                   \\
        100                        & 50                                                        & 0.01                                                                 & 15                                                & 4                     & 20                   & 5                     & 25                   \\
        50                         & 100                                                       & 0.01                                                                 & 15                                                & 0.1                   & 20                   & 0.1                   & 20                   \\
        100                        & 100                                                       & 0.01                                                                 & 15                                                & 0.1                   & 8                    & 0.1                   & 9                    \\
        \hline \hline
    \end{tabular}%
    \label{tab:lambda_values}%
    \begin{tablenotes}
        \small
        \item
        \textit{Notes}: Upper and lower limits for the sequences of candidate $\lambda$ penalty tuning parameter values. The sequences are of length 50 and run from $\underline{\lambda}$ to $\overline{\lambda}$.
    \end{tablenotes}
\end{table}%


\subsection{DGPs with Serially Correlated Errors}
\label{sec:Appendix_corr_errors}

\begin{table}
    \centering
    \caption{Clustering accuracy with serially correlated errors}
    \begin{tabular}{ccccccc}
        \hline \hline
                                  & N   & T   & Freq. $\hat{K} = K_0$ & Freq. $\hat{\mathcal{G}}_{\hat{K}} = \mathcal{G}^0_{K_0}$ & ARI   & $\bar{K}$ \\
        \midrule
        \multirow{4}[2]{*}{DGP 1} & 50  & 50  & 0.680                 & 0.190                                                     & 0.855 & 3.270     \\
                                  & 50  & 100 & 0.927                 & 0.753                                                     & 0.978 & 3.080     \\
                                  & 100 & 50  & 0.680                 & 0.077                                                     & 0.809 & 2.753     \\
                                  & 100 & 100 & 1.000                 & 0.720                                                     & 0.990 & 3.000     \\
        \midrule
        \multirow{4}[2]{*}{DGP 2} & 50  & 50  & 0.140                 & 0.027                                                     & 0.577 & 2.130     \\
                                  & 50  & 100 & 0.757                 & 0.647                                                     & 0.878 & 2.760     \\
                                  & 100 & 50  & 0.073                 & 0.007                                                     & 0.501 & 1.940     \\
                                  & 100 & 100 & 0.512                 & 0.383                                                     & 0.768 & 2.510     \\
        \midrule
        \multirow{4}[2]{*}{DGP 3} & 50  & 50  & 0.567                 & 0.030                                                     & 0.708 & 2.680     \\
                                  & 50  & 100 & 0.903                 & 0.560                                                     & 0.937 & 2.990     \\
                                  & 100 & 50  & 0.160                 & 0.007                                                     & 0.550 & 2.130     \\
                                  & 100 & 100 & 0.933                 & 0.433                                                     & 0.951 & 2.970     \\
        \hline \hline
    \end{tabular}%
    \label{tab:sim_grouping_ar}%
    \begin{tablenotes}
        \small
        \item
        \textit{Notes}: Frequency of \textit{FUSE-TIME} obtaining the correct number of groups $\hat{K} = K_0$ and the correct grouping $\hat{\mathcal{G}}_{\hat{K}} = \mathcal{G}^0_{K_0}$, the ARI, and the average estimated number of total groups $\bar{K}$ based on a Monte Carlo study with 300 replications. The errors are serially correlated with an autoregressive coefficient of 0.3.
    \end{tablenotes}
\end{table}%

\begin{table}
    \centering
    \caption{\textit{RMSE} of coefficient estimates with serially correlated errors}
    \begin{tabular}{ccccccc}
        \hline
        \hline
                                  &                                                 & N   & T   & \textit{PSE} & \textit{post-Lasso} & oracle \\
        \midrule
        \multirow{4}[2]{*}{DGP 1} & \multirow{4}[2]{*}{$\hat{\alpha}_{k,0} (t/T) $} & 50  & 50  & 0.426        & 0.282               & 0.188  \\
                                  &                                                 & 50  & 100 & 0.346        & 0.226               & 0.165  \\
                                  &                                                 & 100 & 50  & 0.470        & 0.290               & 0.163  \\
                                  &                                                 & 100 & 100 & 0.339        & 0.212               & 0.15   \\
        \midrule
        \multirow{8}[2]{*}{DGP 2} & \multirow{4}[2]{*}{$\hat{\alpha}_{k,1} (t/T) $} & 50  & 50  & 0.357        & 0.225               & 0.161  \\
                                  &                                                 & 50  & 100 & 0.275        & 0.158               & 0.137  \\
                                  &                                                 & 100 & 50  & 0.359        & 0.238               & 0.135  \\
                                  &                                                 & 100 & 100 & 0.309        & 0.149               & 0.095  \\
        \cmidrule{2-7}            & \multirow{4}[2]{*}{$\hat{\alpha}_{k,2} (t/T) $} & 50  & 50  & 0.454        & 0.279               & 0.161  \\
                                  &                                                 & 50  & 100 & 0.314        & 0.162               & 0.120  \\
                                  &                                                 & 100 & 50  & 0.465        & 0.315               & 0.121  \\
                                  &                                                 & 100 & 100 & 0.385        & 0.184               & 0.088  \\
        \midrule
        \multirow{4}[2]{*}{DGP 3} & \multirow{4}[2]{*}{$\hat{\alpha}_{k,3} (t/T)$}  & 50  & 50  & 0.419        & 0.269               & 0.216  \\
                                  &                                                 & 50  & 100 & 0.339        & 0.230               & 0.222  \\
                                  &                                                 & 100 & 50  & 0.498        & 0.321               & 0.212  \\
                                  &                                                 & 100 & 100 & 0.364        & 0.226               & 0.219  \\
        \hline
        \hline
    \end{tabular}%
    \label{tab:RMSE_ar}%
    \begin{tablenotes}
        \small
        \item
        \textit{Notes}: \textit{RMSE} of the \textit{FUSE-TIME} \textit{PSE}, the \textit{FUSE-TIME} \textit{post-Lasso}, and an infeasible oracle estimator based on a Monte Carlo study with 300 replication. The errors are serially correlated with an autoregressive coefficient of 0.3.
    \end{tablenotes}
\end{table}%

Tables \ref{tab:sim_grouping_ar} and \ref{tab:RMSE_ar} report the Monte Carlo simulation study results of \textit{FUSE-TIME} when the innovations are constructed as $\epsilon_{it} = 0.3 \epsilon_{it-1} + e_{it}$, where $e_{it} \sim i.i.d. N(0,1)$. Serial correlation in the errors to such an extent does not infringe on Assumption \superref{line:A1}{1(ii)}, as the process still remains stationary. However, as serial correlation introduces additional estimation uncertainty and previous simulation studies in Section \ref{sec:Simulation} show that such uncertainty complicates the clustering mechanism, we choose to reduce the number of interior knots $M^*$ for this exercise. We select $M^* = 2$ for DGP 1 and $M^* = 1$ for DGPs 2 and 3.

When $T = 100$, the simulation results mirror largely the ones reported in Section \ref{sec:Simulation_results}. However, as serially correlated errors effectively reduce the informational value contained in a time series, the performance in small samples is notably reduced. This does not just concern the grouping mechanism and the resulting estimation inaccuracy, but also the \textit{RMSE} of the infeasible oracle estimator.


\subsection{DGPs with Unbalanced Panels}
\label{sec:Appendix_NA}

Unbalanced panel datasets are a frequent occurrence in empirical work. Subsequently, we re-run our simulation study after discarding a randomly drawn 30\% subset of each time series. This exercise gives an understanding how the clustering performance and the estimation of the functional coefficients behave when some observations have to be interpolated. Furthermore, in line with the decreased informational value of an unbalanced panel dataset, we again deviate from the heuristic for the number of internal knots $M^*$ provided in section \ref{sec:Simulation} and set $M^* = 2$ for DGP 1 and $M^* = 1$ for the remaining two DGPs. Table \ref{tab:sim_grouping_na} displays the clustering performance and Table \ref{tab:RMSE_na} the \textit{RMSE} values.

\begin{table}[t]
    \centering
    \caption{Clustering metrics in an unbalanced panel}
    \begin{tabular}{ccccccc}
        \hline \hline
                                  & N   & T   & Freq. $\hat{K} = K_0$ & Freq. $\hat{\mathcal{G}}_{\hat{K}} = \mathcal{G}^0_{K_0}$ & ARI   & $\bar{K}$ \\
        \midrule
        \multirow{4}[2]{*}{DGP 1} & 50  & 50  & 0.907                 & 0.620                                                     & 0.961 & 3.093     \\
                                  & 50  & 100 & 1.000                 & 0.973                                                     & 0.998 & 3.000     \\
                                  & 100 & 50  & 0.990                 & 0.543                                                     & 0.976 & 2.990     \\
                                  & 100 & 100 & 1.000                 & 0.980                                                     & 0.999 & 3.000     \\
        \midrule
        \multirow{4}[2]{*}{DGP 2} & 50  & 50  & 0.477                 & 0.090                                                     & 0.713 & 2.530     \\
                                  & 50  & 100 & 0.983                 & 0.887                                                     & 0.985 & 2.997     \\
                                  & 100 & 50  & 0.337                 & 0.050                                                     & 0.620 & 2.260     \\
                                  & 100 & 100 & 0.990                 & 0.810                                                     & 0.988 & 2.990     \\
        \midrule
        \multirow{4}[2]{*}{DGP 3} & 50  & 50  & 0.700                 & 0.030                                                     & 0.734 & 3.060     \\
                                  & 50  & 100 & 0.910                 & 0.467                                                     & 0.939 & 3.057     \\
                                  & 100 & 50  & 0.200                 & 0.003                                                     & 0.522 & 2.170     \\
                                  & 100 & 100 & 0.853                 & 0.277                                                     & 0.906 & 2.873     \\
        \hline \hline
    \end{tabular}%
    \label{tab:sim_grouping_na}%
    \begin{tablenotes}
        \small
        \item
        \textit{Notes}: Frequency of \textit{FUSE-TIME} obtaining the correct number of groups $\hat{K} = K_0$ and the correct grouping $\hat{\mathcal{G}}_{\hat{K}} = \mathcal{G}^0_{K_0}$, the ARI, and the average estimated number of total groups $\bar{K}$ based on a Monte Carlo study with 300 replications. 30\% of observations randomly discarded to create an unbalanced panel dataset.
    \end{tablenotes}
\end{table}%

\begin{table}[t]
    \centering
    \caption{\textit{RMSE} of coefficient estimates in an unbalanced panel}
    \begin{tabular}{ccccccc}
        \hline
        \hline
                                  &                                                & N   & T   & \textit{PSE} & \textit{post-Lasso} & oracle \\
        \midrule
        \multirow{4}[2]{*}{DGP 1} & \multirow{4}[2]{*}{$\hat{\alpha}_{k,0} (t/T)$} & 50  & 50  & 0.370        & 0.235               & 0.226  \\
                                  &                                                & 50  & 100 & 0.276        & 0.212               & 0.152  \\
                                  &                                                & 100 & 50  & 0.371        & 0.218               & 0.153  \\
                                  &                                                & 100 & 100 & 0.259        & 0.203               & 0.142  \\
        \midrule
        \multirow{8}[2]{*}{DGP 2} & \multirow{4}[2]{*}{$\hat{\alpha}_{k,1} (t/T)$} & 50  & 50  & 0.316        & 0.201               & 0.142  \\
                                  &                                                & 50  & 100 & 0.233        & 0.145               & 0.123  \\
                                  &                                                & 100 & 50  & 0.334        & 0.225               & 0.124  \\
                                  &                                                & 100 & 100 & 0.241        & 0.138               & 0.112  \\
        \cmidrule{2-7}            & \multirow{4}[2]{*}{$\hat{\alpha}_{k,2} (t/T)$} & 50  & 50  & 0.403        & 0.240               & 0.149  \\
                                  &                                                & 50  & 100 & 0.267        & 0.138               & 0.124  \\
                                  &                                                & 100 & 50  & 0.413        & 0.278               & 0.127  \\
                                  &                                                & 100 & 100 & 0.273        & 0.127               & 0.112  \\
        \midrule
        \multirow{4}[2]{*}{DGP 3} & \multirow{4}[1]{*}{$\hat{\alpha}_{k,3} (t/T)$} & 50  & 50  & 0.304        & 0.162               & 0.109  \\
                                  &                                                & 50  & 100 & 0.207        & 0.076               & 0.064  \\
                                  &                                                & 100 & 50  & 0.402        & 0.228               & 0.068  \\
                                  &                                                & 100 & 100 & 0.236        & 0.076               & 0.053  \\
        \hline
        \hline
    \end{tabular}%
    \label{tab:RMSE_na}%
    \begin{tablenotes}
        \small
        \item
        \textit{Notes}: \textit{RMSE} of the \textit{FUSE-TIME} \textit{PSE}, the \textit{FUSE-TIME} \textit{post-Lasso}, and an infeasible oracle estimator based on a Monte Carlo study with 300 replication. 30\% of observations randomly discarded to create an unbalanced panel dataset.
    \end{tablenotes}
\end{table}%

In large samples both the clustering performance and the \textit{RMSE} are very similar to the ones reported in Section \ref{sec:Simulation_results}. However, just like the in the previous subsection, small samples suffer from the reduced number of observations. This dynamic is particularly apparent in DGP 2. Omitting observations, even on the interior of time series, seems akin to reducing $T$ with respect to the clustering performance and estimation accuracy of \textit{FUSE-TIME}.


\subsection{Implementation of the \textit{Time-Varying C-Lasso}}
\label{sec:Appendix_C_lasso}

In Section \ref{sec:Simulation} we employ the \textit{time-varying C-Lasso} by \citet{su2019sieve} as a benchmark, using Matlab replication files kindly provided by the authors. The \textit{time-varying C-Lasso} and our methodology are exposed to the same simulated data, the DGPs following \citet[sec. 6]{su2019sieve}. The settings of the \textit{time-varying C-Lasso} are specified as documented in \citet[sec. 6]{su2019sieve}: We set the polynomial degree to $d = 3$, the number of interior spline knots according to the heuristic $M^* = \lfloor (NT)^{1/6} \rfloor$ and the \textit{C-Lasso} penalty tuning parameter to $\lambda = (NT)^{-(2K + 3)/24}$. Furthermore, as the \textit{C-Lasso} requires an explicit specification of the number of groups, we evaluate their IC for $K = \{2, 3, 4\}$, with $\rho$ in the IC equaling $\rho = M^* \log(NT) / (NT)$ \citep[cf.][eq. 4.9]{su2019sieve}.


\subsection{Benchmarking the \textit{FUSE-TIME} and \textit{C-Lasso} Implementation Speed}
\label{sec:Appendix_implementation_speed}

To provide a feel of the computational load, we benchmark the computational speed of both the \textit{time-varying C-Lasso} and our \textit{FUSE-TIME} routines on the first 100 out of the 300 simulated sample paths of DGP 1 with $N = 50$ and $T = 50$ (see Section \ref{sec:Simulation_DGP}).

The settings for each model are identical to Section \ref{sec:Simulation}. The \textit{FUSE-TIME} is specified as in Section \ref{sec:Appendix_Algo}, using the function \texttt{fuse\_time()} of our companion R-package \texttt{PAGFL} \citep{haimerl2025pagfl} and searching over the 50 $\lambda$ tuning parameter values given in Table \ref{tab:lambda_values}. We refer to the replication files available at \href{https://github.com/Paul-Haimerl/replication-tv-pagfl}{\texttt{github.com/Paul-Haimerl/replication-tv-pagfl}} for the exact implementation. The \textit{time-varying C-Lasso} implementation is detailed in Section \ref{sec:Appendix_C_lasso}.

Table \ref{tab:speed} presents summary statistics of the computing time of each software routine in seconds on a Lenovo Thinkpad T14 (64-bit Windows 11 E) with a quad-core AMD Ryzen 7 Pro 7840U CPU and 16 GB of memory. Since the computational speed depends on whether a convergence criterion is met or the maximum number of iterations is reached, we also report the proportion of sample paths for which convergence was achieved.
\begin{table}[t]
    \centering
    \caption{Implementation speed benchmark}
    \begin{tabular}{lcccccc}
        \hline
        \hline
                                      & Min     & Q25      & Median   & Q75      & Max      & Freq. converged \\
        \hline
        \textit{\textit{FUSE-TIME}}   & 4.606s  & 9.600s   & 15.240s  & 39.345s  & 259.493s & 1.000           \\
        \textit{Time-varying C-Lasso} & 77.769s & 116.855s & 141.435s & 187.787s & 387.950s & 0.060           \\
        \hline
        \hline
    \end{tabular}%
    \label{tab:speed}%
    \begin{tablenotes}
        \small
        \item
        \textit{Notes}: Computing time for each software routine for the first 100 simulated sample paths with $N=50$ and $T=50$ and the percentage amount of sample paths for which numerical convergence of the optimization routines was achieved.
    \end{tablenotes}
\end{table}%

The \textit{FUSE-TIME} implementation is approximately nine times faster than the \textit{time-varying C-Lasso}. This is, although \textit{FUSE-TIME} satisfies the convergence criterion for every sample path, whereas the maximum number of iterations, set by the original authors in \citet{su2019sieve}, is exceeded in 94\% of cases. If we were to allow the numerical algorithm of the \textit{time-varying C-Lasso} to fully converge, the computational load would increase significantly. In addition, the 50 considered $\lambda$ values of \textit{FUSE-TIME} lead to groupings ranging from a homogeneous panel ($\hat{K} = 1$) to an entirely heterogenous one ($\hat{K} = N$), whereas the \textit{time-varying C-Lasso} considers only $K = \{2,3,4\}$.

\section{Details on the Empirical Illustration}
\label{sec:empirical_appendix}

\begin{figure}[t]
        \centering
        \includegraphics[width=15cm]{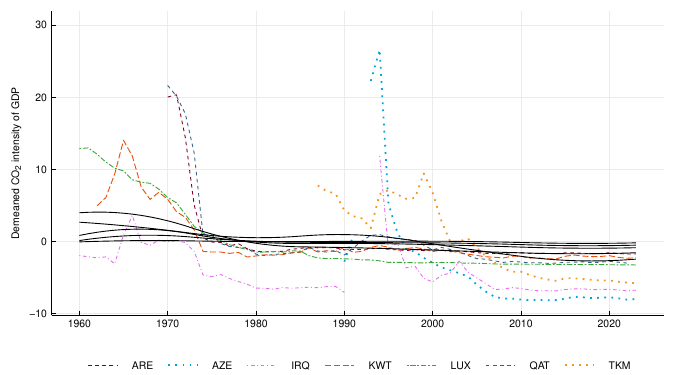}
        \caption[caption]{\small Demeaned CO\textsubscript{2} intensity of countries excluded from the analysis (colored) and the estimated group-specific trend functions (black, solid) based on the remaining panel dataset. The 1991 and 1992 observations for Iraq (1991: 103.51, 1992: 117.79) in addition to the 1992 and 1993 observations for Azerbaijan (1992: 99.06, 1993: 53.88) are omitted from the figure for ease of exposition.}
        \label{fig:outlier_countries}
\end{figure}

The empirical illustration focuses on the 100 largest economies by GDP in 2022. The Global Carbon Budget provides CO\textsubscript{2} emission data for 99 of these economies, not tracking Puerto Rico. Additionally, we omit Azerbaijan (AZE), Iraq (IRQ), Kuwait (KWT), Luxembourg (LUX), Qatar (QAT), Turkmenistan (TKM), and the United Arab Emirates (ARE) from our study. These countries exhibit severe outliers and pronounced idiosyncratic volatility in their CO\textsubscript{2} intensity time series, predominantly attributable to extraordinary geopolitical events, such as conflicts and oil shocks, or unique economic structure, such as tax havens or petro-economies. Figure \ref{fig:outlier_countries} makes it apparent that the CO\textsubscript{2} intensity trajectories of these discarded economies diverge markedly from one another and from the estimated trend functions, further justifying their exclusion.

Table \ref{tab:empirics_summarystats} presents descriptive statistics for the final sample of CO\textsubscript{2} emission intensities. Table \ref{tab:Group_estim} provides a detailed report of the estimated group structure, along with the group compositions in terms of the 92 countries included in the sample. 

When estimating the empirical model, we consider the following candidate values of $d$, $M^*$, and $\lambda$: $d=\{2, 3, 4, 5\}$, $M^* = \{1, 2, \dots, 5\}$, $\lambda \in [0.01, 1.5]$ in increments of 0.01.

\begin{table}[t]
        \centering
        \caption{Descriptive statistics of the CO\textsubscript{2} intensity panel dataset}
        \begin{tabular}{lccccccc}
                \hline \hline
                                              & Mean    & Std.     & Min   & Q25    & Median & Q75     & Max       \\
                \hline
                CO\textsubscript{2}           & 249.916 & 849.137  & 0.026 & 10.414 & 42.67  & 152.551 & 11953.218 \\
                GDP                           & 395.621 & 1521.375 & 0.063 & 11.349 & 49.997 & 217.156 & 27360.935 \\
                CO\textsubscript{2} intensity & 1.381   & 1.696    & 0.037 & 0.358  & 0.784  & 1.675   & 13.685    \\
                Observational horizon         & 56.511  & 12.48    & 29    & 55.5   & 64     & 64      & 64        \\
                \hline \hline
        \end{tabular}%
        \label{tab:empirics_summarystats}%
        \begin{tablenotes}
                \small
                \item
                \textit{Notes}: Summary statistics on the panel dataset employed in the empirical illustration. CO\textsubscript{2} is measured in million tonnes, GDP in billion 2024 U.S. dollars, and the observational horizon in years.
        \end{tablenotes}
\end{table}%

\begin{table}[t]
        \centering
        \caption{Group structure in the CO\textsubscript{2} emission intensity of GDP}
        \begin{tabular}{llllll}
                \hline \hline
                \multirow{4}[0]{*}{Group 1 (18)} & Angola             & China       & Kazakhstan  & Serbia       & Uruguay        \\
                                                 & Belarus            & Czechia     & Poland      & Slovakia     & Uzbekistan     \\
                                                 & Bulgaria           & Estonia     & Romania     & Sweden       &                \\
                                                 & Chile              & Italy       & Russia      & Ukraine      &                \\
                \hline
                \multirow{6}[0]{*}{Group 2 (28)} & Algeria            & Greece      & Libya       & Panama       & Switzerland    \\
                                                 & Argentina          & Hong Kong   & Malaysia    & Philippines  & Tunisia        \\
                                                 & Bolivia            & India       & Mexico      & Portugal     & Turkey         \\
                                                 & Dem. Rep. Congo    & Indonesia   & New Zealand & Saudi Arabia & Vietnam        \\
                                                 & Egypt              & Israel      & Norway      & Singapore    &                \\
                                                 & Finland            & Kenya       & Pakistan    & Spain        &                \\
                \hline
                \multirow{5}[0]{*}{Group 3 (24)} & Australia          & Colombia    & Hungary     & Lithuania    & South Africa   \\
                                                 & Austria            & Croatia     & Iran        & Myanmar      & South Korea    \\
                                                 & Bahrain            & Denmark     & Ireland     & Netherlands  & USA            \\
                                                 & Belgium            & France      & Japan       & Peru         & United Kingdom \\
                                                 & Canada             & Germany     & Latvia      & Slovenia     &                \\
                \hline
                \multirow{2}[0]{*}{Group 4 (8)}  & Bangladesh         & Cameroon    & Ethiopia    & Nepal        &                \\
                                                 & Brazil             & Ecuador     & Ghana       & Tanzania     &                \\
                \hline
                \multirow{3}[0]{*}{Group 5 (14)} & Costa Rica         & El Salvador & Morocco     & Paraguay     & Thailand       \\
                                                 & Côte d'Ivoire      & Guatemala   & Nigeria     & Sri Lanka    & Uganda         \\
                                                 & Dominican Republic & Jordan      & Oman        & Sudan        &                \\
                \hline \hline
        \end{tabular}%
        \label{tab:Group_estim}%
        \begin{tablenotes}
                \small
                \item
                \textit{Notes}: Estimated group structure $\hat{\mathcal{G}}_{\hat{K}}$ in the trends of the CO\textsubscript{2} emission intensity. Group cardinalities are in parenthesis.
        \end{tablenotes}
\end{table}%


\end{document}